\documentclass[aps,prc,amsfonts,twocolumn,superscriptaddress,showpacs,showkeys,nofootinbib]{revtex4-1}
\pdfoutput=1
\usepackage{color}
\usepackage{graphics,epsfig}
\usepackage{verbatim}

\usepackage{amsmath}
\usepackage{amssymb}
\usepackage{amstext}
\usepackage{mathrsfs}
\usepackage{latexsym}
\usepackage{float}
\usepackage{subfigure}
\usepackage{wasysym}
\usepackage{fancyhdr}

\usepackage{epstopdf}

\newcommand {\be} {\begin{equation}}
\newcommand {\ba} {\begin{eqnarray}}
\newcommand {\ee} {\end{equation}}
\newcommand {\ea} {\end{eqnarray}}

\def\etal{\textit{et al.}}

\begin{document}
\hyphenation{RCSLACPOL}

\newcommand*{\ANL}{Argonne National Laboratory, Argonne, Illinois 60439, USA}
\newcommand*{\ANLindex}{1}
\affiliation{\ANL}
\newcommand*{\ASU}{Arizona State University, Tempe, Arizona 85287-1504, USA}
\newcommand*{\ASUindex}{2}
\affiliation{\ASU}
\newcommand*{\CSUDH}{California State University, Dominguez Hills, Carson, California 90747, USA}
\newcommand*{\CSUDHindex}{3}
\affiliation{\CSUDH}
\newcommand*{\CMU}{Carnegie Mellon University, Pittsburgh, Pennsylvania 15213, USA}
\newcommand*{\CMUindex}{4}
\affiliation{\CMU}
\newcommand*{\CUA}{Catholic University of America, Washington, D.C. 20064, USA}
\newcommand*{\CUAindex}{5}
\affiliation{\CUA}
\newcommand*{\SACLAY}{Irfu/SPhN, CEA, Universit\'e Paris-Saclay, 91191 Gif-sur-Yvette, France}
\newcommand*{\SACLAYindex}{6}
\affiliation{\SACLAY}
\newcommand*{\CNU}{Christopher Newport University, Newport News, Virginia 23606, USA}
\newcommand*{\CNUindex}{7}
\affiliation{\CNU}
\newcommand*{\UCONN}{University of Connecticut, Storrs, Connecticut 06269, USA}
\newcommand*{\UCONNindex}{8}
\affiliation{\UCONN}
\newcommand*{\FU}{Fairfield University, Fairfield, Cinnecticut 06824, USA}
\newcommand*{\FUindex}{9}
\affiliation{\FU}
\newcommand*{\FERRARAU}{Universita' di Ferrara , 44121 Ferrara, Italy}
\newcommand*{\FERRARAUindex}{10}
\affiliation{\FERRARAU}
\newcommand*{\FIU}{Florida International University, Miami, Florida 33199, USA}
\newcommand*{\FIUindex}{11}
\affiliation{\FIU}
\newcommand*{\FSU}{Florida State University, Tallahassee, Florida 32306, USA}
\newcommand*{\FSUindex}{12}
\affiliation{\FSU}
\newcommand*{\Genova}{Universit$\grave{a}$ di Genova, 16146 Genova, Italy}
\newcommand*{\Genovaindex}{13}
\affiliation{\Genova}
\newcommand*{\GWUI}{The George Washington University, Washington, DC 20052, USA}
\newcommand*{\GWUIindex}{14}
\affiliation{\GWUI}
\newcommand*{\GEORGIAC}{Georgia College, Milledgeville, Georgia 31061, USA}
\newcommand*{\GEORGIACIindex}{15}
\affiliation{\GEORGIAC}
\newcommand*{\ISU}{Idaho State University, Pocatello, Idaho 83209, USA}
\newcommand*{\ISUindex}{16}
\affiliation{\ISU}
\newcommand*{\INFNFE}{INFN, Sezione di Ferrara, 44100 Ferrara, Italy}
\newcommand*{\INFNFEindex}{17}
\affiliation{\INFNFE}
\newcommand*{\INFNFR}{INFN, Laboratori Nazionali di Frascati, 00044 Frascati, Italy}
\newcommand*{\INFNFRindex}{18}
\affiliation{\INFNFR}
\newcommand*{\INFNGE}{INFN, Sezione di Genova, 16146 Genova, Italy}
\newcommand*{\INFNGEindex}{19}
\affiliation{\INFNGE}
\newcommand*{\INFNRO}{INFN, Sezione di Roma Tor Vergata, 00133 Rome, Italy}
\newcommand*{\INFNROindex}{20}
\affiliation{\INFNRO}
\newcommand*{\INFNTUR}{INFN, Sezione di Torino, 10125 Torino, Italy}
\newcommand*{\INFNTURindex}{21}
\affiliation{\INFNTUR}
\newcommand*{\ORSAY}{Institut de Physique Nucl\'eaire, CNRS/IN2P3 and Universit\'e Paris Sud, Orsay, France}
\newcommand*{\ORSAYindex}{22}
\affiliation{\ORSAY}
\newcommand*{\ITEP}{Institute of Theoretical and Experimental Physics, Moscow, 117259, Russia}
\newcommand*{\ITEPindex}{23}
\affiliation{\ITEP}
\newcommand*{\JMU}{James Madison University, Harrisonburg, Virginia 22807, USA}
\newcommand*{\JMUindex}{24}
\affiliation{\JMU}
\newcommand*{\KNU}{Kyungpook National University, Daegu 41566, Republic of Korea}
\newcommand*{\KNUindex}{25}
\affiliation{\KNU}
\newcommand*{\MISS}{Mississippi State University, Mississippi State, Mississippi 39762-5167, USA}
\newcommand*{\MISSindex}{26}
\affiliation{\MISS}
\newcommand*{\UNH}{University of New Hampshire, Durham, New Hampshire 03824-3568, USA}
\newcommand*{\UNHindex}{27}
\affiliation{\UNH}
\newcommand*{\NSU}{Norfolk State University, Norfolk, Virginia 23504, USA}
\newcommand*{\NSUindex}{28}
\affiliation{\NSU}
\newcommand*{\OAK}{Oak Ridge National Laboratory, Oak Ridge, Tennessee  37830, USA}
\newcommand*{\OAKindex}{29}
\affiliation{\OAK}
\newcommand*{\OHIOU}{Ohio University, Athens, Ohio  45701, USA}
\newcommand*{\OHIOUindex}{30}
\affiliation{\OHIOU}
\newcommand*{\ODU}{Old Dominion University, Norfolk, Virginia 23529, USA}
\newcommand*{\ODUindex}{31}
\affiliation{\ODU}
\newcommand*{\RPI}{Rensselaer Polytechnic Institute, Troy, New York 12180-3590, USA}
\newcommand*{\RPIindex}{32}
\affiliation{\RPI}
\newcommand*{\URICH}{University of Richmond, Richmond, Virginia 23173, USA}
\newcommand*{\URICHindex}{33}
\affiliation{\URICH}
\newcommand*{\ROMAII}{Universita' di Roma Tor Vergata, 00133 Rome, Italy}
\newcommand*{\ROMAIIindex}{34}
\affiliation{\ROMAII}
\newcommand*{\MSU}{Skobeltsyn Institute of Nuclear Physics, Lomonosov Moscow State University, 119234 Moscow, Russia}
\newcommand*{\MSUindex}{35}
\affiliation{\MSU}
\newcommand*{\SCAROLINA}{University of South Carolina, Columbia, South Carolina 29208, USA}
\newcommand*{\SCAROLINAindex}{36}
\affiliation{\SCAROLINA}
\newcommand*{\SPECTRALSCI}{Spectral Sciences Inc., 01803 Burlington, Massachusetts, USA}
\newcommand*{\SPECTRALSCIindex}{37}
\affiliation{\SPECTRALSCI}
\newcommand*{\TEMPLE}{Temple University,  Philadelphia, Pennsylvania 19122, USA}
\newcommand*{\TEMPLEindex}{38}
\affiliation{\TEMPLE}
\newcommand*{\JLAB}{Thomas Jefferson National Accelerator Facility, Newport News, Virginia 23606, USA}
\newcommand*{\JLABindex}{39}
\affiliation{\JLAB}
\newcommand*{\UTFSM}{Universidad T\'{e}cnica Federico Santa Mar\'{i}a, Casilla 110-V Valpara\'{i}so, Chile}
\newcommand*{\UTFSMindex}{40}
\affiliation{\UTFSM}
\newcommand*{\EDINBURGH}{Edinburgh University, Edinburgh EH9 3JZ, United Kingdom}
\newcommand*{\EDINBURGHindex}{41}
\affiliation{\EDINBURGH}
\newcommand*{\GLASGOW}{University of Glasgow, Glasgow G12 8QQ, United Kingdom}
\newcommand*{\GLASGOWindex}{42}
\affiliation{\GLASGOW}
\newcommand*{\VT}{Virginia Tech, Blacksburg, Virginia  24061-0435, USA}
\newcommand*{\VTindex}{43}
\affiliation{\VT}
\newcommand*{\VIRGINIA}{University of Virginia, Charlottesville, Virginia 22901, USA}
\newcommand*{\VIRGINIAindex}{44}
\affiliation{\VIRGINIA}
\newcommand*{\WM}{College of William and Mary, Williamsburg, Virginia 23187-8795, USA}
\newcommand*{\WMindex}{45}
\affiliation{\WM}
\newcommand*{\YEREVAN}{Yerevan Physics Institute, 375036 Yerevan, Armenia}
\newcommand*{\YEREVANindex}{46}
\affiliation{\YEREVAN}

\newcommand*{\NOWISU}{Idaho State University, Pocatello, Idaho 83209, USA}
\newcommand*{\NOWGLASGOW}{University of Glasgow, Glasgow G12 8QQ, United Kingdom}
\newcommand*{\NOWINFNGE}{INFN, Sezione di Genova, 16146 Genova, Italy}

\title{Determination of the proton spin structure functions for $0.05<Q^2<5$ GeV$^2$ using CLAS}

\author{R.G.\ Fersch}
\affiliation{\WM}
\affiliation{\CNU}
\author{N. Guler}
\affiliation{\ODU}
\affiliation{\SPECTRALSCI}
\author{P.~Bosted}
\affiliation{\JLAB}
\affiliation{\WM}
\author{A.~Deur}
\affiliation{\JLAB}
\author{K. Griffioen}
\affiliation{\WM}
\author {C. Keith} 
\affiliation{\JLAB}
\author {S.E.~Kuhn} 
\affiliation{\ODU}
\author{R. Minehart}
\affiliation{\VIRGINIA}
\author {Y.~Prok} 
\affiliation{\ODU}
\author {K.P. ~Adhikari} 
\affiliation{\MISS}
% ...


\collaboration{The CLAS Collaboration}
     \noaffiliation

\date{\today}

\begin{abstract}

We present the results of our final analysis of the full data set of $g_1^p(Q^2)$,  the spin structure function
of the proton, collected using CLAS at Jefferson Laboratory in 2000-2001. Polarized electrons with
energies of 1.6, 2.5, 4.2 and 5.7 GeV were scattered from proton
targets ($^{15}$NH$_3$ dynamically polarized along the beam direction) and detected with CLAS. From the
measured double spin asymmetries, we extracted virtual photon asymmetries $A_1^p$ and $A_2^p$ and 
spin structure functions  $g_1^p$ and $g_2^p$
over a wide kinematic range (0.05~GeV$^2 < Q^2 <$~5~GeV$^2$ and 
1.08~GeV~$< W <$~3~GeV), and
calculated moments of $g_1^p$. We compare our final results with various theoretical
models and expectations, as well as with parametrizations of the world data. Our data, with their precision
and dense kinematic coverage, are able tparametrizationo constrain fits of
polarized parton distributions, test pQCD predictions for quark polarizations at
large $x$, offer a better understanding of quark-hadron duality, and provide more precise values
of higher-twist matrix elements in the framework of the operator product expansion.

\end{abstract}


\maketitle

\section{INTRODUCTION}\label{s1}

Understanding the structure of the lightest stable baryon, the proton, in terms of
its fundamental constituents, quarks and gluons, is a long-standing goal at the intersection of
particle and nuclear physics. In particular, the decomposition of the total spin of the
nucleon, $J = \frac{1}{2}$, into contributions from quark and gluon helicities and orbital
angular momentum still remains an open challenge 30 years after the discovery of the
``spin puzzle'' by the European Muon Collaboration~\cite{Ashman:1987hv}. 
Although deep-inelastic electron and muon scattering (DIS),
semi-inclusive DIS (SIDIS), proton-proton collisions, deeply virtual Compton scattering (DVCS) and deeply virtual meson
production (DVMP), have all been used to understand nucleon spin,
inclusive polarized lepton scattering remains the benchmark for the study of longitudinal nucleon spins. 
The inelastic
scattering cross section can be described in the Born approximation (1-photon exchange) by
four structure functions ($F_1^p, F_2^p, g_1^p$ and $g_2^p$), all of which depend only
on $Q^2$, the 4-momentum transfer squared, and $\nu$, the virtual photon energy.
Two of these, $g_1^p$ and $g_2^p$, carry fundamental information about the
spin-dependent structure of the nucleon. The status of the world data
for $g_1^p$ and $g_2^p$ and their theoretical interpretation are reviewed in 
Refs.~\cite{Kuhn:2008sy,Aidala:2012mv}.
\\
\indent
The new experimental data from Jefferson Laboratory (JLab) reported in this paper,  
expand significantly the kinematic range over which $g_1^p$ for the proton
is known to high precision. In particular, data were collected down to the rather small
$Q^2 \approx 0.05$ GeV$^2$, over a wide range of final-state masses, $W$,
that include the resonance region (1~GeV~$< W < $~2~GeV) and part of the DIS region,
(2~GeV~$< W <$~3~GeV with $Q^2>1$ GeV$^2$). The DIS data can serve as a low-$Q^2$ anchor for the extraction 
(see Ref.~\cite{Sato:2016tuz}) of
polarized parton distribution functions (PDFs) within the framework of the
next-to-leading order (NLO) evolution
equations~\cite{Dokshitzer,GribovLipatov,Altarelli}, 
and they can be used to pin down higher-twist contributions within the framework
of the operator product expansion (OPE)~\cite{OPE1,OPE2,SSFope}. 
They also can test various predictions
for the asymptotic behavior of the asymmetry $A_1^p(x)$ as the momentum fraction $x \rightarrow 1$.
The data in the resonance region reveal
new information on resonance transition amplitudes (and their interference with the 
non-resonant background),  and they can be used to characterize the transition from hadronic
to partonic degrees of freedom as $Q^2$ increases (parton-hadron duality). 
Finally, various sum rules that constrain moments of $g_1^p$ at both high and low
$Q^2$ can be tested.
\\
\indent
All data presented in this paper, referred to as the EG1b experimental run,  were collected with the CEBAF Large
Acceptance Spectrometer (CLAS) \cite{Mecking:2003zu}  in Jefferson Laboratory's Hall B during the time
period 2000--2001. 
Previously, a smaller data set in similar but more restrictive kinematics was
obtained with CLAS in 1998; those proton and deuteron results were published in Refs.~\cite{Fatemi:2003yh}
and ~\cite{Yun:2002td}, respectively.
The present data set was taken with beam energies of 1.6, 2.5, 4.2 and 5.7 GeV on 
polarized hydrogen ($^{15}$NH$_3$) and deuteron  ($^{15}$ND$_3$) targets.
The results on the deuteron are presented in Ref.~\cite{Guler:2015hsw}.
Preliminary proton results from the highest and lowest beam energies were
published previously~\cite{Dharmawardane:2006zd,Bosted:2006gp,Prok:2008ev}.
The present paper includes, for the first time, the full data set collected with CLAS
in 2000-2001 on the proton,
and summarizes all details of the experiment and the final analysis. 
\\
\indent
The first data on spin structure functions at low $W$, including the resonance
region, and at moderate $Q^2$, were measured at SLAC and published in 1980~\cite{Baum:1980mh}, 
followed by more
precise data published by the E143 Collaboration in 1996~\cite{Abe:1996ag}.
A comparable data set to the one presented here,  covering a wide kinematic range, was collected for the
neutron, using polarized $^3$He as an effective neutron target and the spectrometers
in Jefferson Laboratory's Hall A~\cite{Amarian:2003jy,Zheng:2004ce}. 
A more restricted data set on the proton and 
deuteron at an average $Q^2$ of 1.3 GeV$^2$, covering the resonance region
with both transversely and longitudinally polarized targets, was acquired
in Jefferson Lab's Hall C~\cite{Wesselmann:2006mw}.  Precise $g_1^p$ and $g_1^d$ data from the CLAS 
EG1-dvcs experiment were published recently \cite{Prok:2014ltt}. These results provided measurements 
of these structure functions at $Q^2 >$ 1 GeV$^2$, giving results at higher $x$ than accessible in 
EG1b; results from EG1b in this publication complement these results by improving the precision of 
$g_1^p$ at lower $Q^2$ in and near the resonance region.
\\
\indent
In the following, we introduce the necessary formalism and theoretical background
(Sec. II), describe
the experimental setup (Sec. III), discuss the analysis procedures (Sec. IV), 
present the results for all measured and derived quantities, as well as 
models and comparison to theory (Sec. V), and summarize our
conclusions (Sec. VI).


\section{THEORETICAL BACKGROUND}\label{s2}

\subsection{Formalism}

Cross sections for inclusive high energy electron scattering off a nucleon target with 
4-momentum $p^\mu$ and mass $M$ depend, in general,
on the beam energy $E$, the scattered electron energy $E'$ and the scattering angle $\theta$ (all defined in the laboratory frame with the proton initially at rest)\footnote{For beam and target polarization along the beam axis, the azimuth $\phi$ can be ignored since
no observable can depend on it.},
 or, equivalently,
on  the three relativistically invariant variables
\begin{equation}
\label{Q2:eq}
Q^2 = -q^2 = 4EE'\sin^2\frac{\theta}{2} , 
\end{equation}
\begin{equation}
\label{nu:eq}
\nu = \frac{p\cdot q}{M} = E - E^\prime ,
\end{equation}
and
\begin{equation}
\label{y:eq}
y = \frac{p\cdot q}{p\cdot k} = \frac{\nu}{E} ,
\end{equation}
in which  $q^\mu = k^\mu - k'^\mu$ is the four-momentum carried by the virtual photon,
which (in the Born approximation) is equal to the difference between initial ($k$) 
and final ($k'$) electron four-momenta.

The first two variables can be combined 
with the initial four-momentum of the target nucleon to calculate the invariant mass of the final state,
\begin{equation}
\label{W:eq}
W =\sqrt{(p + q)^2}=\sqrt{M^2 + 2M\nu - Q^2} ,
\end{equation}
and the Bjorken scaling variable,
\begin{equation}
\label{x:eq}
x = \frac{Q^2}{2p\cdot q} = \frac{Q^2}{2M\nu} ,
\end{equation}
which is interpreted as the momentum fraction of the struck parton in the infinite momentum frame.

The following combinations of these variables are also useful:
\begin{equation}
\label{gamma:eq}
\gamma = \frac{2Mx}{\sqrt{Q^2}} = \frac{\sqrt{Q^2}}{\nu} ,
\end{equation}
\begin{equation}
\label{tau:eq}
\tau = \frac{\nu^2}{Q^2} = \frac{1}{\gamma^2} ,
\end{equation}
and the virtual photon polarization ratio
\begin{eqnarray}
\label{epsilon:eq}
\epsilon & = & \frac{2(1-y) - \frac{1}{2}\gamma^2y^2}{(1-y)^2 + 1 + \frac{1}{2}\gamma^2y^2}  \nonumber \\
  & = & \left( 1 + 2[1 +
  \tau]\tan^2\frac{\theta}{2}\right)^{-1} .
\end{eqnarray}

\subsection{Cross sections and asymmetries}
\label{SFsAsyms}
In the Born approximation the cross section for inclusive
electron scattering
with beam and target spin parallel ($\uparrow\Uparrow$) or antiparallel ($\uparrow\Downarrow$) 
to the beam direction   
can be expressed in terms of the four structure functions 
$F_1^p, F_2^p, g_1^p$ and $g_2^p$, all of which depend on $\nu$ and $Q^2$:
\begin{eqnarray}
\label{sigfull}
\frac{d\sigma^{\uparrow\Downarrow/\uparrow\Uparrow}}{d\Omega dE'}  & = &
 \sigma_M \left[ \frac{F_2^p}{\nu} + 2\tan^2\frac{\theta}{2} \, \frac{F_1^p}{M}  \pm 2\tan^2\frac{\theta}{2} \right. \nonumber \\
& &   \left.   \times \left( \frac{E + E' \cos\theta}{M \nu} g_1^p - \frac{Q^2}{M \nu^2} g_2^p \right) \right] ,
\end{eqnarray}
where the Mott cross section
\begin{equation}
\label{sigmott}
\sigma_M = \frac{4\alpha^2E'^2 } {Q^4}\cos^2\frac{\theta}{2} ,
\end{equation}
where $\alpha$ is the electromagnetic fine structure constant.
We can now define the double spin asymmetry $A_{||}$ as
\begin{equation}
\label{Aparintro:eq}
A_{||}(\nu,Q^2) = \frac{d\sigma^{\uparrow\Downarrow} - d\sigma^{\uparrow\Uparrow}}{d\sigma^{\uparrow\Downarrow} + 
d\sigma^{\uparrow\Uparrow}} .
\end{equation}

Introducing the ratio $R^p$ of the absorption cross sections for longitudinal over transverse
virtual photons ($\gamma^*$),
\begin{equation}
\label{Rsig}
R^p = \frac{\sigma_L({\gamma^*})}{\sigma_T({\gamma^*})} = 
\frac{F_2^p}{2 x F_1^p} (1+\gamma^2) - 1,
\end{equation}
(where $L$ and $T$ represent longitudinal and transverse polarization, respectively) we can define two additional quantities,
\begin{equation}
\label{eta:eq}
\eta = \frac{\epsilon\sqrt{Q^2}}{E - E'\epsilon}
\end{equation}
and the ``depolarization factor''
\begin{equation}
\label{D1:eq}
D = \frac{1 - E'\epsilon/E}{1+\epsilon R^p} ,
\end{equation}
which allow us to express $A_{||}$ in terms of the structure functions:
\begin{equation}
\label{g2solve:eq}
\frac{A_{||}}{D} =
(1+\eta\gamma) \frac{g_1^p}{F_1^p} +
\gamma(\eta-\gamma) \frac{g_2^p}{F_1^p} .
\end{equation}

Alternatively, the double spin asymmetry $A_{||}$  can also be interpreted
in terms of the virtual photon asymmetries
\begin{equation}
\label{A1:eq}
A_1^p(\gamma^*) \equiv \frac{\sigma^{\frac{1}{2}}_T({\gamma^*}) - \sigma^{\frac{3}{2}}_T({\gamma^*})}{\sigma^{\frac{1}{2}}_T({\gamma^*}) + \sigma^{\frac{3}{2}}_T({\gamma^*})}
= \frac{g_1^p-\gamma^2 g_2^p}{F_1^p}
\end{equation}
and
\begin{equation}
\label{A2:eq}
A_2^p(\gamma^*) \equiv \frac{\sigma_{LT}}{\sigma_T} =
 \frac{2\sigma_{LT}(\gamma^*)}{\sigma^{\frac{1}{2}}_T({\gamma^*}) + \sigma^{\frac{3}{2}}_T({\gamma^*})} = \gamma \frac{g_1^p+g_2^p}{F_1^p} .
\end{equation}
Here, $\sigma^{\frac{1}{2}}_T({\gamma^*})$ and $\sigma^{\frac{3}{2}}_T({\gamma^*})$ represent the transversely polarized photon cross-sections for production of spin-$\frac{1}{2}$ and spin-$\frac{3}{2}$ final hadronic states, respectively, and $\sigma_{LT}(\gamma^*)$ is the interference cross-section between longitudinal and transverse virtual photons.
Note that both unpolarized structure functions $F_1^p$ and $F_2^p$ [as implicitly contained in $D$; see Eqs. (\ref{Rsig}) and (\ref{D1:eq})] are contained in the definition of these asymmetries. Here, $A_1^p$ is the asymmetry for transverse (virtual) photon 
absorption on a nucleon with total final-state spin projection $\frac{1}{2}$ or $\frac{3}{2}$
along the incoming photon direction, and $A_2^p$ is an interference asymmetry 
between longitudinally and transversely polarized
virtual photon absorption.
The relationship to the measured quantity $A_{||}$ is
\begin{equation}
\label{Apar:eq}
A_{||}(\nu,Q^2) = D[A_1^p(\nu,Q^2) + \eta A_2^p(\nu,Q^2)] .
\end{equation}
 $A_{||}$ is the primary observable determined directly from the data described in this paper. 
 The structure functions $g_1^p, g_2^p$ and the virtual photon asymmetries $A_1^p, A_2^p$ are extracted from
 these asymmetries. In particular, given a model or data for $F_1^p$, $R^p$ and $A_2^p$, $A_1^p$ can be
 extracted directly using Eq.~(\ref{Apar:eq}), and $g_1^p$ can be extracted using
 \begin{equation}
 \label{g1fromApar}
 g_1^p = \frac{\tau}{1+\tau} \left(\frac{A_{||}}{D} + (\gamma - \eta) A_2^p \right) F_1^p .
 \end{equation}
 A simultaneous extraction of both asymmetries $A_1^p$ and  $A_2^p$ from measurements of $A_{||}$ alone
 is possible by exploiting the dependence of the factors $D$ and $\eta$ in Eqs.~(\ref{g2solve:eq}) and (\ref{Apar:eq})
 on the beam energy for the same kinematic point $(\nu,Q^2)$. This is the super-Rosenbluth separation 
 of Sec.~\ref{SuperRBasym}.

\subsection{Virtual photon absorption asymmetries}
\label{photonabsorption}
Data on the virtual photon absorption asymmetries $A_1^p$ and $A_2^p$ are of great interest
in both the the nucleon resonance and DIS regions.

For inelastic scattering leading to specific final (resonance) states, $A_1^p$ can be
interpreted in terms of the helicity structure of the transition from the nucleon ground state
to the final state resonance. If the final state has total spin $S = \frac{1}{2}$, the absorption cross section
$\sigma^{\frac{3}{2}}_T({\gamma^*})$ leading to final spin projection $S_z = \frac{3}{2}$ along the 
virtual photon direction obviously cannot contribute, requiring $A_1^p = 1$ [see Eq.~(\ref{A1:eq})]. Vice versa,
excitations of spin $S=\frac{3}{2}$ resonances like the $\Delta(1232)$ receive a strong contribution
from $\sigma^{\frac{3}{2}}_T({\gamma^*})$ and therefore can have a negative $A_1^p$. 
Both $A_1^p$ and $A_2^p$ are directly related to the helicity transition amplitudes,
$A_{\frac{3}{2}}(\nu, Q^2)$ (transverse photons leading to final-state helicity $\frac{3}{2}$), 
$A_{\frac{1}{2}}(\nu, Q^2)$ (transverse photons leading to final-state helicity $\frac{1}{2}$), and
$S^*_{\frac{1}{2}}(\nu, Q^2)$ (longitudinal photons):
\begin{equation}
A_1^p = \frac {|A_{\frac{1}{2}}|^2 - |A_{\frac{3}{2}}|^2} {|A_{\frac{1}{2}}|^2 + |A_{\frac{3}{2}}|^2} \quad\text{and} \\
\end{equation}\label{eq:resamps}
\begin{equation}
A_2^p = \sqrt{2} \frac{\sqrt{Q^2}}{q^*} \frac{S^*_{\frac{1}{2}}A_{\frac{1}{2}}}{|A_{\frac{1}{2}}|^2 + |A_{\frac{3}{2}}|^2} \, .
\end{equation} \label{eq:resamps2} 
Here, $q^*$ is the (virtual) photon three-momentum in the rest frame of the resonance. As an example, the $\Delta(1232)$ is excited by a (nearly pure) $M1$ transition at low $Q^2$, with
$A_{\frac{3}{2}} \approx \sqrt{3} A_{\frac{1}{2}}$ and therefore $A_1^p \approx -0.5$.
In general, the measured asymmetries $A_1^p$ and $A_2^p$ at a given value of $W$ 
provide information on the relative strengths of overlapping resonance transition amplitudes and the
non-resonant background. By looking at the $Q^2$--dependence of the asymmetry for a specific
$S=\frac{3}{2}$ resonance (e.g., the $D_{13}$), one can study the transition from $A_{\frac{3}{2}}$ dominance
at small $Q^2$ (including real photons) to the $A_{\frac{1}{2}}$ dominance 
expected from quark models and perturbative quantum chromodynamics (pQCD) at large $Q^2$. 

In the DIS region,  $A_1^p(x)$ can yield information on the polarization of the valence quarks at
large $x$. In a simple SU(6)-symmetric quark model, with three constituent quarks at rest,
the polarization of valence up and down quarks yields
$A_1^p(x) = 5/9$. 
Most realistic
models predict that $A_1^p(x) \rightarrow 1$ as $x \rightarrow 1$, implying that a valence quark,
which carries nearly all of the nucleon momentum in the infinite momentum frame, will be polarized
along the proton's spin direction. However, the approach to the limit $x = 1$ is quite different for 
different models.
In particular, relativistic constituent quark models~\cite{Isgur:1998yb}
predict a much slower rise towards $A_1^p = 1$ than 
pQCD calculations~\cite{Brodsky:1994kg,Farrar:1975yb} that incorporate helicity conservation. 
Modifications of the pQCD picture to include orbital angular momentum~\cite{Avakian:2007xa}
show an intermediate rise towards $x = 1$.
Precise measurements of $A_1^p$ at large $x$ in the DIS region are therefore of high
importance.

The asymmetry $A_2^p$ is not very well-known in the DIS region,  and it has no simple interpretation.
However, it is constrained by the Soffer inequality~\cite{Soffer:1999zv,Artru:2008cp}
 \begin{equation} \label{eq:A2bound}
 |A_2^p|\leq \sqrt{R^p\,(1+A_1^p)/2} . \end{equation}
 
 Data on $A_1^p$ have been extracted by collaborations at CERN, 
SLAC and DESY
\cite{Ashman:1987hv, Ashman:1989ig, Adeva:1998vv, Ageev:2007du, Alexakhin:2006vx, Ackerstaff:1999ey, Airapetian:2006vy, 
 Abe:1996ag, Anthony:1996mw, Abe:1998wq, Abe:1997cx, Anthony:1999rm, Anthony:2000fn, Anthony:2002hy}
(mostly in the DIS region), as well
 as by collaborations at Jefferson Laboratory~\cite{Zheng:2003un, Zheng:2004ce, Dharmawardane:2006zd, Prok:2008ev}. 
Data on $A_2^p$ from the same labs and MIT Bates are more limited in the $Q^2$ range covered
~\cite{Abe:1998wq,Anthony:2002hy,Airapetian:2011wu,Bates:ref,Kramer:2005qe,Wesselmann:2006mw,Adolph:2015saz,Flay:2016wie,
RondonAramayo:2009zz,KANG:2013lva}.

\subsection{The spin structure function $g_1^p(x,Q^2)$}

In a simple quark-parton model, the structure function $g_1^p(x)$ is independent of $Q^2$, and can be 
interpreted in terms of the
difference $\Delta q(x) = {q\uparrow(x)} - {q\downarrow(x)}$ of parton 
densities for quarks with helicity
aligned versus antialigned
with the overall longitudinal nucleon spin, as a function of the 
momentum fraction $x$
carried by the struck quark. In particular, for the proton 
\begin{equation}
g_1^p(x) = \frac{1}{2}\, \sum_j \,
e^2_j\,[ \Delta q_j(x) + \Delta \bar q_j(x)]
\end{equation}
where the sum goes over all relevant quark flavors (up, down, strange, etc.) for quark densities $q_j$, and
$e_j$ are the corresponding electric charges (2/3, $-$1/3, $-$1/3,~\ldots).

Within QCD, this picture is modified in two important ways:
\begin{enumerate}
\item The coupling of the virtual photon to the quarks is modified by QCD radiative effects (e.g., gluon
emission).
\item The parton densities $\Delta q_j(x,Q^2)$ and $\Delta \bar q_j(x,Q^2)$, and hence $g_1^p(x,Q^2)$,
become (logarithmically) dependent on the resolution $Q^2$ of the probe, 
as described by the 
DGLAP (Dokshitzer-Gribov-Lipatov-Altarelli-Parisi) evolution equations~\cite{GribovLipatov, Altarelli, Dokshitzer}.
 At NLO and higher, these
equations couple quark and gluon PDFs at lower $Q^2$ to those at higher $Q^2$ via
the so-called splitting functions. Therefore, measuring the $Q^2$ dependence of $g_1^p$ with high
precision over a wide range in $Q^2$ can yield additional information on the spin structure of the nucleon,
including the contribution of the gluon helicity distribution $\Delta G(x)$. 
\end{enumerate}

Accurate data are therefore
needed at both the highest accessible $Q^2$ (presently from the 
COMPASS Collaboration at CERN)
and the lowest $Q^2$ that is still consistent with the pQCD description of DIS
(the data taken at Jefferson Laboratory). In the region of lower $Q^2$, additional
scaling violations occur due to higher-twist contributions and target mass corrections, 
leading to correction terms proportional to 
powers of $1/Q$. These corrections can be extracted 
from our data
since they cover seamlessly the transition from $Q^2 \ll 1$ GeV$^2$ to the scaling region
$Q^2 > 1$ GeV$^2$. An additional complication arises because at moderate to high 
$x$, low $Q^2$ corresponds to the region of the nucleon resonances ($W<2$ GeV). 
In this case, one would expect the quark-parton description of $g_1^p$ to break down,
and hadronic degrees of freedom (resonance peaks and troughs) to
dominate the behavior of $g_1^p(x)$, analogous to the asymmetry $A_1^p$ discussed above.

\subsubsection*{Bloom-Gilman duality}
\label{bloom}
Bloom and Gilman 
observed~\cite{Bloom:1970xb}
that the unpolarized structure function $F_2^p(x,Q^2)$ in the resonance region resembles,
on average, the same structure function at much higher $Q^2$, in the DIS region, where
the quark-parton picture applies. This agreement, which improves if one plots the data
against the Nachtmann variable~\cite{Nachtmann:1973mr}
\be
\label{Nachtmann}
\xi = \frac{Q^2}{M(\nu+\sqrt{Q^2+\nu^2})} = \frac{ |\vec{q}| - \nu}{M},
\ee
(where $|\vec{q}|$ is the magnitude of the virtual photon 3-momentum) 
is one example of ``quark-hadron duality," where both quark-parton and hadronic 
interpretations of the same data are possible. 
De Rujula \etal ~\cite{DeRujula:1976tz,DeRujula:1976ke} interpreted this 
duality
as a consequence of relatively small higher-twist contributions to the 
structure functions.
Duality has been observed both for the integral of structure functions 
over the whole
resonance region, $W < 2$ GeV (``global duality''), as well as for averages
over individual resonances (``local duality'')~\cite{Melnitchouk:2005zr}. 

Initial duality data on polarized structure functions from SLAC~\cite{Abe:1998wq}
and HERMES~\cite{Airapetian:1998wi,Airapetian:1999ib}
have been followed by much more detailed examinations of duality in this 
case by experiments
at Jefferson Laboratory~\cite{Fatemi:2003yh, Solvignon:2008hk, Wesselmann:2006mw}, 
including  results from a partial analysis of the
present data set~\cite{Bosted:2006gp}. 
Reference~\cite{Melnitchouk:2005zr} summarizes the conditions under
which duality has been found to hold at least approximately. The complete data set discussed in this
paper increases substantially the kinematic range over which high-precision 
data exist in the resonance region
and beyond, and can be compared to extrapolations from the DIS region. 
A full analysis accounting
for QCD scaling violations and target mass 
effects~\cite{Blumlein:1998nv} can make this 
comparison more rigorous and quantitative.

\subsection{The spin structure function $g_2^p(x,Q^2)$}

The second spin-dependent structure function in inclusive DIS, 
$g_2^p(x,Q^2)$, does not have an 
intuitive interpretation in the quark-hadron picture. The sum 
of $g_1^p + g_2^p = g_T$ is proportional
to $A_2^p$ [Eq.~(\ref{A2:eq})] and has a leading-twist contribution according to the 
Wandzura-Wilczek relation~\cite{Wandzura:1977qf},
\begin{equation}
\label{gTdef}
\bar{g}_T(x,Q^2) = \int _{x}^1 \frac{\bar{g}_1(y,Q^2)}{y} dy,
\end{equation}
and a very small contribution from transverse quark polarization (which is suppressed by the small
quark masses). Here, the notation $\bar{g}$ denotes contributions from leading twist only. The higher twist contributions to $g_T$ (and hence $g_2^p$) can be sizable,  and they are not suppressed by
powers of $1/Q$, which makes $g_T$ or $g_2^p$ a good experimental quantity with which to study quark-gluon
correlations. In particular, the third moment,
\begin{equation}
\label{d2def}
d_2 = 3  \int _{0}^1 x^2 \left[ g_T(x) - \bar{g}_T(x) \right] dx,
\end{equation}
is directly proportional to a twist-3 matrix element that is connected to the so-called
``color polarizabilities'' $\chi_E$ and $\chi_B$ (see Sec.~\ref{moments}) and has
recently been linked to the average transverse force on quarks ejected 
from a transversely polarized nucleon~\cite{Burkardt:2009rf}.
Finally, the Burkhardt-Cottingham sum rule~\cite{Burkhardt:1970ti} predicts that the integral 
\begin{equation}
\label{BCSR}
 \int _{0}^{1+\epsilon} g_2^p(x,Q^2)dx = 0
 \end{equation}
 at {\em all} $Q^2$, in which the upper integration limit $1 + \epsilon$ indicates the inclusion of 
the elastic peak at  $x = 1$.

The EG1b data on $A_{||}$ are not very sensitive to $g_2^p$ or $g_T$, leading to relatively
large statistical uncertainties on their extraction. For this reason,  in this paper we only 
present limited results on $g_2^p$ and no direct evaluations of the 
integrals, Eqs.~(\ref{d2def}) and (\ref{BCSR}). However, we use theoretical constraints
[Eqs.~(\ref{eq:A2bound}) and (\ref{BCSR})], and existing experimental data on $g_2^p$ or $A_2^p$, to
model $A_2^p(x,Q^2)$.  We use this model to
extract $A_1^p$ and $g_1^p$ from our data.

\subsection{Elastic scattering}
\label{elas}
The virtual photon asymmetries $A_1^p$ and $A_2^p$ are also defined for elastic scattering from a nucleon $N$,
$N(e,e')N$,
and Eq.~(\ref{Apar:eq}) applies in this case as well.
Following our discussion in Sec.~\ref{photonabsorption}, 
$A_1^p = 1$ for elastic scattering, since the final state spin
is $\frac{1}{2}$ and hence $\sigma^{\frac{3}{2}}_T({\gamma^*}) = 0$. The elastic asymmetry $A_2^p$ is given by
\begin{equation}
A_2^p(Q^2) = \sqrt{R^p} = \frac{G_E^p(Q^2)}{\sqrt{\tau} G_M^p(Q^2)},
\end{equation}
where $G_E^p$ and $G_M^p$ are the electric and magnetic Sachs form factors of the nucleon.
This relationship can be used to determine the ratio $G_E^p/G_M^p$ from double-polarized scattering;
in our case, we use this ratio, which is well determined by 
JLab experiments~\cite{Gayou:2001qt,Arrington:2007ux},
to extract the product of
beam and target polarization, $P_bP_t$:
\begin{equation}
\label{elasasym}
A_{||}^{meas} = P_bP_t A_{||}^{theo}.
\end{equation}
Here, $A_{||}^{meas}$ is the measured elastic double-spin asymmetry after all corrections
for background contamination have been applied.

One can also extend the definition of $g_1^p(x)$ and $g_2^p(x)$ to include elastic scattering at
$x = 1$ by adding the terms
\ba
\label{elasticgs}
g_1^{pel}(x,Q^2) & = & \frac{1}{2} \frac{G_E^p G_M^p + \tau G_M^{p^2}}{1 + \tau} \delta(x-1)\ {\text{ and}} \nonumber \\ 
g_2^{pel}(x,Q^2) & = & \frac{\tau}{2} \frac{G_E^p G_M^p - G_M^{p^2}}{1 + \tau} \delta(x-1) ,
\ea
which yield finite contributions to the moments (integrals over $x$) that include the elastic contribution. 

\subsection{Moments}
\label{moments}

Moments of structure functions weighted by powers of $x$ are useful quantities for 
investigating the QCD-structure of the nucleon. On the one hand, they can be
connected, via sum rules, to local operators of quark currents or forward Compton
scattering amplitudes. On the other hand, they are currently the only relevant quantities that can be
calculated directly in lattice QCD or in effective field theories like
chiral perturbation theory ($\chi$PT).

The matrix element $d_2$, introduced in Eq.~(\ref{d2def}), is one example
of a moment (the third moment of a combination of $g_1^p$ and $g_2^p$). In the following,
we focus on moments of $g_1^p$ 
 since our data are most sensitive to this structure function. The most important moment is 
\begin{equation}
\label{Gamma1def}
\Gamma_{1}^p(Q^{2})\equiv\int_{0}^{1}g_{1}^p(x,Q^{2})dx.
\end{equation}
In the limit of very high $Q^2$, this moment for the neutron ($n$) and the proton ($p$) is proportional to a
combination of matrix elements of axial quark currents,
\begin{equation}
\label{Gamma1HiQ}
\Gamma_{1}^{p,n}(Q^2 \rightarrow \infty) =\pm \frac{1}{12} a_3 + \frac{1}{36} a_8 + \frac{1}{9} a_0,
\end{equation}
in which $a_3 = g_A = 1.267 \pm 0.004$ (where $g_A$ is the axial vector coupling constant) and $a_8 = F+D \approx 0.58 \pm 0.03$ (where $F$ and $D$ are SU(3) coupling constants)
 \cite{Hagiwara:2002fs} are the isovector and flavor-octet 
axial charges of the nucleon, which have been determined from nucleon and hyperon $\beta$ decay, and
$a_0$ is the flavor-singlet axial charge, which measures the total contribution of quark helicities
to the (longitudinal) nucleon spin,
\begin{equation}
S_z^{quarks} = \frac{1}{2} \Delta \Sigma =  \frac{1}{2} a_0 .
\end{equation}
Combining Eq.~(\ref{Gamma1HiQ}) for the proton and the neutron yields the famous
Bjorken sum rule~\cite{Bjorken:1966jh, Bjorken:1969mm}:
\begin{equation}
\label{BjHiQ}
\Gamma_{1}^{p} - \Gamma_{1}^{n} = \frac{1}{6} a_3 = 0.211 .
\end{equation}

At high but finite $Q^2$, these moments receive pQCD corrections due to gluon radiative effects.
At leading twist, this yields
\begin{eqnarray}
\label{Gamma1pQCD}
 & &\mu_{2}^{p}(Q^{2})  \equiv  \Gamma_1^{p [LT]}(Q^{2}) = \nonumber \\
 & & =  C_{ns}(Q^{2})\left(  \frac{1}{12} a_3 + \frac{1}{36} a_8 \right)
+ C_{s}(Q^2)\frac{1}{9}a_{0}(Q^2) 
\end{eqnarray}
and
\begin{equation}
\label{BjpQCD}
 \mu_{2}^{p-n}(Q^{2}) \equiv
\Gamma_1^{p [LT]}(Q^{2}) - \Gamma_1^{n [LT]}(Q^{2}) =C_{ns}(Q^2)\frac{1}{6} a_3 . 
\end{equation}
Here, $C_{ns}$ and $C_{s}$ are flavor non-singlet and
singlet Wilson coefficients~\cite{Larin:1997qq}
that can be expanded in powers of the strong coupling constant $\alpha_S$
and hence depend mildly on $Q^2$, while the $Q^2$ dependence of
the matrix element $a_0$ reflects the $\overline{MS}$ renormalization
scheme that is used here, in which $a_{0}=\Delta\Sigma$, the contribution
of the quarks to the nucleon spin.

At the even lower $Q^2$ of the present data, additional 
corrections due to higher-twist
matrix elements proportional
to powers of $1/Q$ become important. These matrix elements are discussed in the next section.

In addition to the leading first moment, odd-numbered higher moments of $g_1^p$ can be defined as
$\int_0^1 x^{n-1} g_1^p(x) dx, \, n = 3,5,7,$~\ldots . These moments are dominated by high $x$
(valence quarks) and are thus particularly well determined by Jefferson Laboratory data. They
can also be related to hadronic matrix elements of local operators
or (in principle) evaluated using lattice QCD.
In the following, we will make explicit use of the third moment,
$a_{2}(Q^2) = \int_0^1  x^2 g_1^p(x,Q^2) dx$.

\subsubsection*{Higher twist and OPE}

Higher-twist matrix elements reveal information about quark-gluon and quark-quark interactions,
 which are important
for understanding quark confinement. A study of higher-twist matrix elements can be carried out in the
OPE formalism, which describes
the evolution of structure functions and their moments in the pQCD domain. 

In OPE, the first moment of $g_{1}^p(x,Q^{2})$ can be written as \footnote{
In this case, the elastic contribution Eq.~(\ref{elasticgs})
 to the moment must be included; i.e., the integral
must go over the range $[0 \ldots 1+\epsilon]$.}

\begin{equation}
\label{twistexpansion}
\Gamma_{1}^p(Q^{2}) =\sum_{\tau=2,4...}\frac{\mu_{\tau}(Q^{2})}{Q^{\tau-2}} ,
\end{equation}
in which $\mu_{\tau}(Q^{2})$ are sums of twist elements
up to twist $\tau$. The twist is defined as the mass dimension
minus the spin of an operator. Twist elements greater than $2$ can
be related to quark-quark and quark-gluon correlations. Hence, they
are important quantities for the study of quark confinement. The leading
twist contribution is given by the twist-2 coefficient $\mu_{2}$ defined
in Eq.~(\ref{Gamma1pQCD}).
The next-to-leading-order twist
coefficient is
\begin{equation}
\label{mu4}
\mu_{4}(Q^{2})=\frac{M^{2}}{9}\left[a_{2}(Q^2)+4d_{2}(Q^{2})+4f_{2}(Q^{2})\right] , 
\end{equation}
in which $a_{2}$ ($d_{2}$) is a twist 2 (3) target
mass correction that can be related to higher moments of $g_{1}^p$ ($g_1^p$ and $g_2^p$).
The matrix element  $f_{2}$ (twist-4)  \cite{OPE1} 
can be extracted from
the $Q^2$-dependence of $\Gamma_1^p$.
The matrix elements $d_2$ and $f_2$ are related to the color polarizabilities,
which are the responses of the color magnetic and electric fields
to the spin of the proton \cite{Stein:1995si,Ji:1995qe},
\begin{equation}
\label{colorpol}
\chi_E = \frac{2}{3}(2d_2 + f_2) \qquad {\rm and} \qquad \chi_M = \frac{1}{3}(4d_2 - f_2) .
\end{equation}
Theoretical values for $f_2$ and the color polarizabilities have been 
calculated using
quark models~\cite{Signal:1996ct}, QCD sum rules~\cite{Balitsky:1989jb},  
and lattice QCD~\cite{Dolgov:1998js}.

\subsubsection*{Moments at low $Q^2$}
The first moment
 of $g_1^p$ is particularly interesting since there is not only a sum rule for its
high--$Q^2$ limit [Eq.~(\ref{Gamma1HiQ})], but its approach to $Q^2 \rightarrow 0$
is governed by the Gerasimov-Drell-Hearn (GDH) sum 
rule~\cite{Gerasimov:1965et, Drell:1966jv}.
For real photons ($Q^2 = 0$) and nucleon targets, the GDH sum rule reads
\be
 \label{eq:GDH}
\int_0^\infty \frac{d\nu}{\nu} \, [\sigma^{\frac{3}{2}}_T(\nu) 
-\sigma^{\frac{1}{2}}_T(\nu)] = - \frac{2\pi^2\alpha}{M^2}\, \kappa^2,
\ee
in which $\kappa$ is the anomalous magnetic moment of the  nucleon. This sum rule
was based on a low-energy theorem for the forward spin-flip Compton amplitude $f_2(\nu)$
as $\nu \rightarrow 0$ which is connected to the left-hand side of Eq.~(\ref{eq:GDH})  through a dispersion relation.
The photon absorption cross sections
$\sigma^{\frac{3}{2},\frac{1}{2}}_T$ enter into $A_1^p$, $A_2^p$, 
$g_1^p$, and $g_2^p$ [Eq.~(\ref{A1:eq})], and consequently the GDH sum rule constrains the slope of the first moment\footnote{
In the present context, all moments {\em exclude} the elastic contribution since it does not 
contribute to real photon absorption. Hence, $\Gamma_1^p(Q^2) \rightarrow 0$ as $Q^2 \rightarrow 0$.}
 of $g_1^p$ as $Q^2 \rightarrow 0$:
\be
\label{Gamma1slope}
\frac{d \Gamma_1^p(Q^2)}{d Q^2} = - \frac{\kappa^2}{8 M^2} .
\ee
After generalizing the spin-dependent  Compton amplitude to
virtual photons, $S_1(\nu,Q^2)$, one can extend the GDH sum rule to
non-zero $Q^2$ using a similar dispersion relation~\cite{Ji:1999pd},
\be
\label{genGDH}
\frac{M^3}{4} S_1(0,Q^2) = \frac{2M^2}{Q^2} \Gamma_1^p(Q^2),
\ee
with $(M^3/4)S_1(0,Q^2) = -\kappa^2 / 4$ as $Q^2\to 0$. 
$S_1(0,Q^2)$ can be expanded in a power series in $Q^2$ around $Q^2=0$.
The coefficients of this expansion have been calculated up to NLO 
in  $\chi$PT~\cite{Ji:1999pd}, yielding predictions for both the first and second derivative of 
$\Gamma_1^p$ near the photon point. Since  $\chi$PT can be considered as the
low-energy effective field theory of QCD, $\Gamma_1^p$ can extend our understanding of 
the strong interaction to lower $Q^2$ values inaccessible to pQCD.

Extending the analysis of low-energy Compton amplitudes to higher powers
in $\nu$, one can get additional sum 
rules~\cite{Gorchtein:2004jd}. In
particular, one can generalize the forward spin polarizability, $\gamma_0^p$,
to include virtual photons:
\be
\label{gamma0}
\gamma_0^p(Q^2) = \frac{16 \alpha M^2}{Q^6} 
\int_0^{1} x^2 \, \big[g_1^p(x,Q^2) - \gamma^2 g_2^p(x,Q^2)\big]\,dx .
\ee
This too can be calculated  using
$\chi$PT~\cite{Prok:2008ev,Deur:2008ej}.

\section{THE EXPERIMENT}\label{s3}

The experiment was carried out at the Thomas Jefferson National Accelerator 
Facility  (Jefferson Laboratory or JLab for short), using a longitudinally polarized electron beam with energies from 1.6 to 
5.7 GeV, 
a longitudinally polarized solid ammonia target (NH$_3$ or ND$_3$), and the CEBAF Large Acceptance 
Spectrometer (CLAS). In this section, we present a brief overview of the experimental setup 
and methods of data collection.

\subsection{The CEBAF polarized electron beam}

The continuous-wave  
electron beam accelerator facility (CEBAF) at Jefferson Laboratory produced electron beams with energies ranging 
from 0.8 GeV to 5.7 GeV, polarizations up to 85\%,
and currents up to 300 $\mu$A.
Detailed descriptions of the accelerator are given in  
Refs.~\cite{CEBAF:ref,Sinclair:2007ez,Kazimi:2004zv,Stutzman:2007ny}.

Polarized electrons are produced by band-gap photoemission from a strained GaAs cathode.
The  circularly polarized photons for this process~\cite{Schultz:1992zy} are supplied by
master-oscillator-power-amplifiers (MOPAs) or titanium:sapphire lasers
configured in an ultra-high-vacuum system~\cite{Sinclair:2007ez}.
The circular polarization of the laser light can be reversed electronically by signals sent to a Pockels cell.
A half-wave plate (HWP) can be inserted into the laser beam to
change the polarization phase by 180$^\circ$. The HWP was inserted
and removed periodically throughout the experiment,  
to ensure that no polarity-dependent bias from the laser is present in the 
measured asymmetry.

The 100 keV electrons emerging from the GaAs entered the injector line 
\cite{Sinclair:2007ez,Liu:1997ue}, 
where their energies were boosted prior to injection 
into the main accelerator, which consists of two superconducting linacs connected by  
recirculation arcs. Each linac segment contains a series of superconducting niobium radio frequency (RF) cavities, 
driven by 5 kW klystrons~\cite{CEBAF:ref}.

A harmonic RF separator
system splits the interleaved beam bunches and delivers them to the appropriate
experimental hall (A, B, or C) \cite{CEBAF:ref}. 
The electron current in Hall B ranged from 0.3 to 10 nA, selected according to
the beam energy, the target type, and the spectrometer torus polarity.

\subsection{Beam monitoring and beam polarimetry}
The Hall B beam line 
incorporated several instruments to measure the intensity, position, and profile of the beam.
A Faraday cup at the end of the beam line
measured the absolute electron flux.
A M\o ller polarimeter was inserted periodically into the beam to measure its polarization.

Three beam position monitors (BPMs) were located 36.0, 24.6, and 8.2 m
upstream from the CLAS center. They measured the beam intensity and its
position in the transverse $xy$ plane. Each BPM was composed of three RF cavities. The
BPM position measurements were cross-calibrated using the ``harp'' beam profile scanners---thin
wires that were moved transverse to the beam direction---which 
also determined beam width and halo.  One-second averages of the BPM outputs 
were used in a feedback loop to keep the beam
centered on the target \cite{Mecking:2003zu}.

The beam electrons were collected by the Faraday Cup (FC) located
29.0 m downstream from the CLAS center. The FC was
used to integrate the beam current. 
The FC was a lead cylinder with diameter of 15 cm and thickness of 75 radiation lengths (r.l.) placed co-axially to the beam line. Its weight was 4000 kg.

The charge collection in 
the FC  \cite{Mecking:2003zu} was coupled to the CLAS data acquisition system using
a current-to-pulse rate converter. Both the total (ungated)
and detector live-time-gated counts  were recorded. 
The FC readout was also tagged by a helicity signal to normalize
the current for different helicity states.
The beam position monitors were periodically calibrated with
the Faraday cup.

The M\o ller polarimeter, located at the entrance of Hall B, 
was used to measure the beam polarization. M\o ller polarimetry requires a target of
highly magnetizable material in the beamline.
Therefore, dedicated M\o ller data runs
of approximately 30 min each
were taken periodically throughout the experiment.
The polarimeter consisted of a target chamber with a 25-$\mu$m-thick
Permendur (49\% Fe, 49\% Co, 2\% Va) foil oriented at 
$\pm$20$^\circ$ with respect to the beam line, longitudinally polarized
to 7.5\% by a 120 G Helmholtz magnet \cite{polarimeters:ref}. Two
quadrupoles separated the scattered electrons from the beam.
Elastic electron-electron scattering coincidences were used to determine the beam polarization, 
from the well-known double spin asymmetry \cite{Wagner:1990sn}.
The M\o ller measurements typically had a statistical
uncertainty of 1\% and a systematic uncertainty of $\sim$2$-$3\% 
\cite{Mecking:2003zu}. The average beam polarization was about 70\%.
Since we determined the product of beam and target polarization directly from our data,
the M\o ller polarimeter served primarily to 
ensure that the beam remained highly polarized during the beam exposures, as well as to check
the consistency of the polarization during the data analysis.


\subsection{The polarized target \cite{Keith:2003ca}}

 \indent
Cylindrical targets filled with solid ammonia beads immersed in liquid $^4$He were located at the center of CLAS, co-axial with the beam
line. The protons in the ammonia beads
were polarized using the method of dynamic nuclear
polarization (DNP), described in Refs.~\cite{DNP:ref,orientation:ref, Crabb:1997cy}.  
The required magnetic field was provided by a superconducting axial 5~T magnet (Helmholtz
coils) whose field was uniform over the target,
varying less than a factor of $10^{-4}$ over a cylindrical volume of 20 mm in length and 
diameter \cite{Keith:2003ca}. The target material was immersed in 
liquid helium (LHe) cooled to $\sim 1-1.5$ K  using
a $\sim$ 0.8-W $^4$He evaporation refrigerator.
The target system was contained in a cryostat designed to fit inside the central 
field-free region of CLAS, accessible for the insertion of the target material,
and allowing detection of particles scattered into a 48$^\circ$ forward cone over the majority of the CLAS acceptance. 

\begin{figure}
\centering
\includegraphics[width=8.7cm]{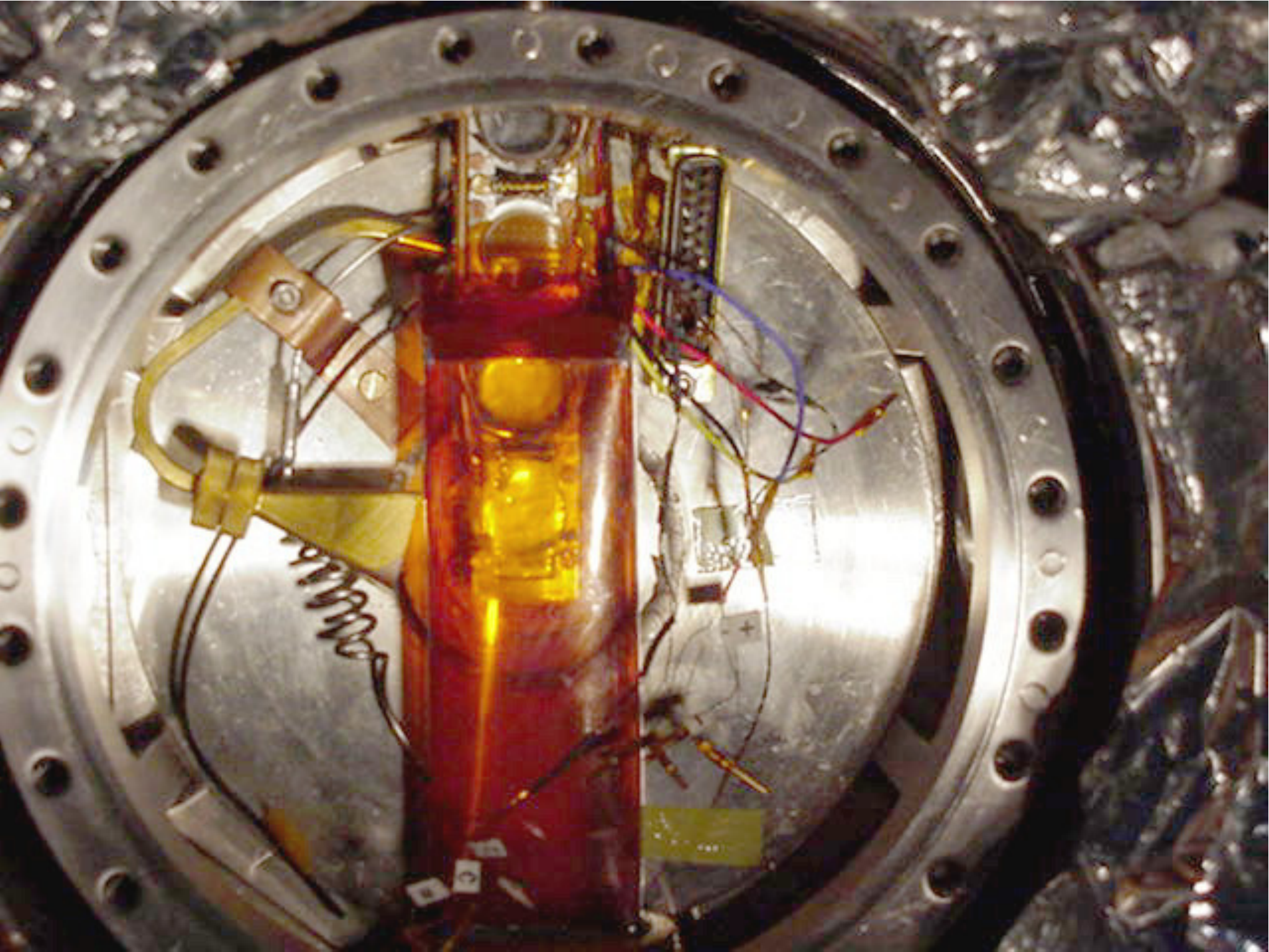}
\caption[Photograph of the target chamber interior and minicup]
{(Color online) An 
internal view of the target chamber, viewed from upstream, showing the orange transparent Kapton
cylindrical LHe minicup into which the target stick was inserted. Note the metal
``horn,'' the source of microwave emission, on the left side.}
\label{banjo:fig} 
\end{figure}

The cryostat contained four cylindrical target cells with axes parallel to the beam line, made of  2-mm-thick polychlorotrifluoroethylene (PCTFE), 15~mm in diameter and 10~mm in length,
with 
0.02-cm aluminum entrance windows and 0.03~cm Kapton exit windows. Tiny holes in the exit windows of the cells allow LHe to 
enter and cool the ammonia beads contained in two of the cells. A third cell contained a 
2.2-mm-thick (1.1 \% r.l.) disk of amorphous carbon, and  the fourth was left empty. The carbon and 
empty cells were used for estimating nuclear backgrounds and for systematic checks.
These target cells were mounted on a vertical target stick that could be removed from the cryostat
for filling the ammonia cells and moved up and down to
center the desired cell on the beam line.
The targets were immersed in LHe inside a vertically oriented cylindrical 
container called the ``minicup.'' The minicup and the target chamber are shown in 
Fig. \ref{banjo:fig}. Thin windows in the cryostat allowed
scattered particles to emerge in the forward and side directions.

The DNP method of proton (or deuteron) polarization uses a hydrogenated 
(or deuterated) compound (e.g., $^{15}$NH$_3$) in which a dilute assembly of paramagnetic centers was
produced by pre-irradiation with a low-energy
electron beam. 
During the experiment the target material was exposed constantly to microwave
radiation of approximately 140 GHz to drive the  hyperfine transition that polarizes
the proton spins. The microwave radiation was
supplied by an extended interaction oscillator (EIO) that generated about 1 W
of microwave power with a bandwidth of about 10 MHz. 
The microwaves were transmitted to whichever target cell was in the electron 
beam through a system of waveguides connected to a gold-plated 
rectangular ``horn" (visible in Fig. \ref{banjo:fig}).
The microwave frequency could be adjusted over a bandwidth of 2 GHz
 to match the
precise frequency required by the DNP. The negative and positive nuclear spin states 
were separated by $\sim$ 400 MHz, so that either polarization state could be achieved by
selecting the appropriate microwave 
frequency. Throughout the experiment, the sign of the nuclear polarization was periodically reversed to 
minimize the effects of false spin asymmetries.

During the experiment, the target polarization was monitored with an NMR system,
which includes a coil 
wrapped around the outside of the target cell in a resonant RLC (tank) circuit. The circuit was
driven by an RF generator tuned to the proton Larmor frequency (212.6 MHz). Depending on the sign of the 
target polarization, the coil either absorbed or emitted energy with a corresponding gain or loss 
in the resonant circuit. The induced voltage in the RLC circuit was measured and 
translated into the corresponding polarization of the sample.

To avoid depolarization from local heating,
the beam was rastered over the face of the target in a spiral pattern, using two pairs 
of perpendicular electromagnets upstream from the target.
Radiation damage to the target material from the 
electron beam was repaired by a periodic annealing process in which the target material 
was heated to 80-90~K. Annealing was done  approximately once a week. After several annealing cycles, the
maximum polarization tended to decrease, requiring the loading of fresh target material
several times during the experiment. NH$_3$ material was replaced when the polarization reached a level of approximately
10\% less than previous anneals. 
Target material was typically replaced after receiving a cumulative level of charge equivalent 
to that delivered by 2$-$3 weeks of 5 nA beam time.

The polarized target was operated for seven months during the EG1b experiment. 
The typical proton polarization maintained during the run was $\sim 70 - 75$\%, with a maximum 
value of 96\% without beam on target, and always remaining above 50\% during production running 
(more details on the target and its operation can be found in  Ref.~\cite{Keith:2003ca}). 

\subsection{The CLAS spectrometer}
\begin{figure}
\centering
\includegraphics[width=8.0cm]{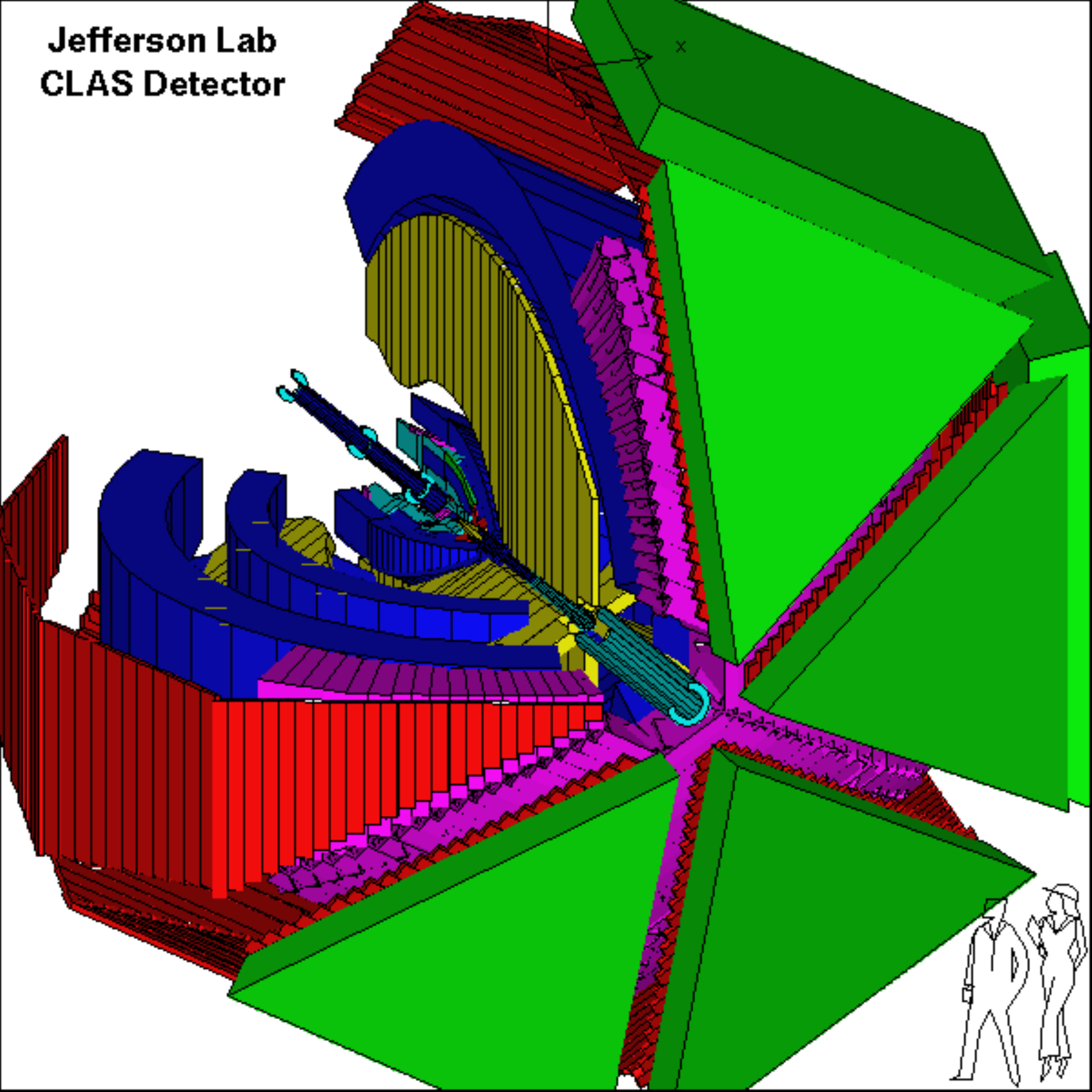}
\caption[CLAS Spectrometer]{(Color online) The CLAS spectrometer. Different colors represent different components of
the detector (from the central target outward): three layers of drift chambers (DCs) (blue) and the torus magnet (yellow), 
Cerenkov counters (CCs) (magenta), TOF counters (SCs) (red) 
and electromagnetic calorimeters (ECs) (green).  The electron beam travels through the central axis
from upper left to lower right.}
\label{CLAS:fig}
\end{figure}

The CEBAF Large Acceptance Spectrometer (CLAS), described in detail in Ref.~\cite{Mecking:2003zu},
was based on a six-coil toroidal superconducting magnet. 
Figure \ref{CLAS:fig} shows a cutaway view of the detector along the beam line.
Charged particles are tracked
through each of the six magnetic field regions (hereby labeled ``sectors'') between its 
coils, with three layers of multi-wire
drift chambers (DC), numbered 1 to 3 consecutively from the target outward.  \cite{Mestayer:2000we}.

Beyond the magnetic field region, charged particles were detected in a combination of gas 
Cherenkov counters, scintillation counters, and 
total absorption electromagnetic calorimeters. There was one set of scintillation
counters  (SC) \cite{Smith:1999ii} for each of the six sectors. These were used for triggering and for
time-of-flight (TOF) measurements, with a typical time resolution of
0.2$-$0.3 ns.  
In the forward region of the detector, the SC was preceded by gas-filled Cherenkov counters 
(CC) \cite{Adams:2001kk} designed to distinguish electrons and pions. Finally,
each sector included a total absorption sampling electromagnetic
calorimeter (EC) \cite{Amarian:2001zs} made of alternating layers of lead and 
plastic scintillator with a combined thickness of 15 r.l. 
The EC was used to measure the energy of the scattered electrons and to detect neutral particles.

 Torus currents of 1500 A (at low beam energies)
 or 2250 A (at high beam energies) were employed in
 this experiment.  For positive (negative) current,
 forward-going negative particles were bent 
 inward (outward) with respect to the beam line. The two conditions were referred to as ``inbending" and ``outbending," respectively. 
 Inbending allowed for larger acceptance of electrons at
 large scattering angles (high $\theta$) and higher luminosity, whereas outbending allowed for
 larger acceptance at small scattering angles (low $\theta$).
 The reversibility of the magnet current also allowed
 systematic studies of charge-symmetric backgrounds. 

\subsection{Trigger and data acquisition}
All analog signals from CLAS were digitized by FASTBUS and
VME modules in 24 crates. The data acquisition could be triggered by a variety of
combinations of detector signals.  Our event trigger
required signals exceeding minimum thresholds 
in both the EC and CC \cite{vipuli:ref}. All photomultiplier-tube (PMT) time-to-digital-converter (TDC) and analog-to-digital-converter (ADC) signals (i.e., SC, EC, and CC signals) generated
within 90 ns of the trigger were recorded, along with drift-chamber TDC 
signals \cite{Mecking:2003zu}. The trigger supervisor (TS)
generated busy gates and necessary resets, and directed all the signals to the data acquisition system (DAC). 
The DAC accepted event rates of 2 kHz and data rates of 25 MB/s
 \cite{Mecking:2003zu}.
\\
\indent
The simple event builder (SEB), used for offline reconstruction of an event, used
geometric parameters and calibration constants to convert
the  TDC and ADC data into kinematic and particle identification data.
The SEB cycled through particles in the event to search for a
single trigger electron---a negatively charged particle that produced a shower in the EC.
If more than one candidate was found, the one
  with the highest momentum was selected. This particle was traced along its geometric path
back to its intersection in the  target to determine the path length, which, with the assumption
that its velocity $v=c$, determined the event start time. From this start time, the 
TOF of other particles could then be determined from the SC TDC values. The TDC values
from the EC were used when SC values were not available for a given particle.

\section{DATA ANALYSIS}\label{s4}

\subsection{Data and calibrations}
\begin{figure}
\centering
\includegraphics[width=8.8cm]{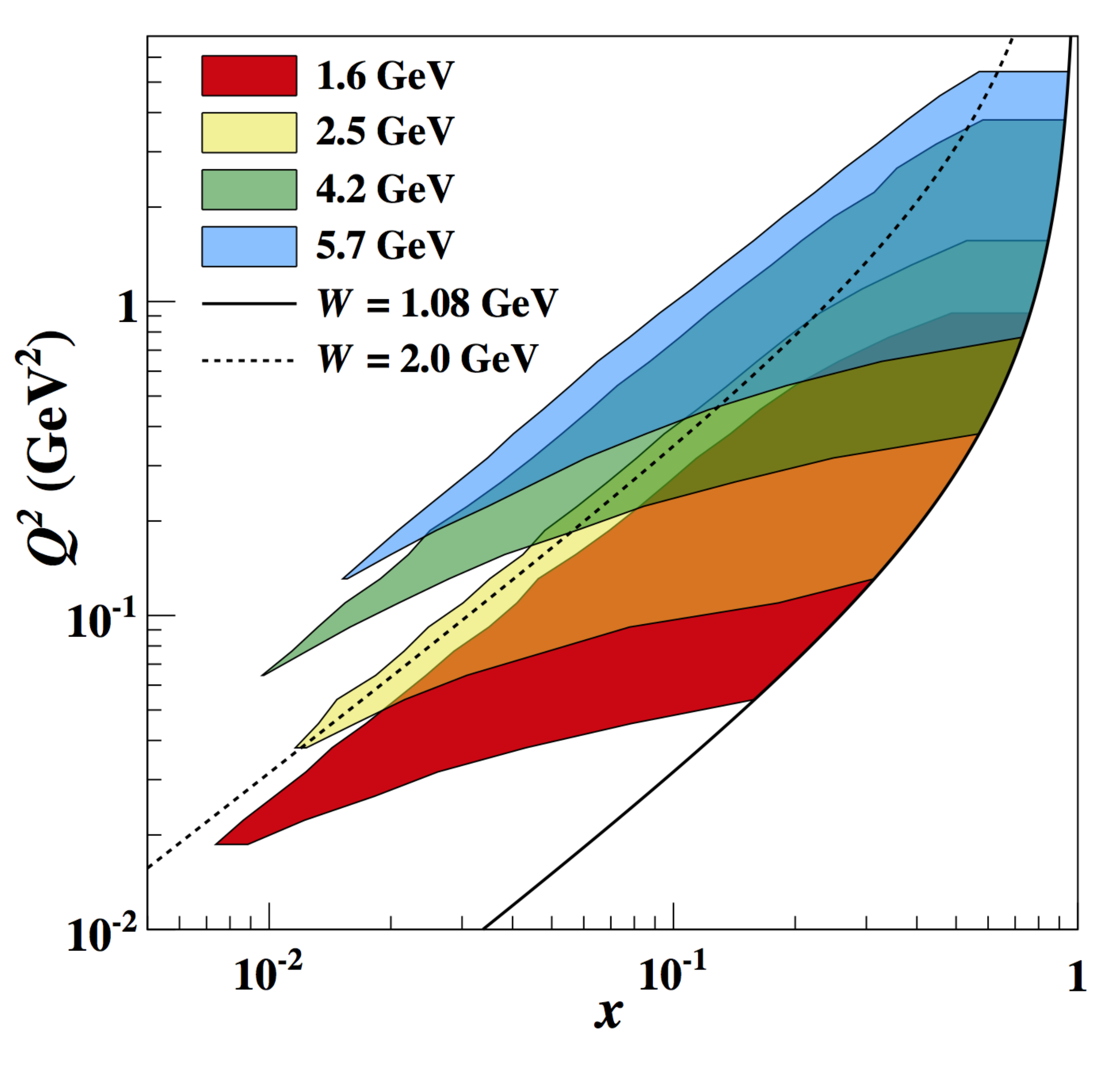}
\caption[Kinematic coverage of EG1b]{(Color online) Kinematic coverage in $Q^2$
  vs\ $x$ for each of the 4 electron 
 beam-energy groups in the EG1b experiment. The solid and dotted lines
denote the $W = $ 1.08 and 2.00 GeV thresholds, respectively.}
\label{kinrange:fig} 
\end{figure}

The EG1b data were collected over a 7-month period from 2000 to 2001.
More than 1.5$\times$10$^{9}$ triggers from the NH$_3$ target
were collected in 11 specific combinations ($1.606+$, $1.606-$, $1.723-$,
$2.286+$, $2.561-$, $4.238+$, $4.238-$, $5.616+$, $5.723+$,
$5.723-$, and $5.743-$) of beam energy (in GeV) and main torus
polarity ($+$,$-$), hereby referred to as ``sets.'' Sets with similar
beam energies comprise four groups with nominal average energies of
1.6, 2.5, 4.2 and 5.7 GeV.
The kinematic coverage for each of these four energy groups  
is shown in Fig.~\ref{kinrange:fig}.

Calibration of all detectors was completed offline according to 
standard CLAS procedures. 
These procedures use a subset of ``sample'' runs for each beam energy
and torus polarity to determine calibration constants for all ADC and TDC
channels. 
During analysis, these data were checked using these
constants, and additional calibrations were performed whenever 
necessary.

The calibration of the TOF system 
(needed for accurate time-based tracking)
resulted in an overall
timing resolution of $<$0.5 ns~\cite{Smith:1999ii}. 
Minimization of the distance-of-closest-approach (DOCA) residuals
in the DC led to typical values of 500 $\mu$m 
for the largest cell sizes (in region 3)~\cite{Mestayer:2000we}. The
EC provided a secondary timing
measurement for forward-going particles, and played a role
for the trigger and for particle identification \cite{Amarian:2001zs}. The mean timing difference
between the TOF and calorimeter signals was minimized,
yielding an overall EC timing resolution of $<$0.5 ns.

After calibration, all raw data were converted into particle
track information and stored (along with other essential run and event data) 
on data-summary tapes (DSTs).

\subsection{Quality assessment}

Quality checks were done to minimize potential bias introduced by malfunctioning detector 
components, changes in the target, and false asymmetries.
DST data that did not meet the
minimal requirements outlined in this section were eliminated from the analysis.

The electron count rate in each sector (normalized by the Faraday cup charge)
was monitored throughout every run. DST files with
count rates outside a prescribed range ($\pm$5\% and $\pm$8\%
for beam energies $<$3 GeV and $>$3 GeV, respectively)  were 
removed from the analysis in order to eliminate temporary problems, such as 
drift chamber trips, encountered during the experiment.

In order to minimize false asymmetries, 
the beam charge asymmetry 
$(Q_{\uparrow}-Q_{\downarrow})/(Q_{\uparrow}+Q_{\downarrow})$ for ungated cumulative  
charges $Q_{\uparrow} (Q_\downarrow)$ for positive (negative) helicities
was monitored. A cut of $\pm$0.005 on this asymmetry 
ensured that the false physics asymmetry due to this effect was much smaller than 10$^{-4}$.

\begin{figure}[h!tb]
\centering
\includegraphics[width=8.0cm]{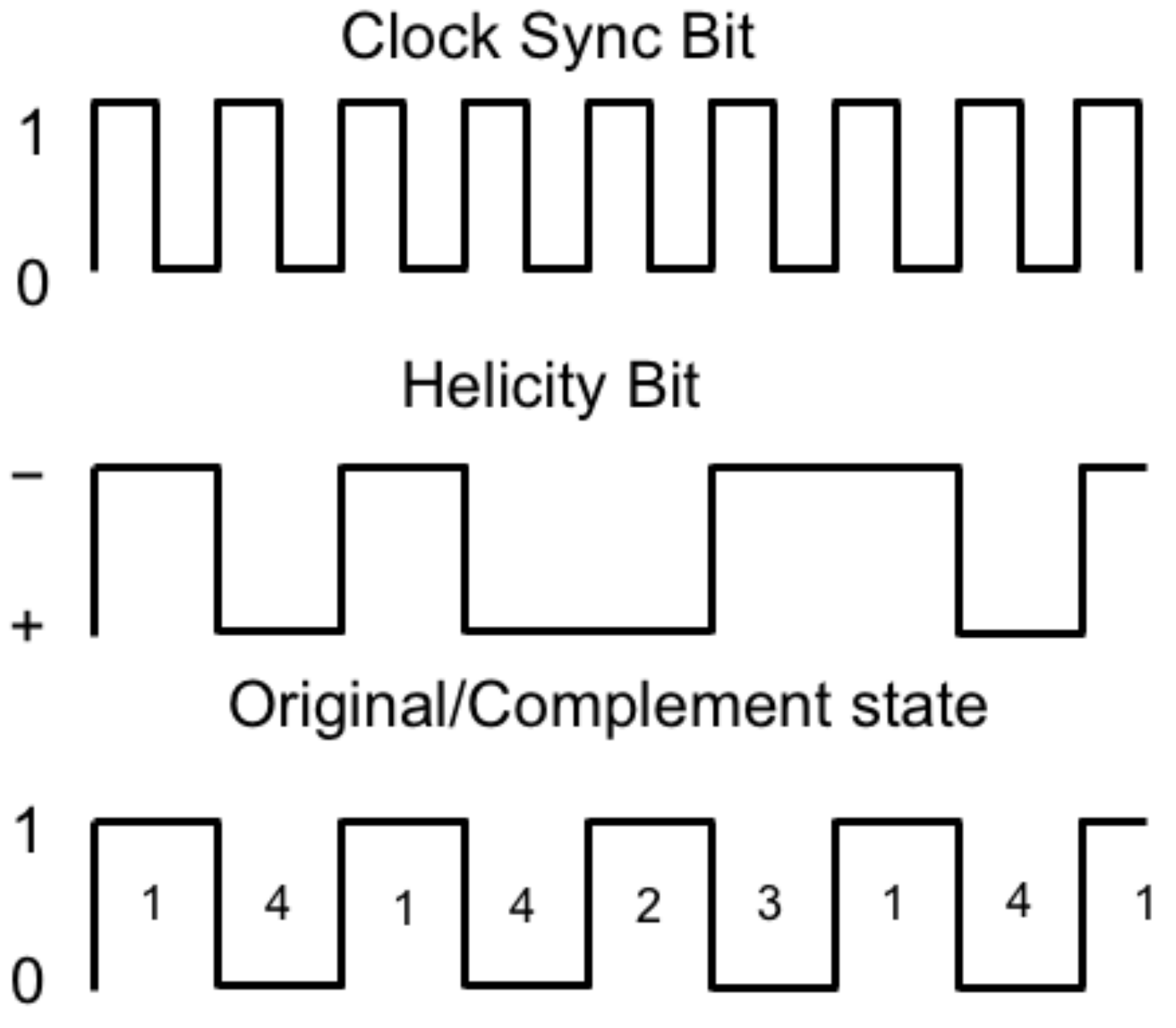}
\caption[Helicity labeling example]{Helicity signal logic.  
The clock signal (top) provided a rising edge every 30 ns. The helicity bit train (middle) was a pseudo-random
stream of opposite bit pairs.  The logic analyzed each helicity bit into four categories (bottom): 1, negative first bit
followed by its complement; 4, positive second bit preceded by its complement; 2, positive first bit followed by its 
complement; and 3, negative second bit preceded by its complement.
Buckets without a complementary partner were removed from the analysis.
}
\label{helclock:fig} 
\end{figure}

Electron helicities were picked pseudorandomly at 30 Hz,  always
in opposite helicity pairs to minimize non-physical
asymmetries. A synchronization clock bit with
double the frequency identified missing bits due to
detector dead-time or other uncertainties, allowing 
ordering of the
pairs (see Fig.~\ref{helclock:fig}). All unpaired helicity states were
removed from the analysis.

\begin{figure}
\centering

\includegraphics[width=9cm, angle=0]{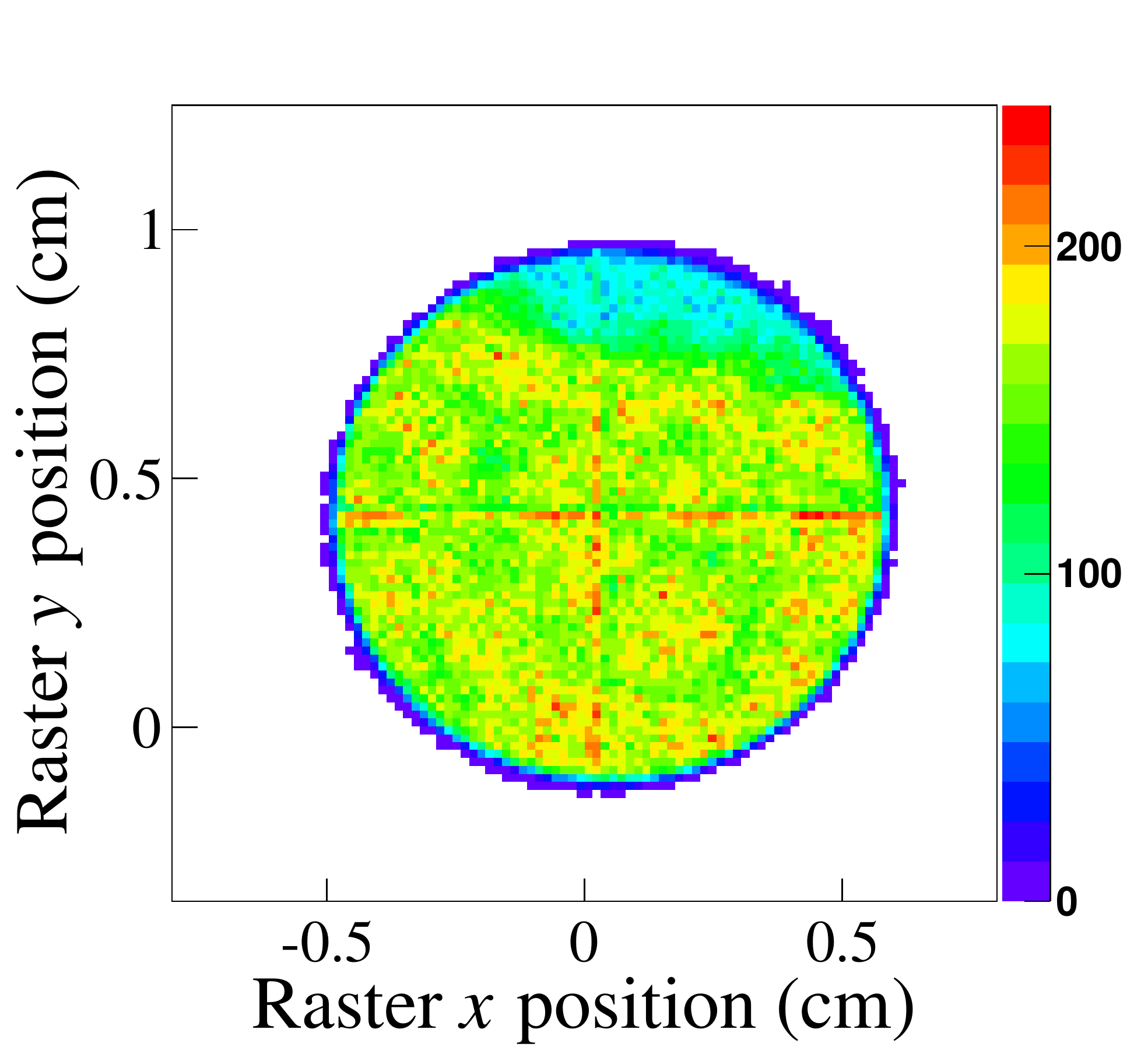}
\caption[Raster event rate density]{(Color online) Raster pattern for a sample run,
  demonstrating some temporary settling of the target material. (The ``crosshair'' pattern is a non-physical relic of the coordinate reconstruction.)
}

\label{rastersample:fig}
\end{figure}

Plots of beam raster patterns were used to monitor target density and beam quality
(see Fig.~\ref{rastersample:fig}). 
Data obtained when raster patterns
exhibited elevated count rates in regions where the beam was grazing the target cup
were also excluded entirely from analysis.\footnote{In one unique case where empty-target runs meeting our selection criteria runs were not available, only data corresponding to anomalous raster regions were removed. A systematic normalization uncertainty of ~2\% on event counts from these runs, obtained from comparison to unaffected runs, is incorporated into our analysis.}

\subsection{Event selection}
As a starting point for the selection of events, particles with momentum $p\ge 0.20E_{beam}$
that fired both the CC and EC triggers were treated as electron candidates. Additional criteria, discussed below, were then applied to minimize background from other particles, primarily $\pi^-$.

\subsubsection{Cherenkov counter cuts}

The CCs use perfluorobutane (C$_4$F$_{10}$) gas, 
and have a threshold of $\sim$9 MeV/$c$ for electrons and 
$\sim$2.8 GeV/$c$ for pions. 
Between these two momenta, the CC
efficiently separated pions from electrons. A minimum of 2.0 
detected photoelectrons (p.e.) in the CC PMTs was required for electron candidates 
with $p < 3.0$ GeV/$c$. For particles with higher momentum, a minimum cut
of 0.5 p.e. was used only to eliminate contributions from
internal PMT noise.

\begin{figure}[h!tb]
\centering
     \includegraphics[width=9.5cm]{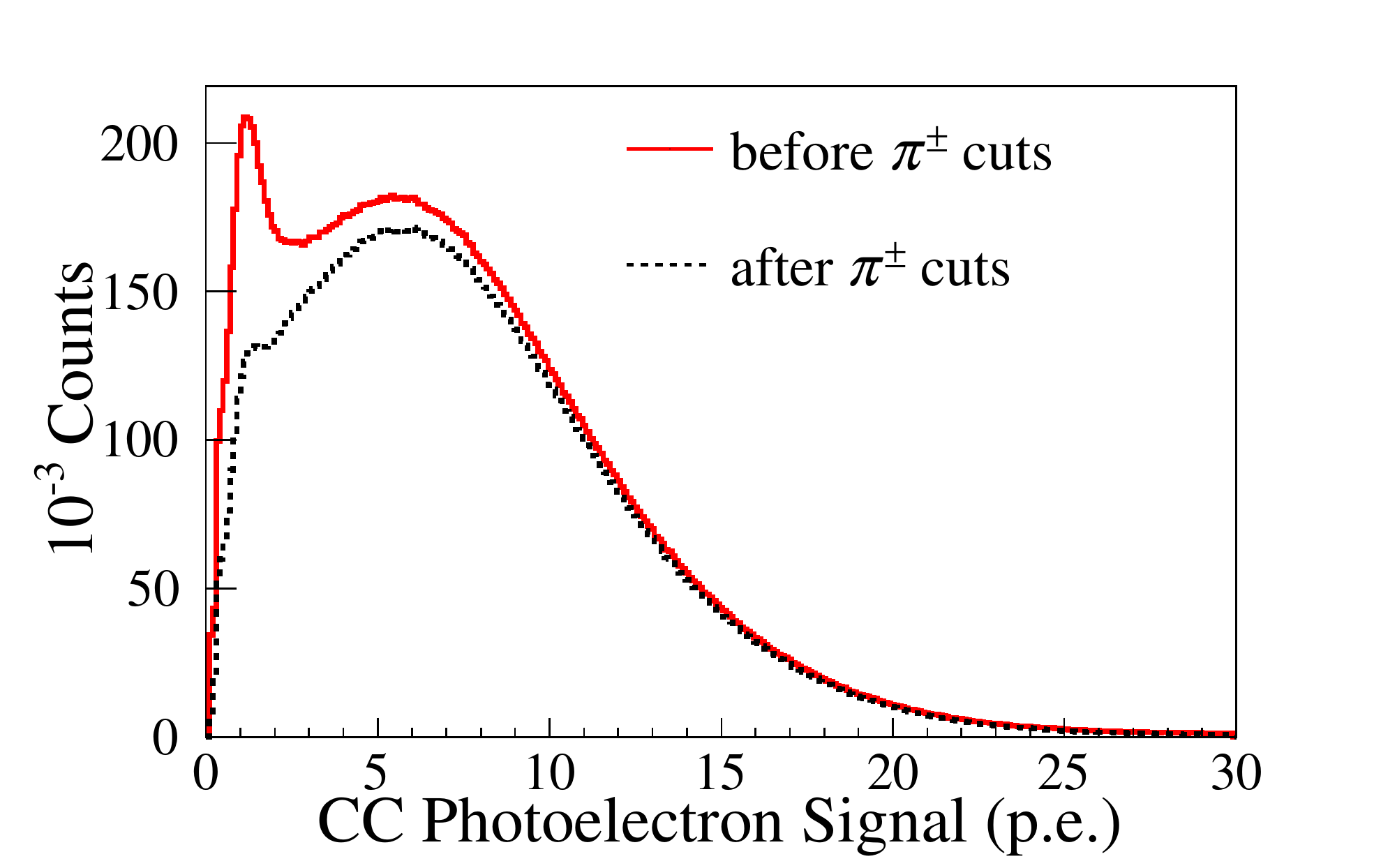}
\caption[Effect of precision Cherenkov cuts on photoelectron spectra]
{(Color online) Cherenkov signal distributions before (red, solid line) and after (black, dotted line) requiring track matching.
}
\label{osibin:fig}
\end{figure}

Geometric and time matching
requirements 
between CC signals and measured tracks were used
to reduce background.
These cuts on the correlation of the CC signal with the triggering
particle track removed the majority of the contamination
dominating the lower part of the CC signal spectrum. The
effect of these cuts is shown in Fig.~\ref{osibin:fig}.  Pion contamination at 
low signal heights was reduced substantially with little loss of good events.

The determination of dilution factors (see Sec.~\ref{DFcorr}) 
required a precise comparison of count rates for different
targets. Therefore, detector acceptance
and efficiency for runs on different targets had to remain constant.
Inefficiencies in the CC were the main source of uncertainty in electron detection efficiency
for CLAS. Therefore, tight fiducial cuts were developed to select  the region
where the CC was highly efficient. These cuts were used only for the 
dilution factor analysis.

\begin{figure}[h!tb]
\centering
 \includegraphics[width=9.0cm]{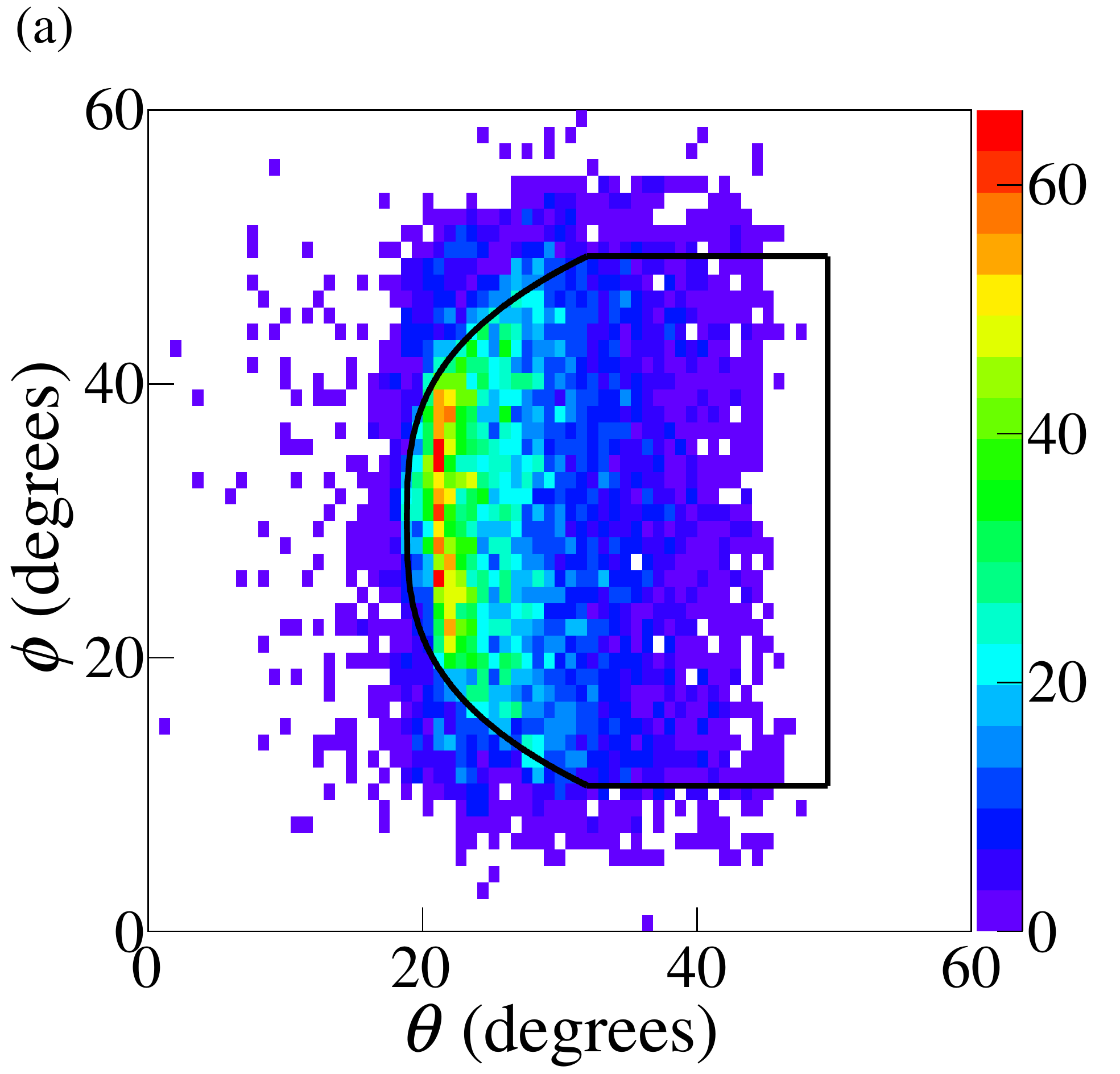}
 \includegraphics[width=9.0cm]{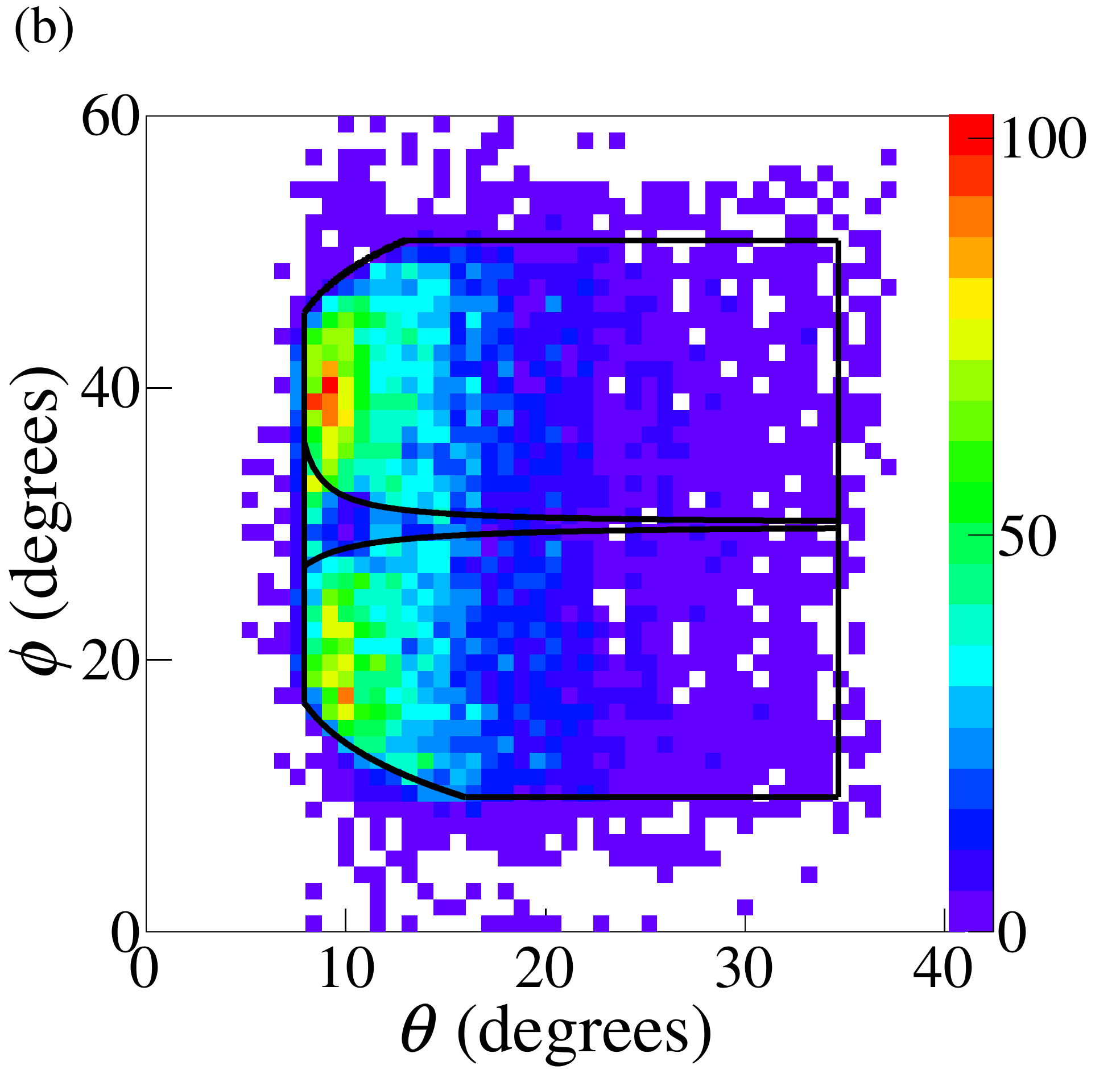}
\caption[In/outbending fiducial cuts (low momentum)]
{(Color online) Sample fiducial cuts for (a) inbending and (b) outbending
  electrons, shown in $\phi$ vs\ $\theta$ for one CLAS sector. 
}
\label{fidin:fig}
\end{figure}

The CC efficiency is defined by the integral of an assumed Poisson distribution yielding
the percentage of electron tracks generating signals above the  2.0
p.e. threshold.
It varied as a function of kinematics due to the CC mirror geometry. 
The mean value of the signal
distribution was determined as a function of
electron momentum $p$ and angles  $\theta$ and $\phi$ using $ep$ 
elastic events from several CLAS runs at beam energies of 1.5$-$1.6 GeV.
The deduced efficiency  map has  a plateau of high efficiency in 
the center of each
sector, which rapidly drops off to zero at the
sector edges. For the fiducial cut, we developed a function 
of $p$, $\theta$, and $\phi$ to define a boundary enclosing events 
with more than 80\% CC efficiency in each 0.5 GeV momentum interval (see Fig. \ref{fidin:fig}). Fiducial cuts were specific to each CLAS torus
setting. Additional center-strip cuts in each sector were required to
remove regions with inefficient detector elements.

\subsubsection{Electromagnetic calorimeter cuts}

\begin{figure}[h!tb]
\centering
\includegraphics[width=9cm, angle=0]{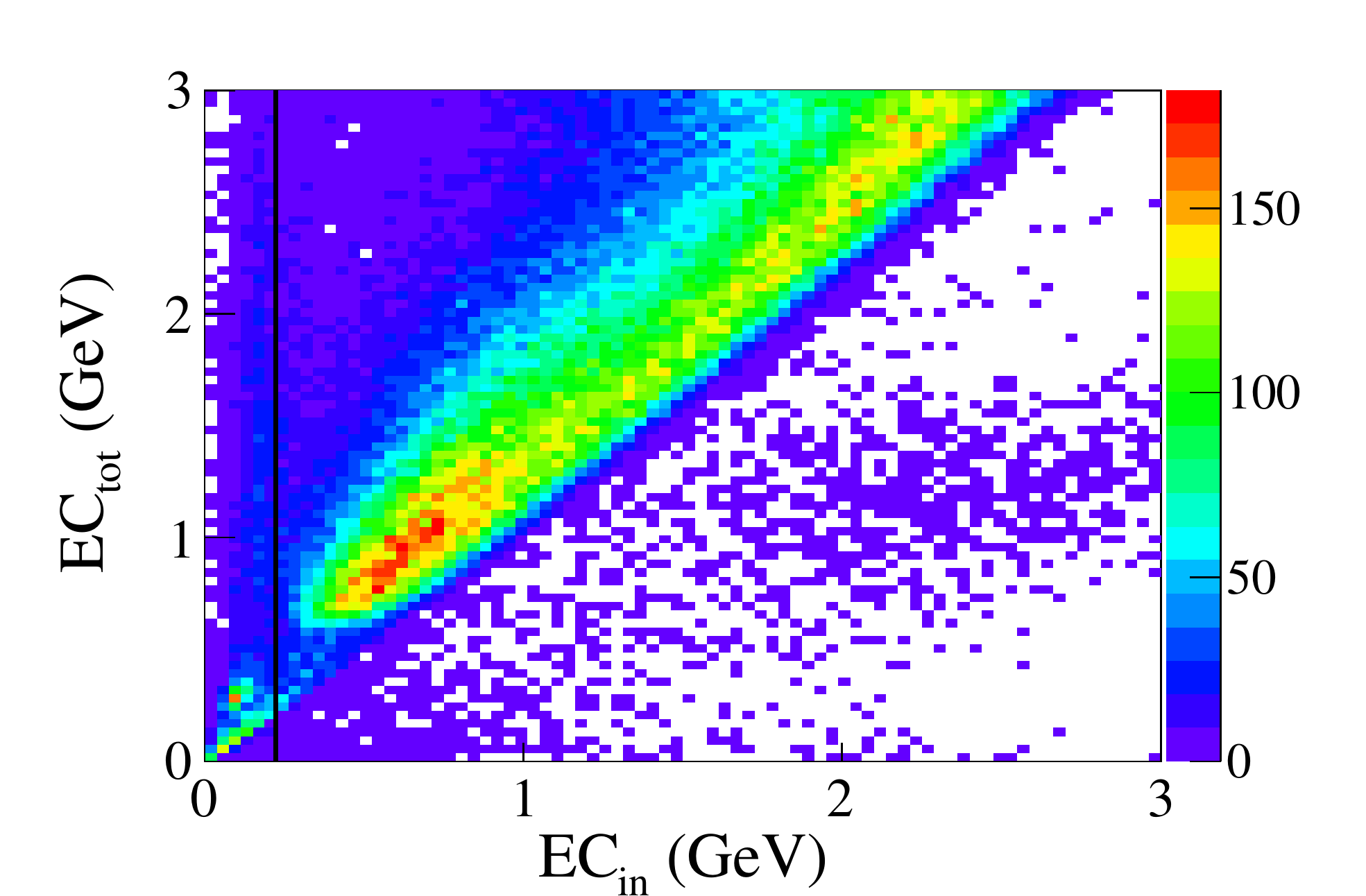}
\caption[Particle Identification by Electromagnetic Calorimeter energy]
{(Color online) The total energy  $EC_{tot}$ deposited in the EC vs\ the energy $EC_{in}$ deposited in
  the inner (front) layer of the EC only for electron candidates. 
  The plot shows a clear separation of
  electrons from light hadrons (bottom left corner). A cut on $EC_{in}$ (shown by the
  vertical line) removes most of the hadron background.
}
\label{ECin:fig}
\end{figure}

Further suppression of pion backgrounds was provided by the EC,
in which minimum ionizing particles (hadrons) deposited far less energy
than showering electrons. A base cut was developed by
observing the energy $EC_{tot}$ deposited in the entire EC  and
the energy  $EC_{in}$ deposited only in the first  5 of 13 layers
(see Fig.~\ref{ECin:fig}). A loose cut of $EC_{in}<$ 0.22 GeV
(including the sampling fraction \cite{Amarian:2001zs}) was used as a first step in
separating pions from electrons in the calorimeter.

\begin{figure}[h!tb]
\centering
    \includegraphics[width=9cm]{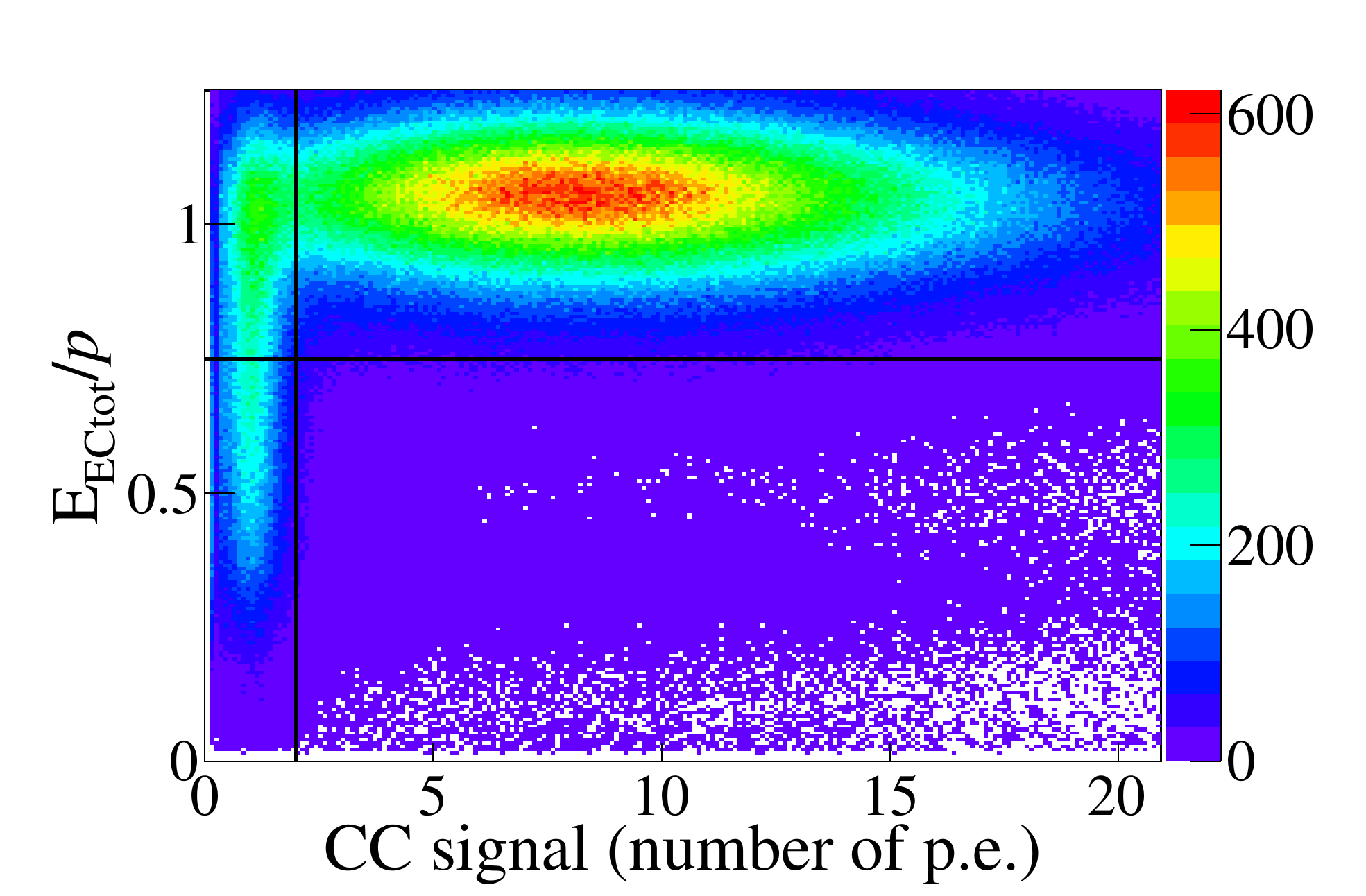}
\caption[Effects of fiducial cuts on pion contamination]{(Color online)
  Scatter plot of $EC_{\rm tot}/p$ vs\ CC signal, at $p<$ 3 GeV/$c$, after fiducial cuts. Only events to the right and above the straight lines are kept as inclusive electrons.
}
\label{PID:fig}
\end{figure}

The EC cuts were further refined by taking into account the relationship between
the momentum of the particle and the energy deposited in the calorimeter.
Since electrons deposited practically
all of their energy in the calorimeter, a lower bound on $EC_{tot}/p$
further reduced contributions from pions. 
For $p> 3$ GeV, where the CC spectrum fails to differentiate
pions and electrons, a strict cut of $EC_{tot}/p>$0.89 was applied,
while a looser cut of $EC_{tot}/p>$0.74 is used at $p<$3 GeV. Figure \ref{PID:fig} 
shows these cuts for events plotted in $EC_{tot}/p$ versus the CC photoelectron signal.

\subsubsection{Remaining $\pi^-$ contamination}

The remaining pion contamination was determined as a function of $\theta$ (5$^\circ$ bins) 
and $p$ (0.3 GeV bins) as follows in each $p$, $\theta$ bin: A modified, extrapolated 
Poisson distribution fit to our CC p.e. spectrum was subtracted from 
the pion ``peak'' seen at low p.e. values (see Fig. \ref{osibin:fig}) to get 
a low p.e. contamination estimate. Then, we analyzed only runs without 
the CC trigger in use, inverting all the electron selection cuts on the EC, 
resulting in a test sample composed nominally of pions. This sample was then normalized to 
the low p.e.\ contamination estimate at p.e. $<$ 2.0.  The normalized nominal 
pion data provided an estimate of the $\pi^-$ contamination present at p.e. $>$ 2.0, 
where the inclusive electrons lie. Dividing by the total number of inclusive electrons 
yielded the contamination fraction $R^p(\theta,p)$.

\begin{figure}[h!tb]
\centering
\includegraphics[width=8cm]{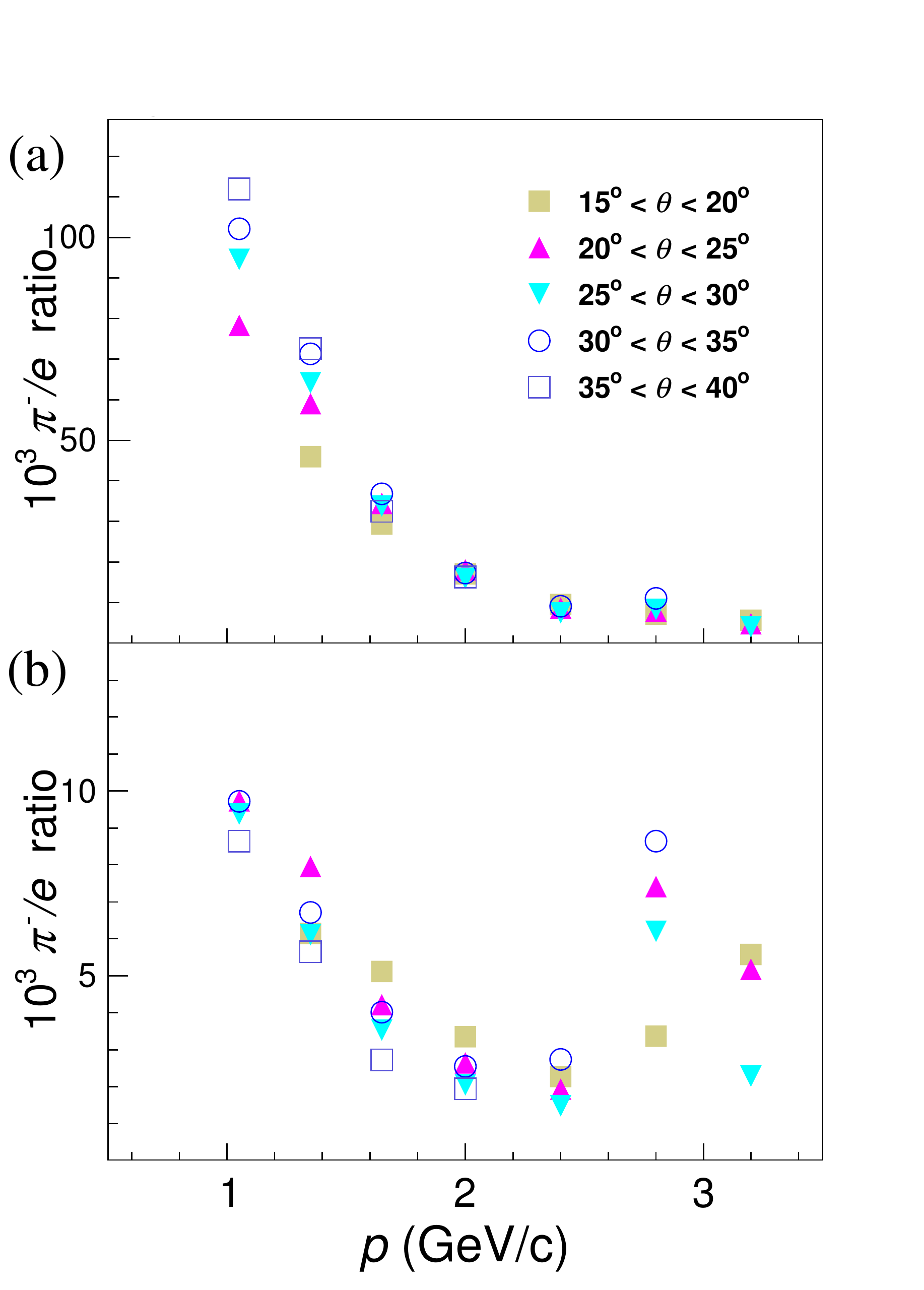}
\caption[Particle Identification by Electromagnetic Calorimeter energy and momentum]
{(Color online) Pion contamination fraction (a) before and (b) after track-matching
  cuts for the 5.7 GeV beam energies, as a function of polar angle and momentum.
  The increase beyond $p=2.8$ GeV/$c$ indicates the threshold beyond which pions start to produce a signal in the CC.
}
\label{pioncon:fig}
\end{figure}

Plots of the pion contamination fractions as a function of $p$ and $\theta$
are shown in Fig.~\ref{pioncon:fig}. These were seldom more
than 1\% of the total electron count. An exponential function 
\begin{equation}
\label{contam1:eq}
R(\theta,p) = e^{a+b\theta+cp+d\theta p}
\end{equation}
was then fit to these points. 
Pion contamination corrections could be made by adding
\begin{equation}
\label{contam2:eq}
\Delta A_{raw} = \frac{ R(\theta,p)(A_{raw}-A_\pi) }{ 1 -
  R(\theta,p) } 
\end{equation}
to the raw asymmetry $A_{raw}$. Since the effect is very small, and
the inclusive pion asymmetry $A_\pi$ is not well known, we 
applied no correction and instead treat $\Delta A_{raw}$ with $A_\pi = 0$ as the 
systematic uncertainty. 

\subsubsection{Background subtraction of pair-symmetric electrons}
Dalitz decay of neutral pions \cite{Dalitz:1951aj} and Bethe-Heitler
processes~\cite{Gehrmann:1997qh} 
can produce $e^+e^-$ pairs at or near the vertex, contaminating the
inclusive $e^-$ spectrum. To determine this contamination, we
assumed that the event reconstruction and detector acceptances for $e^+$
production were identical to those for their paired $e^-$ when the main
torus current was reversed, and that the overall cross-section is small
enough that small differences in beam energy (e.g. 2.286 versus 2.561 GeV)
minimally affected the production rate.

\begin{figure}
\centering
  \includegraphics[width=8.5cm]{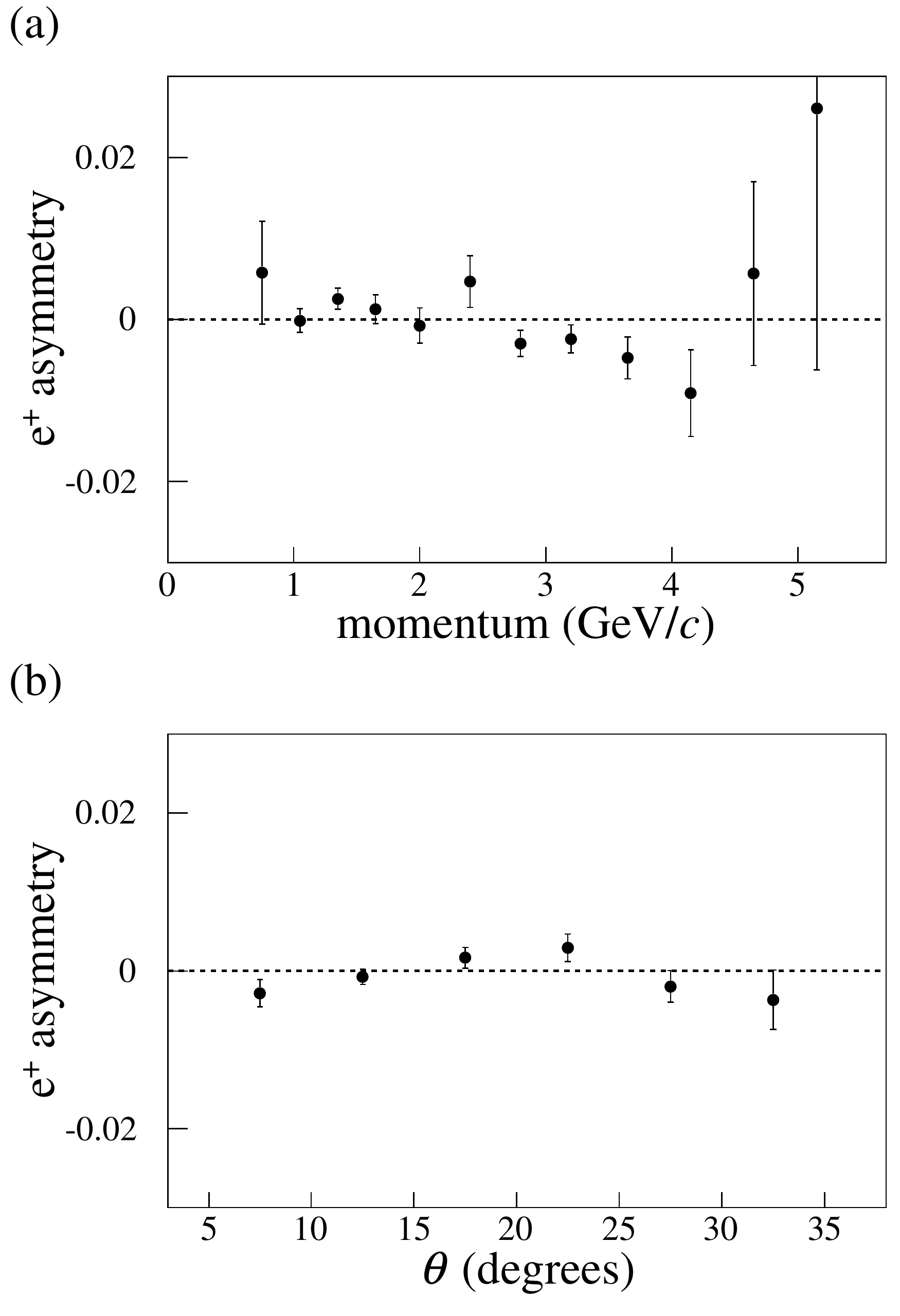}
\caption[Positron asymmetries as a function of $p$]
{Average positron asymmetries for the 5.7 GeV data set 
  as a function of (a) momentum and (b) scattering angle $\theta$. 
}
\label{epemasym:fig}
\end{figure}

\begin{figure}
\centering
  \includegraphics[width=8.3cm]{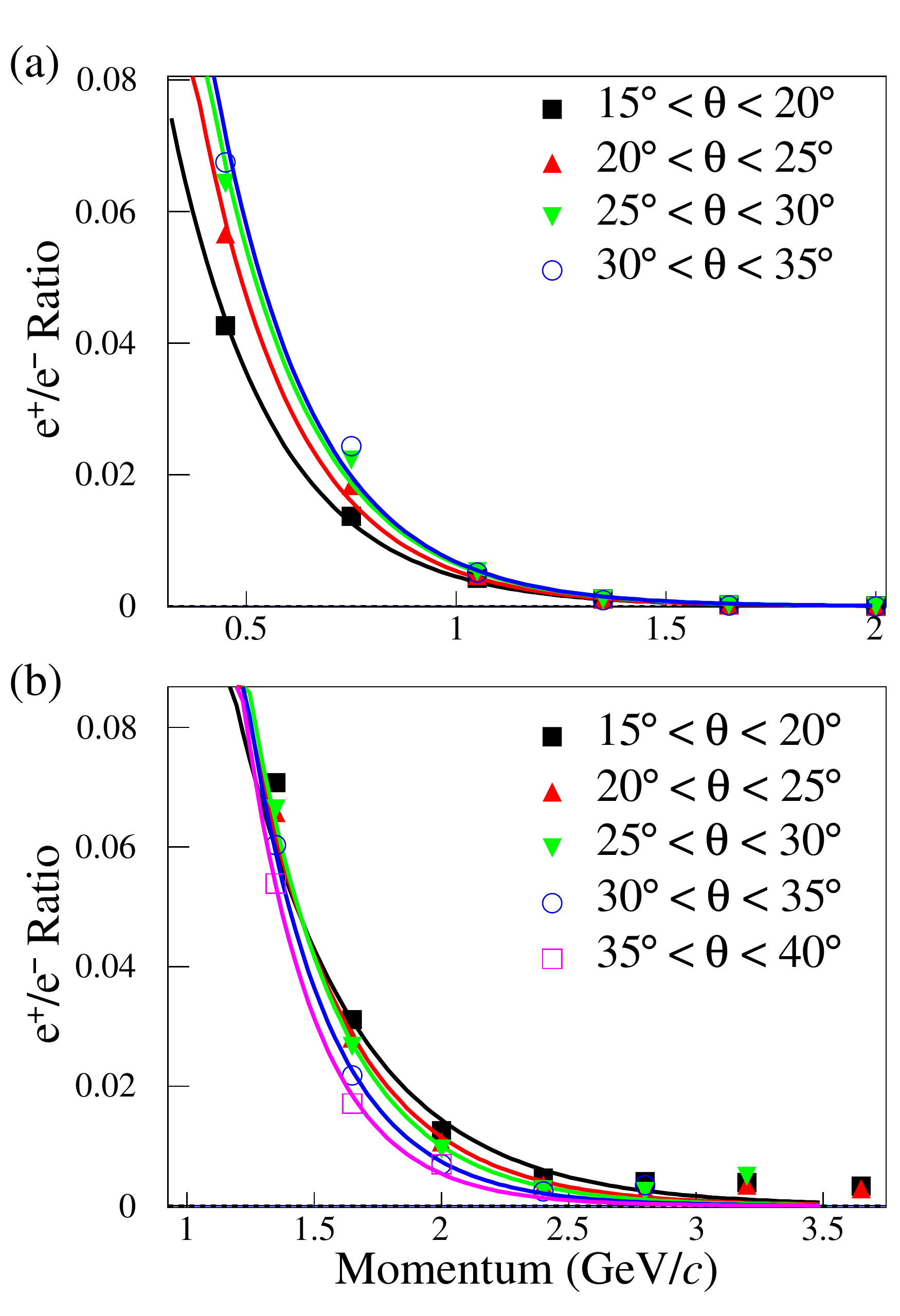}
\caption[Ratios of $e^+/e^-$ vs/ $p$]{(Color online) Ratios of $e^+/e^-$ as a function of 
electron momentum $p$, at various
  $\theta$ angles, for the (a) 2.561$-$ and (b) 5.727+ data sets.
}
\label{epem1:fig}
\end{figure}

Each data set was correlated with another having a similar beam energy but
opposite torus polarity. Events with leading positron triggers
were analyzed identically to those with electron triggers. The overall
double-spin asymmetry for $e^+$ triggers was small (see Fig. \ref{epemasym:fig}). The $e^+/e^-$ contamination ratios $R^p$, which were largest
at low momenta (Fig.~\ref{epem1:fig}), were fit with the
parametrization of Eq.~(\ref{contam1:eq}). Then, Eq.~(\ref{contam2:eq}) (with
$A_\pi\rightarrow A_{e^+}$) was used to determine a multiplicative background 
correction factor 
$C_{\rm back} \equiv (A_{raw}+\Delta A_{raw})/A_{raw}$ to convert the raw
asymmetry to the background-free physics asymmetry.  Here we assumed that $A_{e^+} = 0$,
consistent with the average from our measurements (see Fig.~\ref{epemasym:fig}).

To estimate the systematic uncertainty from this background, two changes
were made to $C_{\rm back}$ in the reanalysis. $R^p$ was changed by half the difference between two
equivalent determinations: one using outbending electrons and inbending positrons, 
and the other using the opposite torus polarities for either particle.
Also, $A_{e^+}$ was set to a nonzero value equal to
3 times the statistical uncertainty of the averaged positron asymmetry.

\subsubsection{Elastic $ep\rightarrow e'p$ event selection}

Both the momentum corrections (Sec.\ref{momentum:sec}) and the determination
of beam polarization $\times$ target polarization (Sec. \ref{pbpt:sec}) required 
identified elastic $ep$ scattering events. For this purpose, we selected
two-particle events containing an electron and one 
track of a positively charged particle. Electron PID cuts were relaxed to require only a minimum of 0.5 CC
p.e. The $E/p$ EC cut thresholds were
lowered to  0.56 for $p<3$ GeV/$c$ and 0.74 for $p>3$ GeV/$c$.
These relaxed cuts increased the statistics
while the exclusivity cuts discussed below removed all pion background.

\begin{figure}
\centering
  \includegraphics[width=7.25cm]{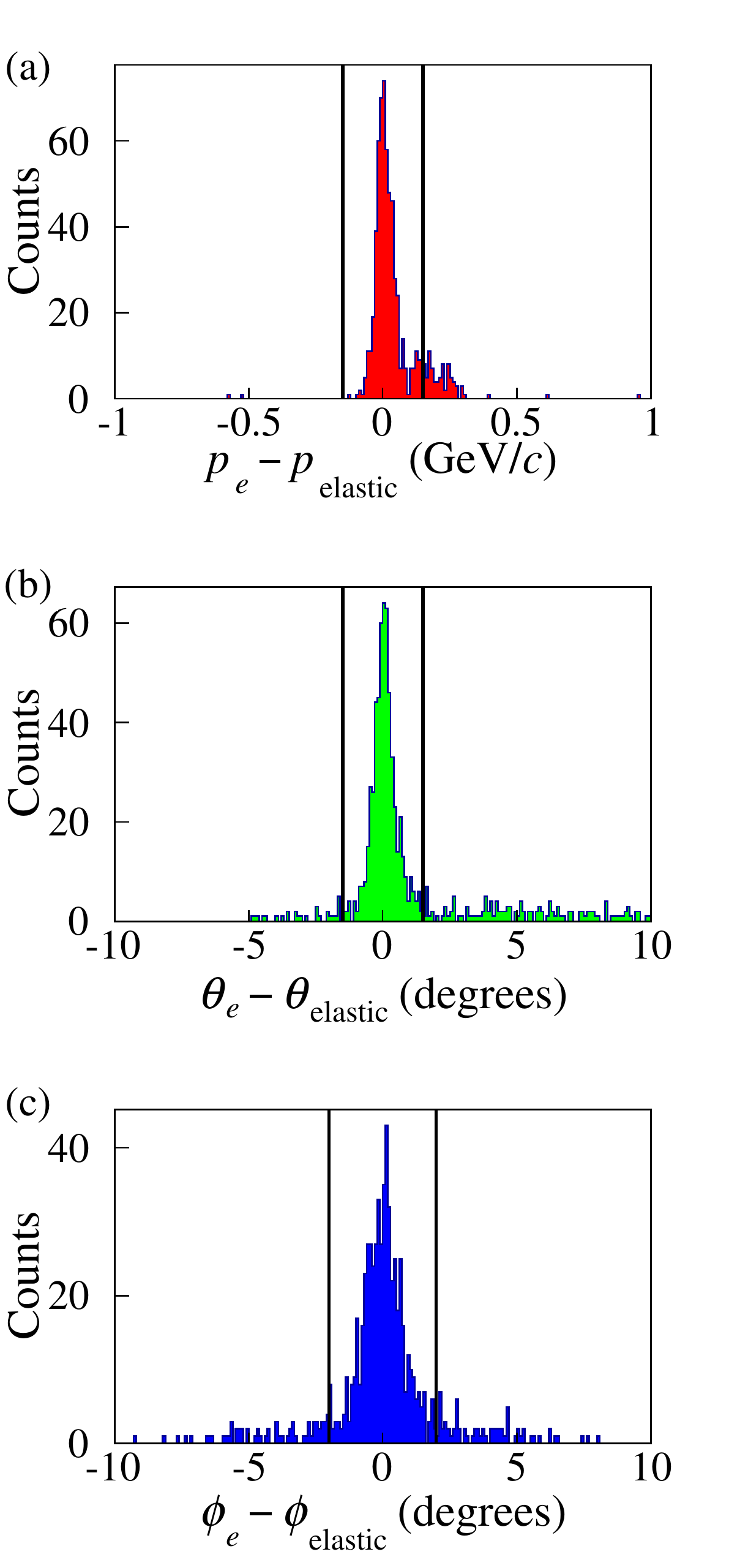}
\caption[Elastic $ep$ event kinematic cuts]
{(Color online) Kinematic cuts on (a) the difference between measured and expected momentum, (b) polar angle, and (c) azimuthal
  angle of elastic $ep$ events. Each of the distributions has the other two cuts applied.
}
\label{epcut:fig}
\end{figure}

A beam-energy-dependent cut on $|M_p - W|$ (where $M_p$ is the proton mass), which ranged from 30 MeV at 1.6 GeV to 50 MeV at 5.7 GeV,
suppressed inelastic contributions.
Further kinematic constraints
were applied on deviations of the missing momentum $p$, the proton polar angle
$\theta$, and the difference between the azimuthal proton and electron angles $\Delta\phi$,
from those expected for elastic $ep$
kinematics (see Fig.~\ref{epcut:fig}). Final
cuts of $\Delta p<$ 0.15 GeV, $\Delta\theta<$ 1.5$^\circ$ and
$\Delta\phi<$ 2.0$^\circ$ identify elastic $ep$ events, with
typically less than 5\% nuclear background (see Fig.~\ref{phiscale:fig}).


\subsection{Event corrections}

The reconstructed track parameters of each event were corrected
for various distortions to extract the  correct kinematic variables
at the vertex.
These kinematic corrections are explained in the
following two subsections.

\subsubsection{Phenomenological kinematics corrections}
Kinematic corrections were implemented to 
account for the effects of energy loss from ionization, multiple scattering,
and geometrical corrections to the 
reconstruction algorithm (for target
rastering and stray magnetic fields).

\begin{figure}
\centering
 \includegraphics[width=9.5cm]{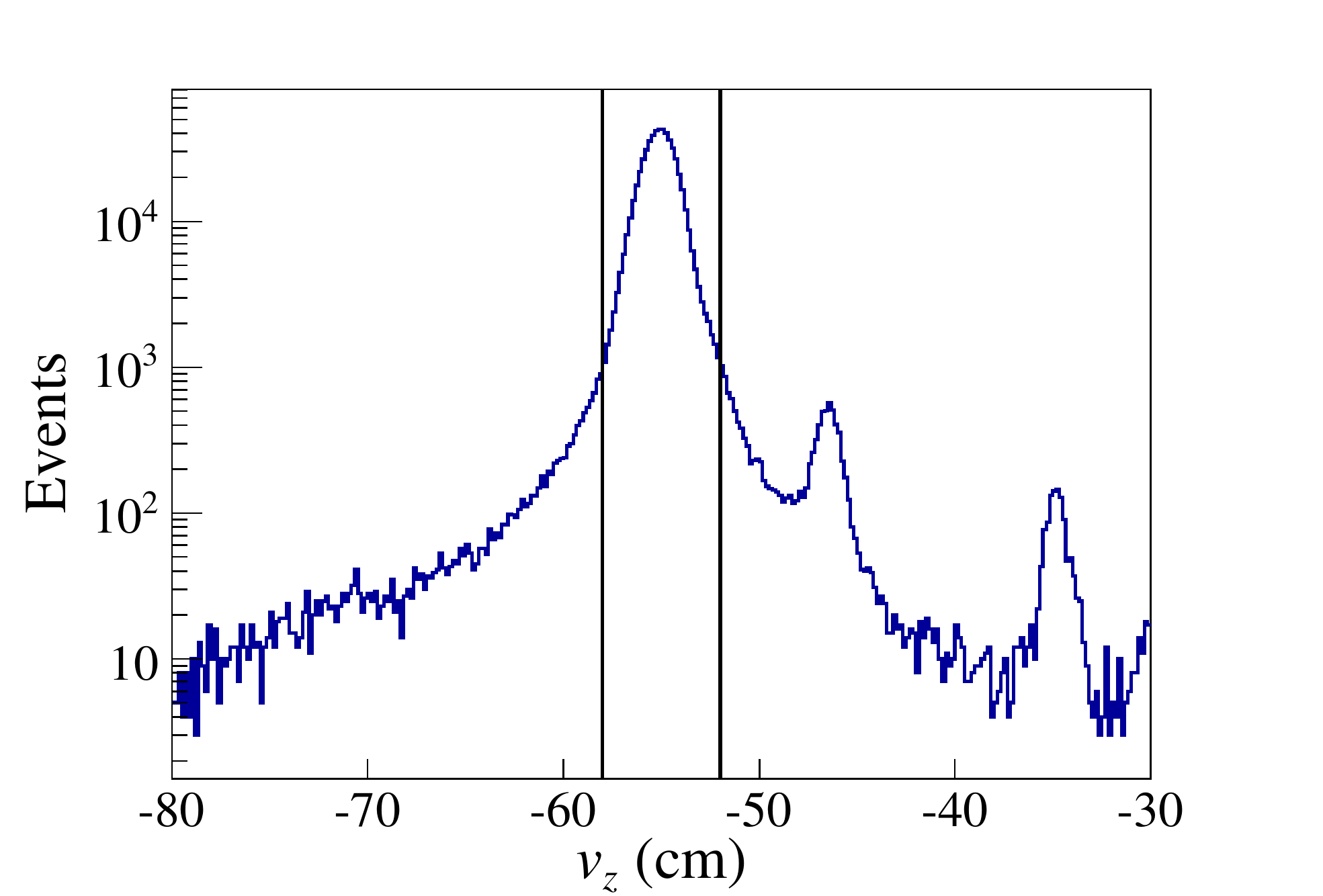}
\caption[Vertex z positions.]{(Color online) 
Vertex $z$ positions for electrons after corrections for the raster.
Secondary peaks correspond to target windows.
A vertex cut of ($-$58 $< v_z <$ $-$52 cm) was applied as shown to select
events from the target.
}
\label{vertexCut:fig}
\end{figure}

Rastering varies the $xy$ position of the beam over the target in a
spiral pattern with a radius of
$\sim$0.5 cm (see  Fig.~\ref{rastersample:fig}).
The instantaneous beam position can be reliably extracted from the raster 
magnet current.
The reconstructed $z$-vertex position (the $z$ axis is along the beam line)
and the ``kick'' in $\phi$
were corrected for
this measured displacement of the interaction point from the nominal
beam center \cite{rastercorr:ref}, 
prior to the application of a nominal ($-58<v_z<-52$ cm) vertex cut 
(see Fig.~\ref{vertexCut:fig}).

Collisional energy loss of both incident and scattered electrons 
within the target was accounted for by assuming a 2.8 MeV/(g/cm$^2$)
energy loss rate $dE/dx$ for electrons \cite{Yao:2006px}. The calculation, incorporating the
target mass thickness, vertex position, and polar scattering angle
$\theta$, yielded typical energy losses of $\sim$2 MeV before and after
the event vertex. The energy loss of scattered hadrons was similarly
estimated using the Bethe-Bloch formula \cite{Leobook}.
\\
\indent
Determination of the effects of multiple scattering on kinematic reconstruction was more
complex, and was studied with the {\footnotesize GEANT} CLAS simulation package {\footnotesize GSIM }
\cite{GSIM:ref}. For multi-particle events,
an average vertex position was determined by calculating a
weighted 
average of individual reconstructed particle vertices. 
Comparing each particle vertex with this average gives a best estimate
for the effect of multiple scattering on that particle on its way to the 
first drift chamber region. The {\footnotesize GSIM} model was then used to
generate an adjustment $d\theta(\theta,1/p)$ \cite{ms:ref} to the measured scattering angle. 

The {\footnotesize GSIM} package was also used to provide a leading-order correction due to 
magnetic field effects not incorporated
into the main event reconstruction software. Particularly important is the
extension of the target solenoid field into the inner layer DC. This study
resulted in corrections applied to the polar angle
$d\theta(\theta,1/p)$ and the azimuthal angle $d\phi(\theta,1/p)$ \cite{ms:ref}.

\subsubsection{Empirical momentum corrections}
\label{momentum:sec}

Imperfect knowledge of the field map of the CLAS magnet, 
misalignment of the drift chamber wires or the drift chambers themselves 
relative to their nominal positions, effects of wire sag, and other possible 
distortions in the drift chamber wire positions used in the tracking code
lead to deviations in the reconstructed kinematics of the scattered 
particles. An empirical method was developed \cite{clasnote2003-005} to 
correct the measured momenta of the particles, using
parameters that were determined by exploiting 
the four-momentum ($p_\mu$) conservation for both elastic $ep$  and  
two-pion production $ep\rightarrow ep\pi^+\pi^-$ events.  

The overall correction function depends on the momentum $\vec{p}$, 
the polar angle $\theta$ and the azimuthal angle $\phi$. It includes 
16 parameters for each sector, totaling 96 parameters, and 7 additional
parameters to improve the fit in the case of negative
torus magnet polarities. 
Corrections in the momentum and polar angle were calculated relative to the
region 1 drift chamber. The azimuthal angle, having a larger intrinsic
uncertainty, was kept fixed since it
was shown to be correct within this uncertainty for elastic events.
\\
\indent
The parameters were optimized by minimization of 
\begin{equation}
\chi^2 = \sum_{i}\sum_{\mu} 
\frac{\Delta p^2_{\mu}}{\sigma^2_{p_{\mu}}} 
+ \sum_{e} 
\frac{(W_{c} - M_{p})^2}{(0.020\text{ GeV})^2},
\label{chi2_A:eqn}
\end{equation}
over $i$ total events and $e$ elastic events. Here, $p_\mu$ are
the components of the missing four-momentum and $\sigma_{p_{\mu}}$ are the expected
resolutions of each component, $\sigma_{p_x} = \sigma_{p_y} =$ 0.014 GeV 
and $\sigma_{p_z} = \sigma_E =$ 0.020 GeV, $M_{p}$ is proton mass, and
$W_{c}$ is the missing mass of the inclusive elastic event.

\begin{figure}
\centering
\includegraphics[width=8.8cm]{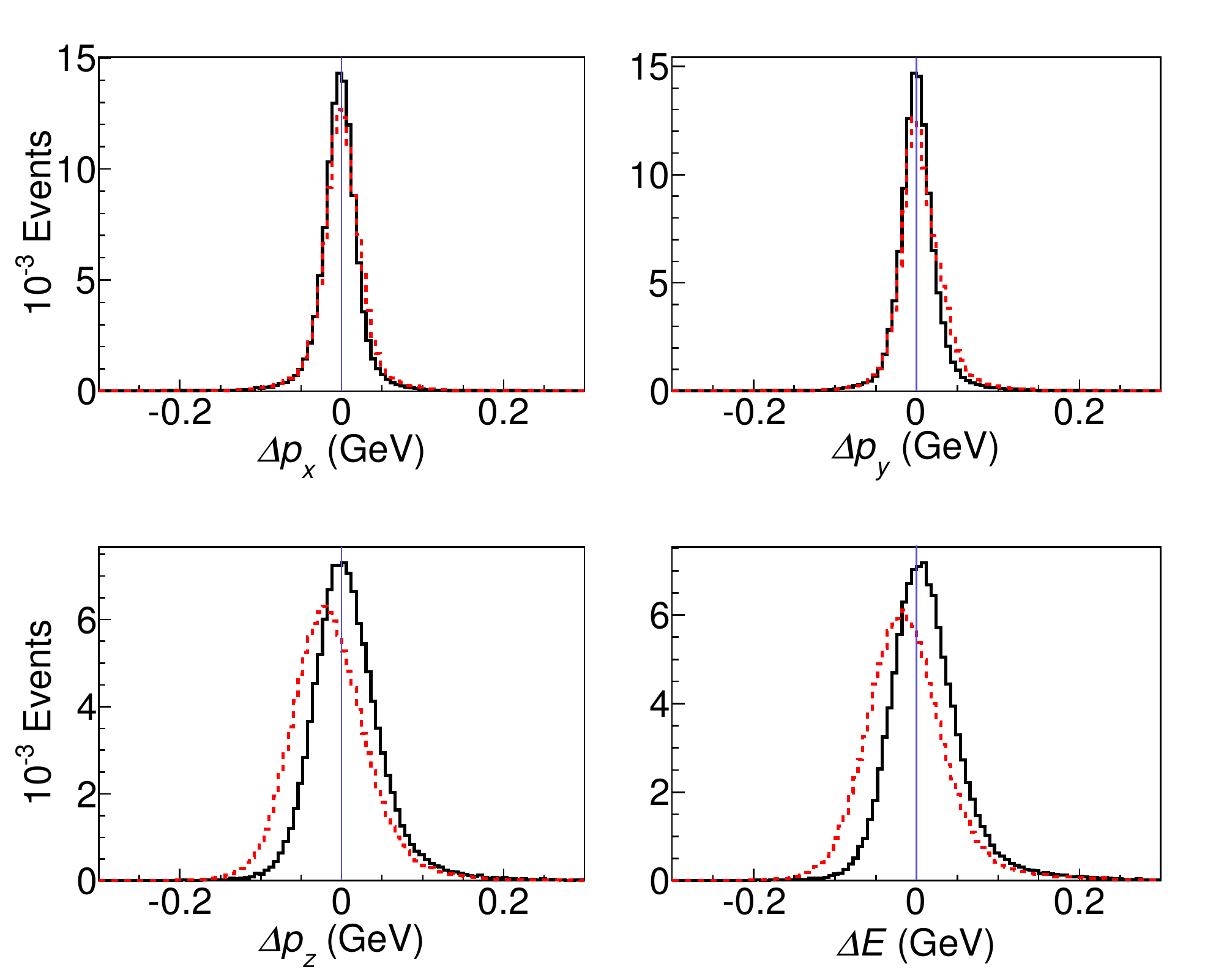}
\caption[Missing energy and momentum distributions for the elastic events.] 
{(Color online)  Missing energy and momentum distributions from elastic events in the
4.238 GeV inbending data set before (dashed red line) and after
(black solid line) momentum corrections.
}
\label{momentum:fig}
\end{figure}

\begin{figure}
\centering
\includegraphics[width=9cm]{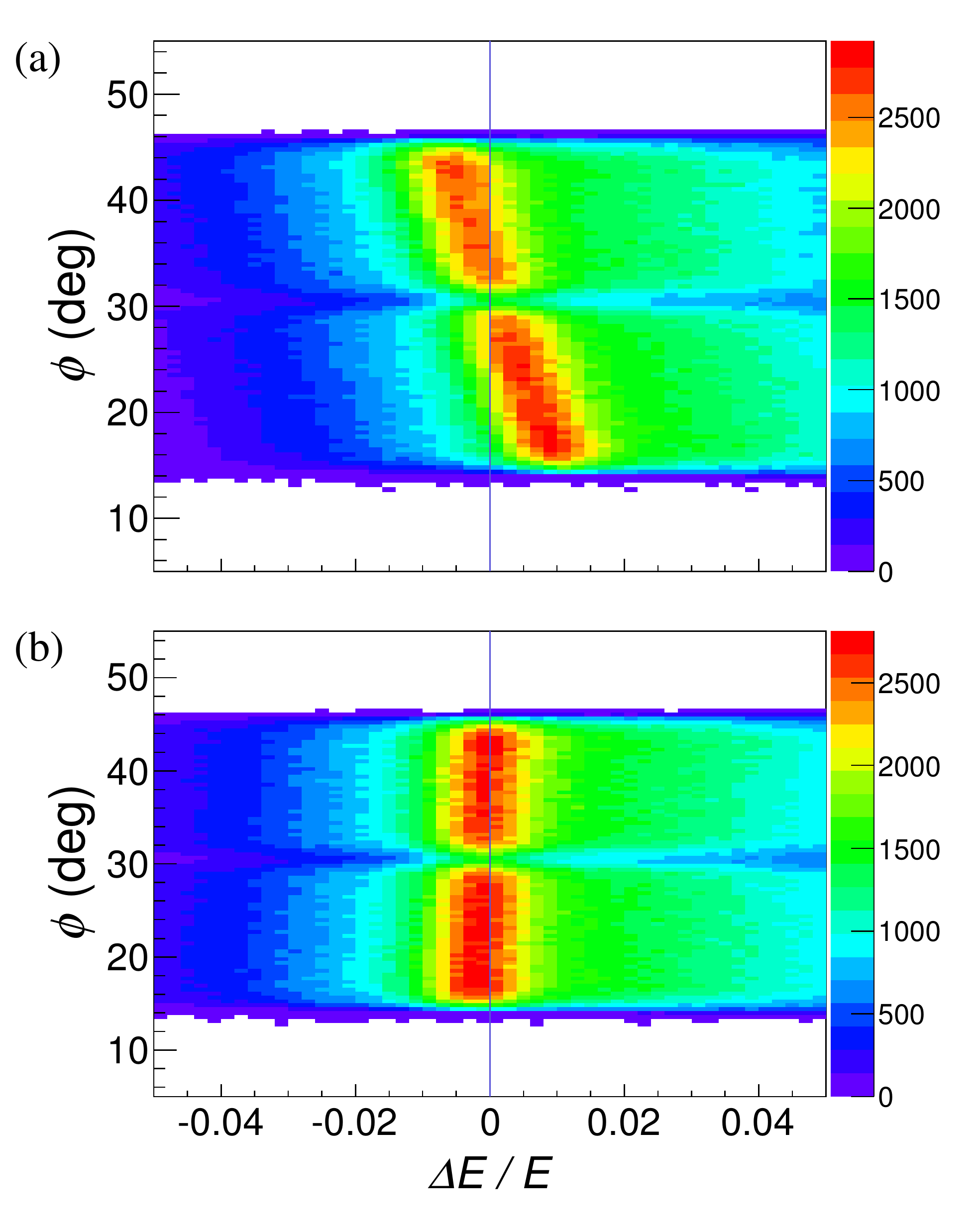}
\caption[$\phi$ vs\ $\Delta{E}/E$ before and after the kinematic corrections for the 1.723 GeV data set.]{(Color online)
Measured energy mismatch $\Delta{E'}/E'$ versus $\phi$ for elastically scattered electrons 
(a) before and (b) after the kinematic corrections for the 1.723$-$ data
  set. After corrections, $\Delta{E'}/E'$ is centered on zero for all azimuthal angles.
}
\label{momentumhigh:fig}
\end{figure}

\begin{figure}
\centering
\includegraphics[width=9cm]{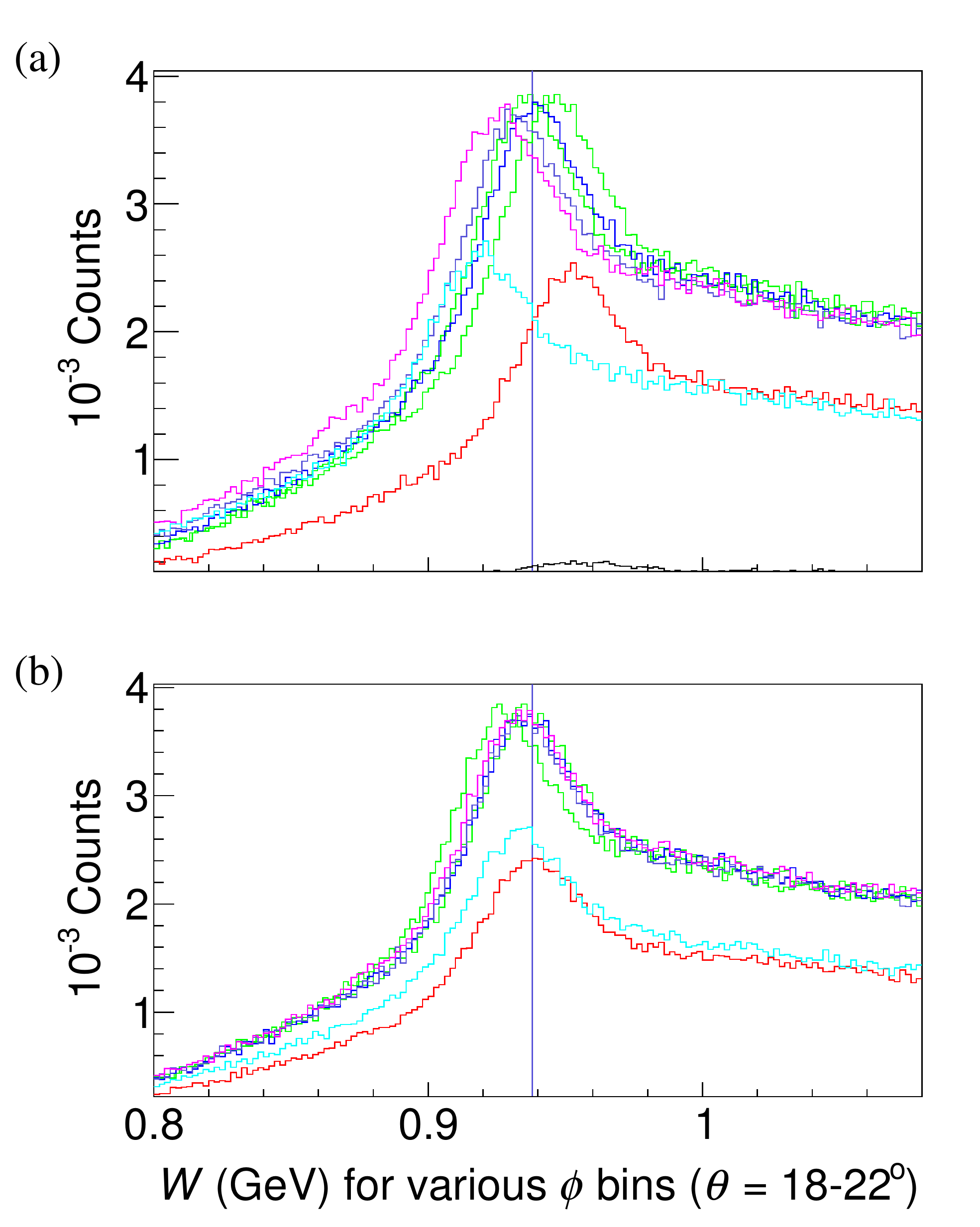}
\caption[Elastic $W$ peak for various $\phi$ bins before and after the kinematic corrections for 
2.286 GeV inbending data set.]
{(Color online) Elastic $W$ peaks (different colors/shades) for seven $\phi$ bins spanning the detector acceptance,
(a) before and (b) after kinematic corrections to the 2.286-GeV data set. The plots represent one sector and one polar-angle
bin. The spurious $\phi$ dependence of the elastic $W$ peak location is removed by these corrections.  
}
\label{Wpeak_Phi1:fig}
\end{figure}

\begin{figure}
\centering
\includegraphics[width=8.8cm]{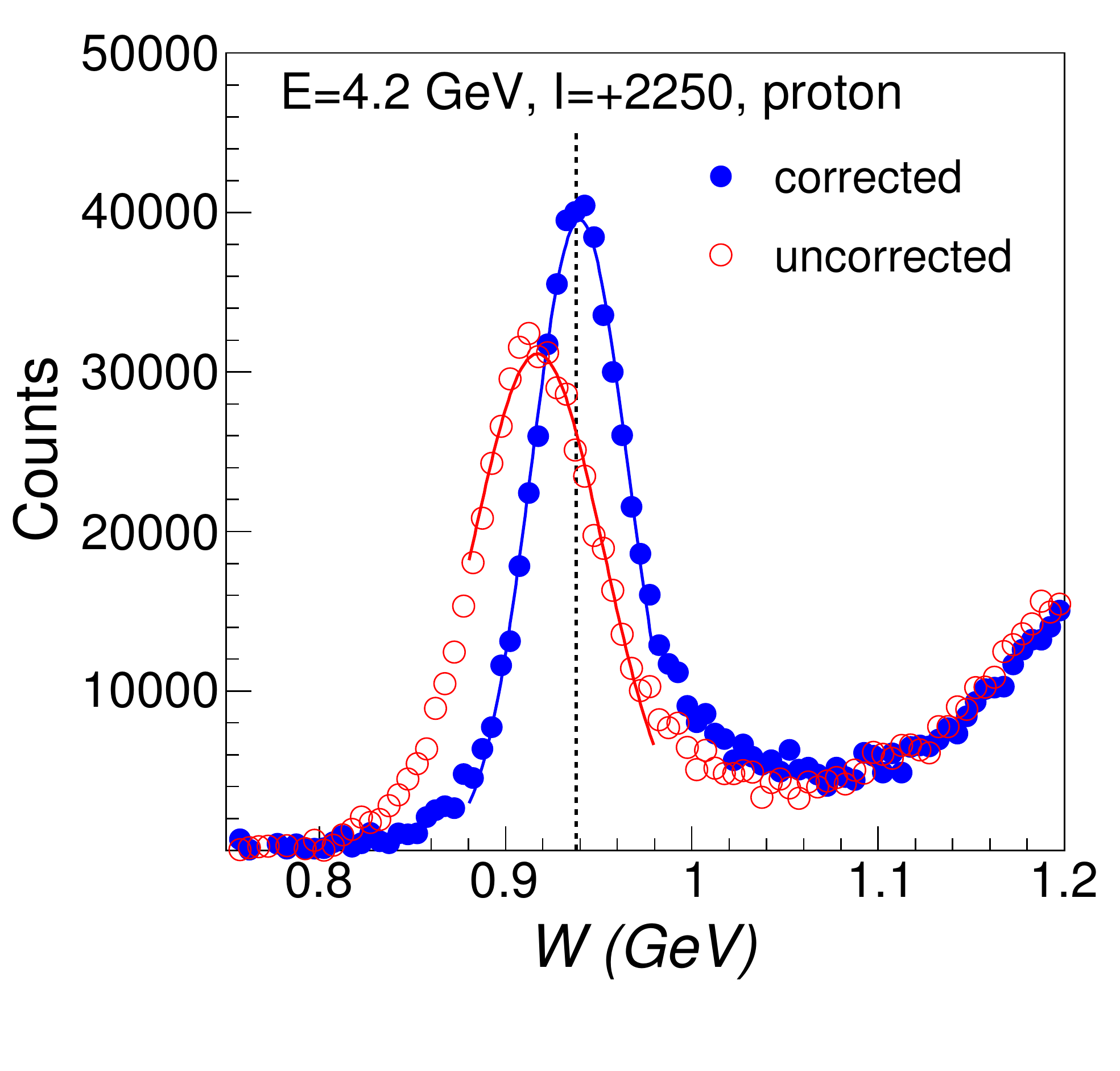}
\caption[$W$ spectrum before and after the kinematic corrections for the 4.231 GeV data set.]{(Color online)
Missing mass $W$ before (red, open circles) and after (blue, solid dots) the kinematic
corrections for the 4.238+ data set. 
The corrections decrease the distribution width and center the mean
on the 0.938 GeV proton mass, which is indicated by the vertical black dashed line.
}
\label{WPeak:fig}
\end{figure}

After looping over all events, an additional term 
$\sum_{par}{par^2}/{\sigma_{par}^2}$,
with estimated intrinsic uncertainties $\sigma_{par}$ for each
parameter $par$, was added to the total $\chi^2$ for each parameter. 
This limited parameters to reasonable ranges, avoiding ``runaway'' 
solutions anywhere in the parameter space.
\\
\indent
In order to avoid preferential weighting due to detector acceptances, elastic
$ep$ events were divided into 1$^\circ$ $\theta$ bins and given a
relative weighting proportional to their distribution in
$\theta$. Inclusion of $ep\pi^+\pi^-$ events ensured that the corrections
maintained validity over the full  space of $\theta$ and $p$.
{\footnotesize MINUIT}-based minimization of $\chi^2$ \cite{MINUIT} was iterated until stable values were
reached, and the width of the missing momenta and energy
distributions was reduced as shown in Fig. \ref{momentum:fig}.
\\
\indent
The relative absence of exclusive scattering events at $\theta\lesssim
12^\circ$ necessitated an additional forward scattering correction
using inclusive elastic scattering data.  Therefore, an additional
adjustment $\Delta p(\theta,\phi)$ containing three more fit parameters was 
applied in a similar manner, except that only the difference $W - M_p$ was minimized,
leading to even better resolution in the elastic peak.
\\
\indent
Application of the kinematic corrections resulted in final $ep$
accuracy of $\sim$ 1.0 MeV/$c$ for spatial momentum coordinates, with
distribution widths $\sigma_{p_x}\approx\sigma_{p_y}\approx$ 17 MeV/$c$ and
$\sigma_{p_z}\approx$ 30 MeV/$c$.  Overall momentum and angle corrections were generally a few
tenths of a percent in electron momentum $p$ and less than one milliradian in polar angle $\theta$.
The overall effect of all kinematic 
corrections can be seen in  Figs. \ref{momentumhigh:fig}, 
\ref{Wpeak_Phi1:fig}, and \ref{WPeak:fig}.
Systematic uncertainties due to the kinematic
inaccuracies of $p_z$, $\sqrt{p_x^2+p_y^2}$, and
$E_{beam}$ were determined by using the smoothly parameterized models
of the asymmetries and structure functions as a proxy for the actual
data, shifting each bin center by an amount equal to its
uncertainty, and subtracting the difference. ``Bin smearing''
uncertainties due to the distribution widths 
were estimated by determining the uncertainty in $W$ corresponding to the momentum uncertainty, 
smearing each bin in the modeled $A_{||}$ by a corresponding Gaussian distribution, and
subtracting the difference from the unsmeared model.

\subsubsection{Charge normalization correction}
The calculation of the dilution factor (nominally $\frac{3}{18}$) required a comparison of the
normalized counts
from the ammonia, carbon, and empty (LHe) targets.
Multiple scattering in the target, as well as changes in beam focusing,
could affect the measurement of beam charge determined
by the Faraday cup, which was  29 m downstream from the target. The 
contribution of multiple
scattering in the target on beam divergence can be estimated with a 
Molliere distribution \cite{Yao:2006px}. At the lowest energies
the size of the beam at the FC exceeded its 5.0~cm aperture.

The (ungated) upstream BPMs were used to establish a
relative correction to the FC signal for different targets.
The BPM to FC ratio at 5.7 GeV
(with multiple scattering suppressed) provided the overall
normalization. For beam energies $E<$ 3 GeV, this ratio provided a correction
factor with an approximate accuracy of 0.001.

The difference in the FC correction factors for the ammonia target
and the empty target was especially large because of the significant
difference in their radiation lengths. The relative factor was
$\sim$1.14 at 1.6 GeV and $\sim$1.05 at 2.4 GeV. These corrections
were needed for dilution factor extractions from data (see below) but
played no role in the extracted physics asymmetries.

\subsection{Asymmetries and corrections}

The raw asymmetry
\begin{equation}
\label{rawasym:eq}
A_{raw} = \frac{n^+-n^-}{n^++n^-}
\end{equation}
was determined, where $n^+$($n^-$) is the live-time gated,
FC-normalized, inclusive electron count rate for (anti-)aligned beam and
target polarizations. Except for a few small corrections, $A_{||}$ is derived from $A_{raw}$ by
dividing out the 
dilution factor $F_{DF}$ (which accounts for unpolarized backgrounds), the
electron beam polarization $P_b$, and the proton target polarization
$P_t$, such that 
\begin{equation}
\label{calcasym:eq}
A_{||} \approx \frac{1}{F_{DF}P_bP_t}\frac{n^+-n^-}{n^++n^-}.
\end{equation}
Smaller contributions due to radiative corrections
and other possible backgrounds were also taken into account. The modeled 
radiative contribution to the polarized and unpolarized 
cross-sections was characterized by an additive term $A_{RC}$ and a
``radiative dilution factor'' $f_{RC}$. Contributions due to
misidentified inclusive electrons ($C_{back}$) and polarized $^{15}$N ($P^*_{^{15}N}$)
were also taken into account, yielding 
\begin{equation}
\label{fullasym:eq}
A_{||} = \frac{C_{back}}{F_{DF}P_b(P_t+P^*_{^{15}N})f_{RC}}A_{raw} +
A_{RC}
\end{equation} 
as the final experimental measurement.
$C_{\rm back}$ has already been described;  the remaining terms will be discussed in sequence.

\subsubsection{Dilution factor}\label{DFcorr}
$F_{DF}\equiv n_p/n_A$ is defined as the ratio of
scattering rates for the proton ($n_p$) and the
whole ammonia target ($n_A$). It varies as a function of $Q^2$ and $W$, and
was calculated directly
from the radiated cross sections. In terms of densities ($\rho$), 
material thicknesses ($\ell$), and cross-sections ($\sigma$), 
\begin{equation} 
n_p \propto \frac{3}{18}\rho_A\ell_A\sigma_p
\end{equation} 
\begin{multline}
\label{nA:eq} 
n_A \propto \rho_{Al}\ell_{Al}\sigma_{Al} +
 \rho_{K}\ell_{K}\sigma_{K} \\+   \rho_A\ell_A(\frac{3}{18}\sigma_p +
 \frac{15}{18}\sigma_N) + \rho_{He}(L-\ell_A)\sigma_{He}
\end{multline}
with the subscripts $A$, $p$, $Al$, $K$, $N$, and $He$ denoting
ammonia ($^{15}$NH$_3$), proton, aluminum foil, kapton foil, nitrogen ($^{15}$N),
and helium ($^4$He), respectively. The acceptance-dependent 
proportionality constant is identical in both of the above relations.
Inclusive scattering data from the
empty (LHe) and $^{12}$C
targets were analyzed to determine the total target cell length ($L$)
and effective NH$_3$ thickness ($\ell_A$).
Scattering rates from the carbon ($n_C$) and empty
($n_{MT}$) targets were expressed as 
\begin{multline}
\label{nC:eq}
n_c \propto \rho_{Al}\ell_{Al}\sigma_{Al} + \rho_K\ell_K\sigma_K \\+ \rho_C\ell_C\sigma_C +
\rho_{He}(L-\ell_C)\sigma_{He}
\end{multline}
and
\begin{equation}
\label{nMT:eq}
n_{MT} \propto \rho_{Al}\ell_{Al}\sigma_{Al} + \rho_K\ell_K\sigma_K 
 + \rho_{He}L\sigma_{He}
\end{equation}
with again the same proportionality constant assumed.
\\
\indent
The inelastic scattering model employed Fermi-smeared 
cross sections calculated 
for each nucleus 
\cite{DeJager:1987qc}, which included
(unpolarized) radiative corrections and corrections for 
the nuclear EMC effect. Free proton
cross sections were calculated from a fit to world data for $F_1^p$ and
$F_2^p$ \cite{Christy:2007ve}. 
For cross sections on heavier nuclei, a Fermi convolution 
of the smearing of free nucleon Born cross sections was fit to 
inclusive scattering data, including EG1b data from
$^{12}$C, solid $^{15}$N, and empty (LHe) targets \cite{Bosted:2007hw}.
The nuclear EMC effect was parameterized using SLAC data
\cite{Norton:2003cb}. Radiative
corrections used the treatment of Mo and Tsai \cite{Mo:1968cg}; external
Bremsstrahlung probabilities incorporated all material thicknesses in CLAS from
the target vertex through the inner layer DC. Radiated cross-sections
(relative to that of $^{12}$C)
were calculated for each target material for radiation length
fractions 0.01$X_0$ and 0.02$X_0$, and were linearly interpolated to
correspond to the fraction $\rho\ell/X_0$ for each material in the appropriate target.

\begin{figure}
\centering
    \includegraphics[width=7cm]{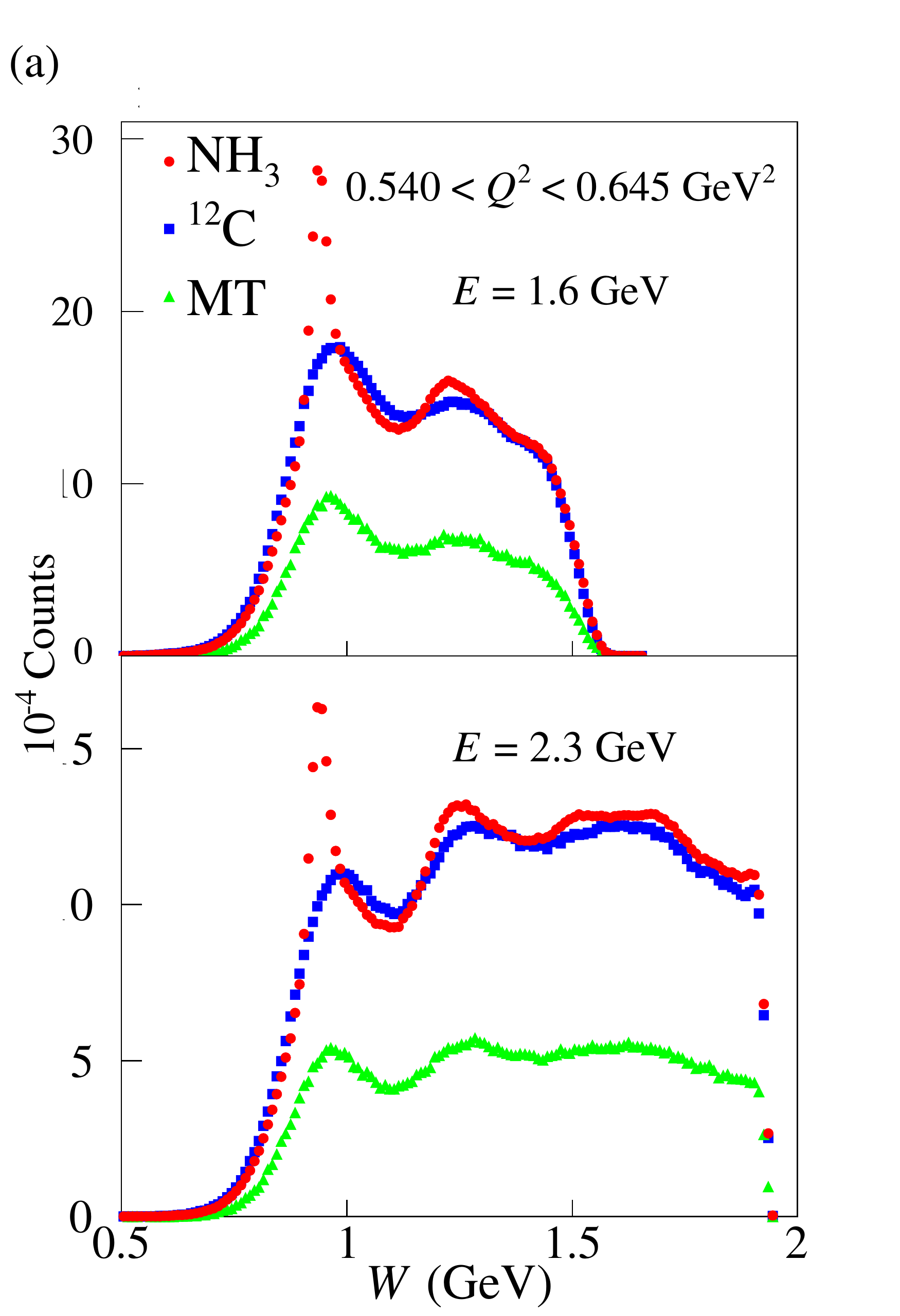}
    \includegraphics[width=7cm]{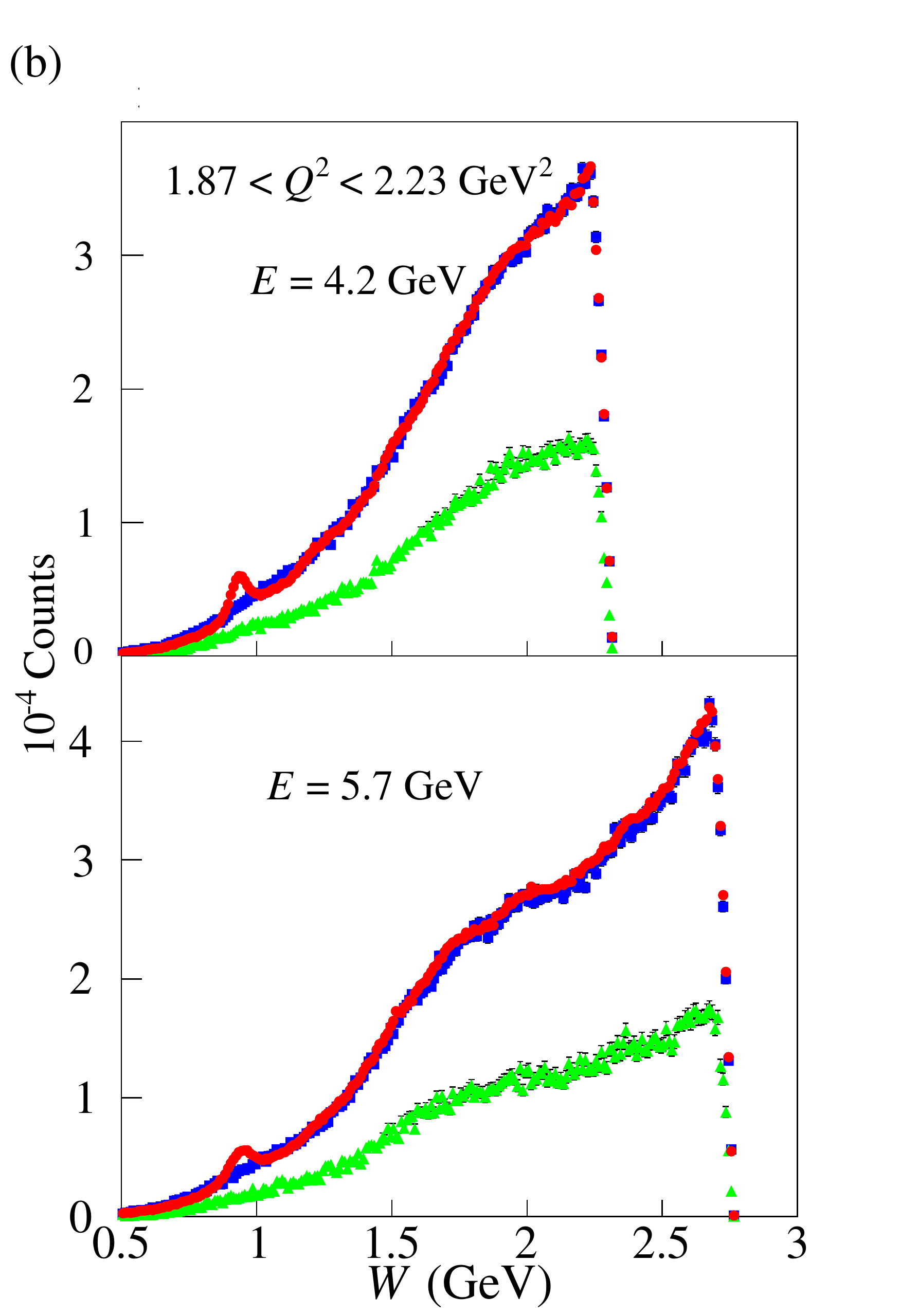}
\caption[]{(Color online)
Inclusive $W$ spectra normalized to the integrated
Faraday cup charge 
for each target (ammonia, red circles; carbon, blue squares; and empty (MT), green triangles) in a
  selected $Q^2$ bin, at (a) the lower two beam energies and (b) the higher two beam energies.}
\label{inc:fig}
\end{figure}

To apply the model, FC charge-normalized inclusive electron
counts were first binned in $Q^2$ and $W$ for all runs in each of the 
11 data sets (see Fig. \ref{inc:fig}). From these sums, the ratios $n_{MT}/n_C$ 
and $n_A/n_C$ were formed. The ratio $n_{MT}/n_C$ then determines $L$ through
solution of Eqs.~(\ref{nC:eq}) and (\ref{nMT:eq}). With $L$ determined,
the ratio $n_{A}/n_C$ determines $\ell_A$ through solution 
of Eqs. (\ref{nA:eq}) and (\ref{nC:eq}). $L$ and $\ell_A$ were
statistically averaged in the inelastic region ($W>1.10$ GeV)
over all $Q^2$ values,
with $1.75<L<2.05$ cm and $0.55<\ell_A<0.65$ cm over the 11 data sets. 
Upper bounds
in $W$ used in calculating the average were $Q^2$-dependent. To evaluate the
effect of the choice of the cutoff on the measurement of $L$($\ell_A$),
the $W$-averaging range was increased (decreased) by
approximately 33\% in a reanalysis (to account for small variations in our measurement at high-$W$) and the resulting difference in $F_{DF}$ was used to estimate the systematic uncertainties
due to these parameters.

\begin{figure}
\centering
  \includegraphics[width=9.1cm]{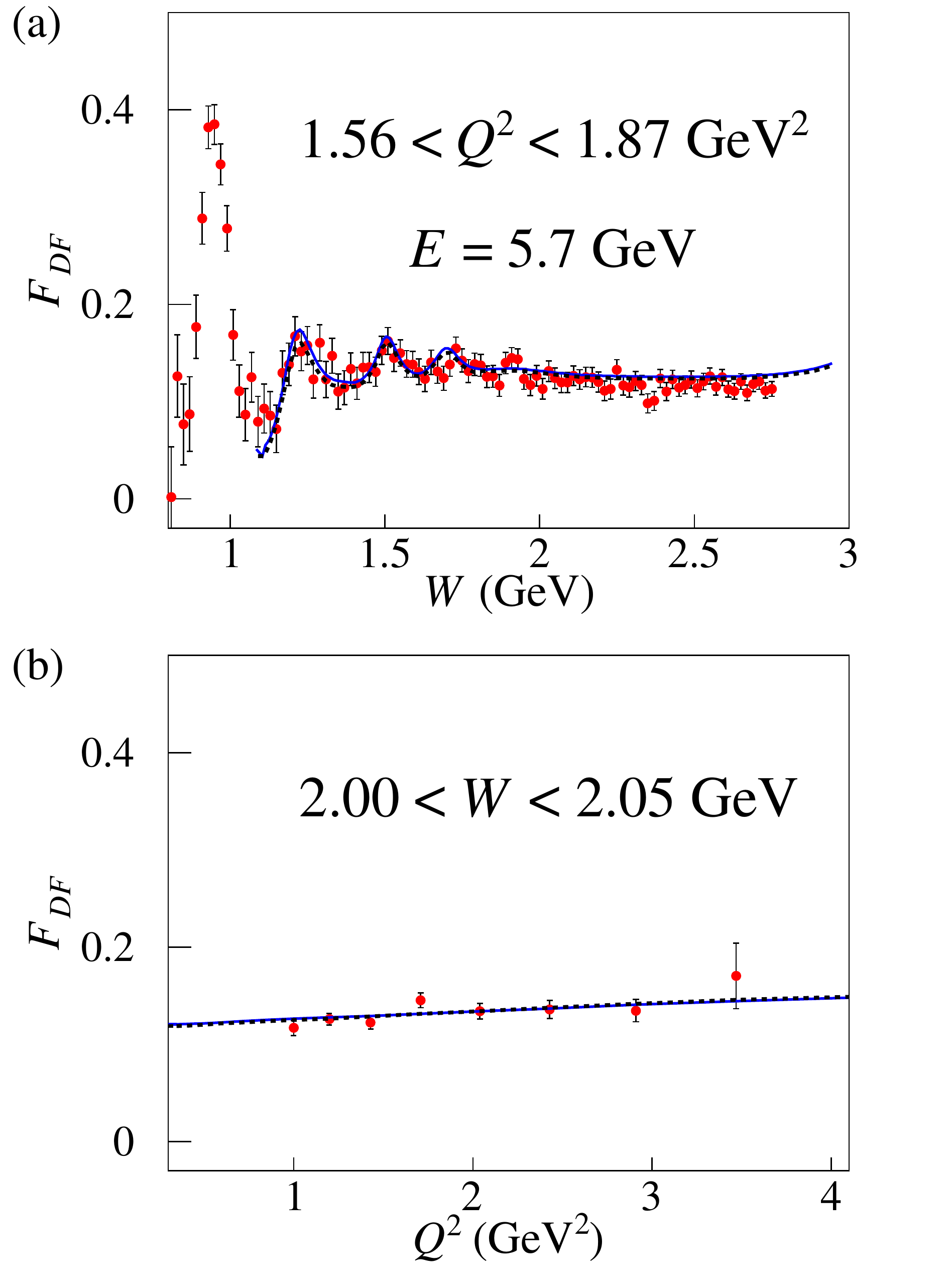}

\caption[Dilution factors from data and model compared]{(Color online) Dilution
  factors (a) $F_{DF}$ vs\ $W$ and (b)  $F_{DF}$ vs\  $Q^2$, for the 5.7 GeV beam energy.
  The solid blue line shows the modeled dilution factor used in the analysis, 
and the dotted black line (most visible in plot (a) at low $W$) is a 
 two-dimensional polynomial fit (in $Q^2$ and $W$) to the red points from the data-driven method. The difference
  between the solid blue and black dotted lines is an estimate of the model systematic uncertainty.
  Over much of the kinematics, $F_{DF}$ is close to the naive ratio 3/18 of
  polarized to unpolarized nucleons in the target.
}
\label{F_DFmodel:fig}
\end{figure}

\begin{figure}
\centering
\includegraphics[width=8cm]{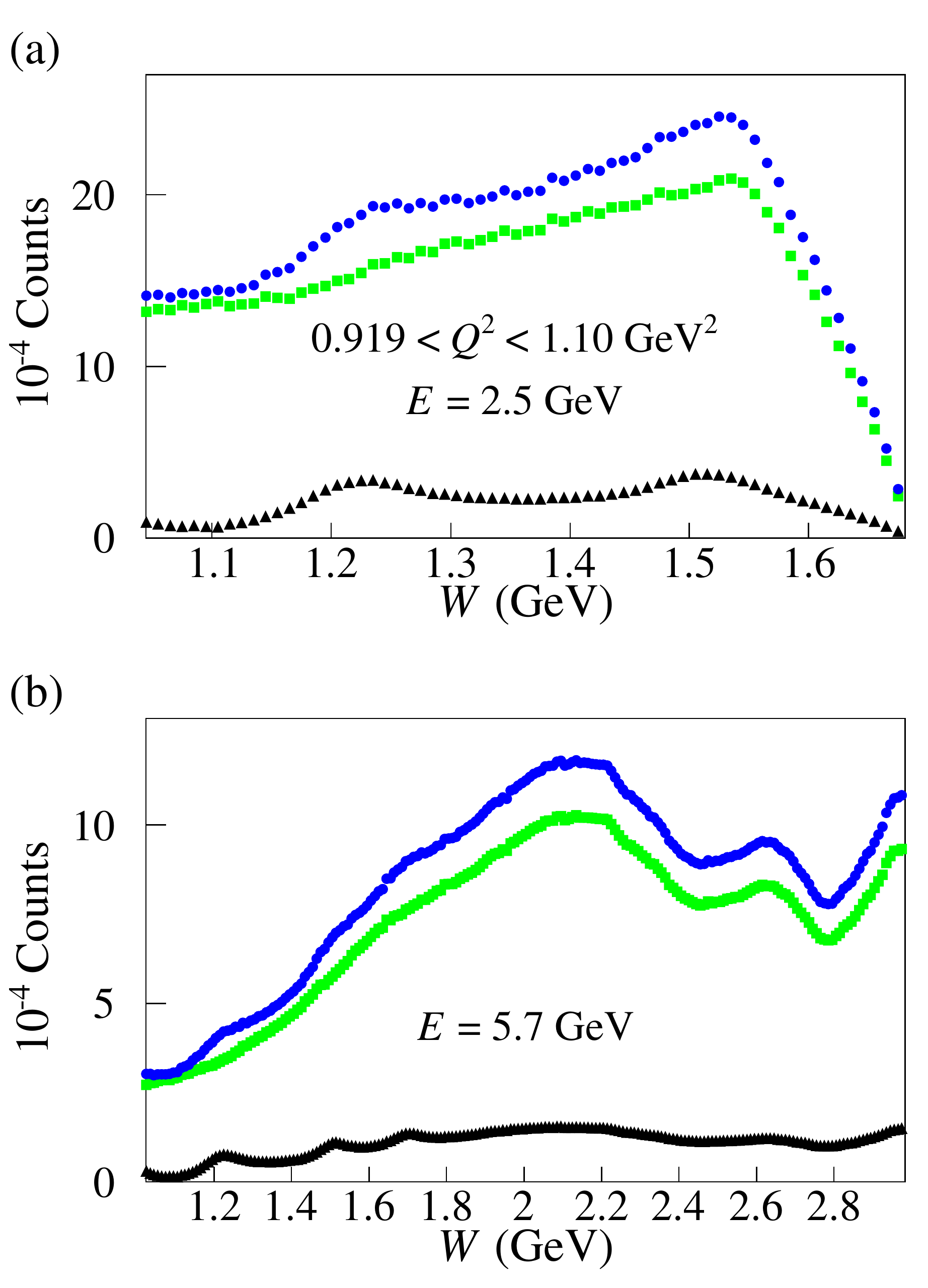}
\caption[Background-subtracted inclusive elastic spectra]{(Color online) Background
  subtraction (using dilution factors) for inclusive inelastic $W$ spectra for a selected $Q^2$ bin at (a) 2.5 GeV and (b) 5.7 GeV beam energies.
  The blue circles are the inclusive counts for ammonia.  The green squares show the subtracted background, 
as determined by the dilution factors, and the black triangles are the difference, which represents the 
free proton counts in the ammonia target. 
}
\label{inc_back:fig}
\end{figure}

Dilution factors $F_{DF}\equiv n_p/n_A$ were then calculated for each
data set. This model was checked against an older
data-driven method
\cite{Fatemi:2003yh,Dharmawardane:2006zd,Prok:2008ev} 
that used the three target count rates,
only one (unradiated) model for the ratio of neutron/proton cross-sections,
and the assumption that $\sigma_C = 3\sigma_{He}$ (see Fig.~\ref{F_DFmodel:fig}).
Values of $L$ and $\ell_A$ varied by less than 2\% between the two
methods. Division of $A_{raw}$ by $F_{DF}$ removes the contributions of the
$^{15}$N, LHe and target foil materials, leaving only the
contribution from scattering
by the polarized protons (see Fig.~\ref{inc_back:fig}).

The densities and thicknesses of all target materials were varied
within their known tolerances to determine systematic uncertainties. Only
the variations of $\rho_C\ell_C$ and $\rho_{He}$ had any significant ($>$0.1\%)
effect on $F_{DF}$. Uncertainties due to the cross section model were
estimated by comparing $F_{DF}$ to a third-degree polynomial
fit to the data-based dilution factors determined using the alternate method.

\subsubsection{Beam and target polarizations ($P_bP_t$)}
\label{pbpt:sec}
Because NMR measurements are dominated by the material near the edge of the target cell
\cite{Keith:2003ca}
(which was not exposed to the beam and therefore had higher polarization than the bulk of the target), the
polarization product $P_bP_t$ was determined experimentally using the
double-spin asymmetry of elastic $ep$ events, taking advantage of the
low background levels for these exclusive events. 
The asymmetry $A_{||}$ for elastic scattering corresponds to the case when $A_1^p=1$, 
$A_2^p = \sqrt{R^p}$, and $R^p = G_E^{p^2}/(\tau G_M^{p^2})$, as given in Eqs. (\ref{D1:eq}) and (\ref{Apar:eq}).
The proton's electric and magnetic form factors $G_E^p(Q^2)$ 
and $G_M^p(Q^2)$ (see Section~\ref{elas}) were calculated using parametrizations of world data \cite{Arrington:2003qk}. 
The polarization
product $P_bP_t$ was determined by dividing the measured elastic
$ep$ asymmetry by the calculated elastic $A_{||}(W=M_p,Q^2)$. 

\begin{figure}
\centering
  \includegraphics[width=9.0cm]{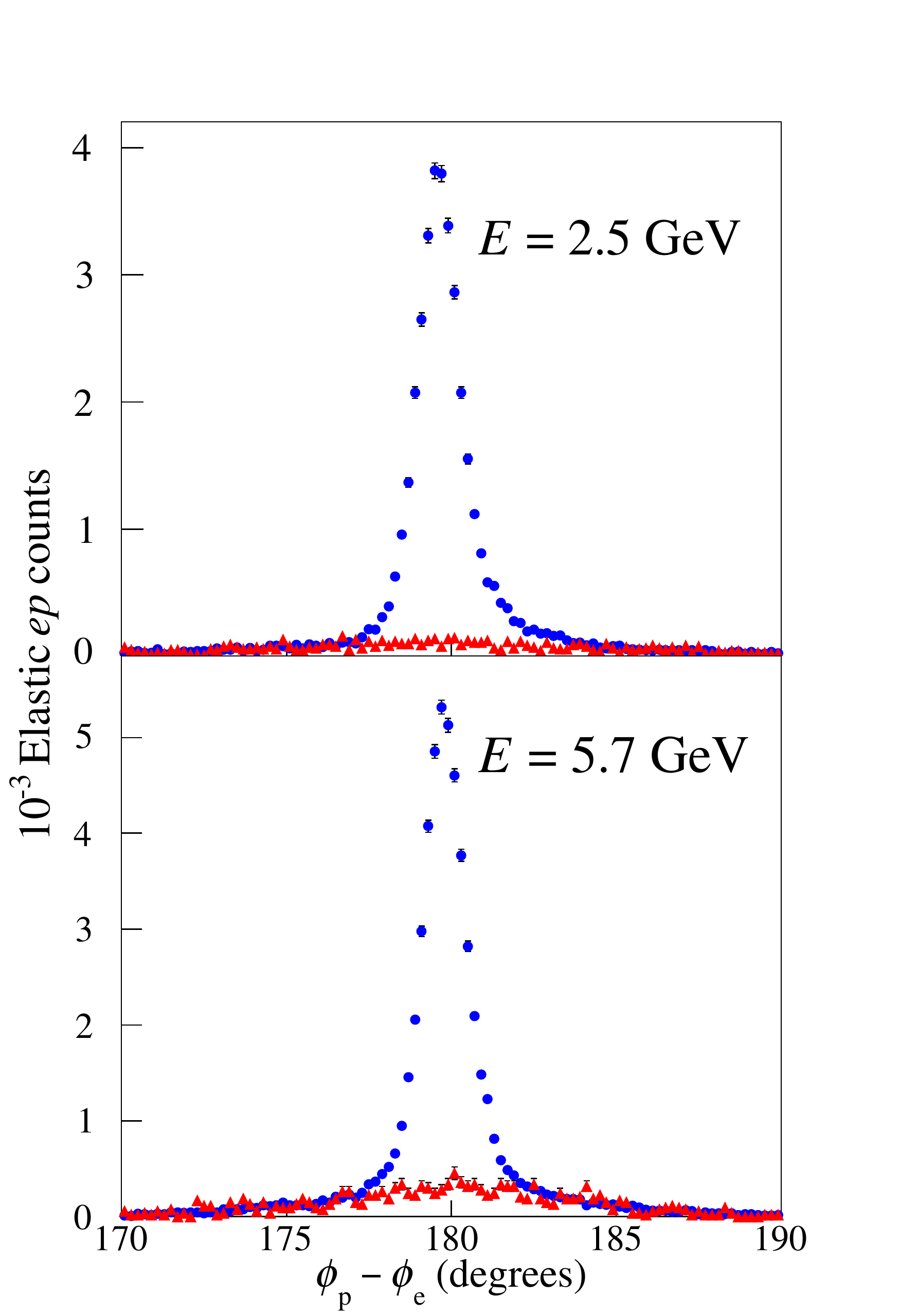}
\caption[Elastic $ep$ events in terms of $\phi_p - \phi_e$] {(Color online) Histogram of
the azimuthal angular difference $\phi_p - \phi_e$ for elastic scattering events from the NH$_3$ target (blue circles) overlaid
with the scaled distribution from the carbon target (red triangles) for two different data sets.  
}
\label{phiscale:fig}
\end{figure}

Background contamination in elastic $ep$ events was determined by
scaling the scattering spectra of the carbon target to match that of the
ammonia target away from the vicinity of the free proton peak. 
Scattering events were selected from $^{12}$C
using all elastic $ep$ cuts except the $\Delta\phi$ cut, and were
normalized to the $ep$ $\Delta\phi$ spectrum in the region
$2^\circ<|\Delta\phi|<6^\circ$ (Fig.~\ref{phiscale:fig}). Nuclear background contributed less than 5\% of the events;
systematic effects due to
miscalculating this background were tested by shifting the
normalization region by 2$^\circ$ and reevaluating.

\begin{figure}
\centering
  \includegraphics[width=8.5cm]{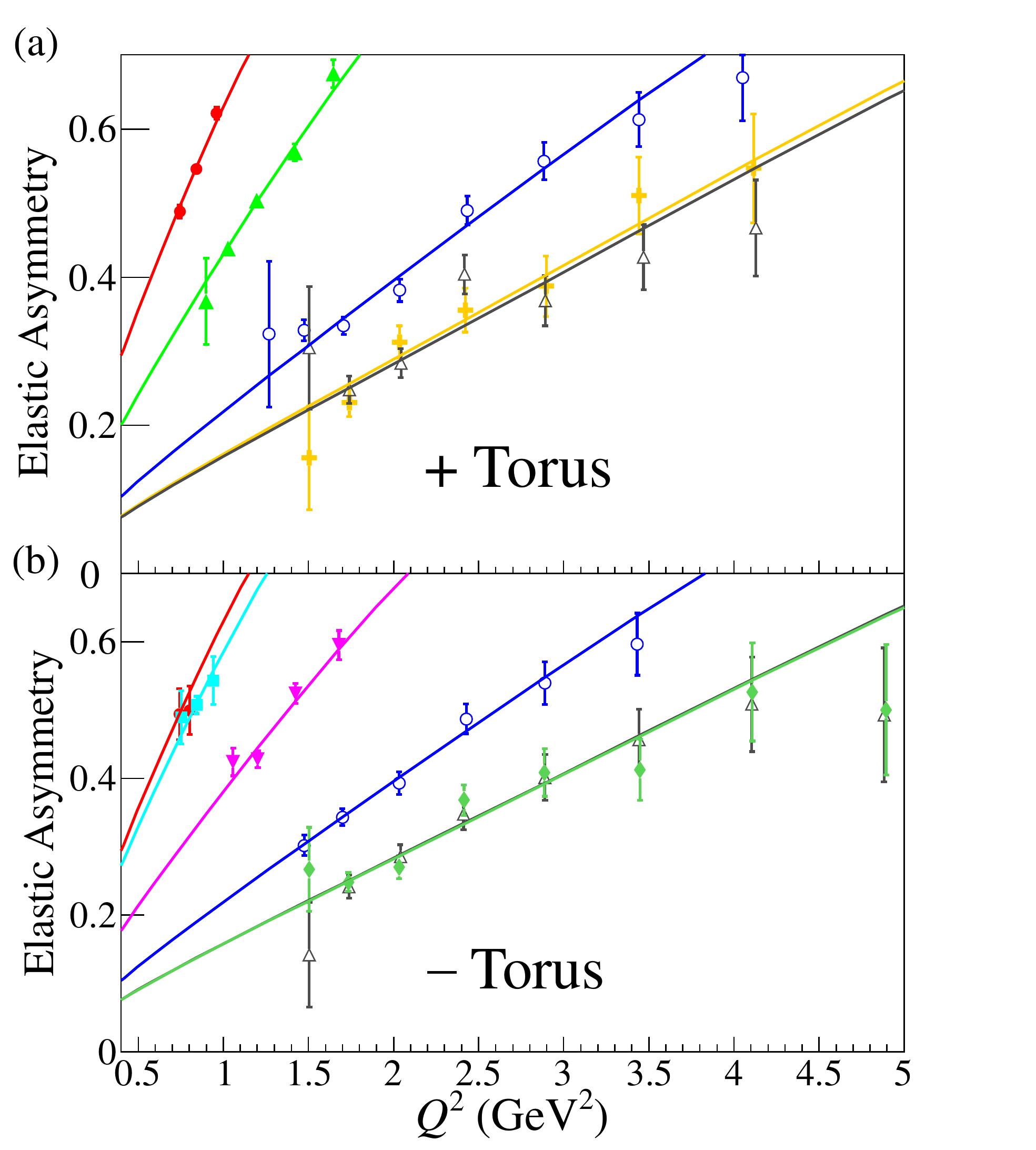}
\caption[$P_bP_t$ values divided by target polarization]
{(Color online) Comparison of the elastic asymmetry $A_{||}(W=M_p,Q^2)$ (solid
  lines) to the measured elastic asymmetries for
  all data sets, normalized by $P_bP_t$. (a) Inbending and (b) outbending sets are shown separately.
  Each line represents a specific beam energy, increasing in energy with descending order from the upper left.
  Each color and marker style (red circles, cyan squares, light green triangles, magenta inverted triangles, blue open circles, 
  orange crosses, gray open triangles, dark green diamonds) represents a different beam energy 
  (1.606, 1.723, 2.286, 2.561, 4.238, 5.615, 5.725 and 5.743 GeV, respectively).  
}
\label{pbpt:fig}
\end{figure}

The derived $P_bP_t$ values were checked for consistency across $Q^2$ for
each beam energy, torus current and target polarization direction.
As a comparison check, a less accurate method using inclusively
scattered electrons in the elastic peak was also employed to measure
$P_bP_t$. This method required the subtraction of much larger
backgrounds and did not incorporate radiative corrections. Within its
larger uncertainty, this second method agreed with the first.
\\
\indent
The calculated elastic asymmetry 
is plotted against
the $P_bP_t$-normalized measured elastic asymmetry 
for each of the 11 data sets 
in Fig.~\ref{pbpt:fig} 
to demonstrate
the precision of the elastic $ep$ data.
Older parametrizations of $G_E$ and $G_M$ \cite{Bosted:1994tm} 
were substituted to
evaluate the systematic uncertainty due to 
the $A_{||}(W=M_p,Q^2)$ model. The $W$ cut
on allowed elastic $ep$ events was also widened by 10 MeV on each side
to test for systematic effects due to $ep$ event selection. The
systematic uncertainty due to the statistical uncertainty on $P_bP_t$ was
determined by adding one standard deviation to $P_bP_t$ for one of the
data sets, and repeating the full analysis; this was repeated
independently for each set.

\subsubsection{Polarized nitrogen correction}
EST (equal spin temperature) theory predicts the relative polarization
ratios between two spin-interacting atoms in a homogeneous medium as the ratio
of their magnetic moments ($P_{^{15}N}/P_{^1H} \approx
\mu_{^{15}N}/\mu_{^1H} \approx -0.09$) at
small polarizations, with higher order terms increasing the magnitude
of this ratio at larger polarizations \cite{Crabb:1997cy}. An empirical fit for
$^{15}$N polarization as a function of proton polarization,
\begin{equation}
\label{15N:eq}
P_{^{15}N} = -(0.136P_p-0.183P^2_p+0.335P^3_p),
\end{equation}
derived in the SLAC E143 experiment for $^{15}$NH$_3$ \cite{Abe:1998wq},
was applied to
determine the nitrogen polarization.

The 3:1 $^1$H/$^{15}$N ratio and the relative alignment of the proton
and $^{15}$N polarizations in the nuclear shell model 
\cite{RondonAramayo:1999da} require factors
of $\frac{1}{3}$ and $-\frac{1}{3}$, respectively, on this
polarization, such that $P_{^{15}N}^* = -\frac{1}{9}P_{^{15}N}$ in
Eq. (\ref{fullasym:eq}). 
Systematic uncertainties were estimated by replacing
the fit of Eq. (\ref{15N:eq}) with the leading-order EST estimate
($P_{^{15}N} = 0.09P_p$) and reanalyzing. 

Elastic $ep$ events were also affected by the nuclear polarization,
though the effect was less, due to the smearing of the $^{15}$N
quasi-elastic peak. We estimated $P_{^{15}N\text{elastic}}\approx
\frac{1}{2} P_{^{15}N}$, and set
$P_{^{15}N\text{elastic}}= 0$ to determine the uncertainty of this effect.


\subsubsection{Radiative corrections}
\label{radcorr:sec}
Radiative corrections to the measured asymmetries $A_{||}$ were computed using the program 
{\footnotesize RCSLACPOL}, which was developed at SLAC for the spin structure function experiment 
E143~\cite{Norton:2003cb}.
Polarization-dependent internal and external corrections were calculated according to the prescriptions
in Refs.\ \cite{Kukhto:1983pv} and \cite{Mo:1968cg}, respectively.  

The polarized and unpolarized radiated cross sections can be expressed as
\begin{equation}
\Delta\sigma_r = \Delta\sigma_B(1 + \Delta\delta_v) + \Delta\sigma_{el} + \Delta\sigma_{qe} + \Delta\sigma_{in}
\end{equation}
and
\begin{equation}
\sigma_r = \sigma_B(1 + \delta_v) + \sigma_{el} + \sigma_{qe} + \sigma_{in}
\end{equation}
respectively, in which $\sigma_B$ is the Born cross section;
$\delta_v$ is the combined electron vertex, vacuum polarization, and internal bremsstrahlung
contributions; and $\sigma_{el}$, $\sigma_{qe}$, and $\sigma_{in}$ are the nuclear elastic, quasi-elastic, and inelastic
radiative tails (the quasi-elastic tail is, of course, absent for a proton target).  
The radiated asymmetry is given by
\begin{equation}
A_r = \frac{\Delta\sigma_r}{\sigma_r}.
\end{equation}

For a given bin, one can write the Born asymmetry as
\begin{equation}
A_B = \frac{A_r}{f_{RC}} + A_{RC}
\end{equation}
in which $f_{RC} = 1 - \sigma_{el} / \sigma_r$ is a radiative dilution factor 
(accounting for the ``dilution'' of the denominator of the asymmetry due to the 
radiative elastic tail) and $A_{RC}$ is an additive correction accounting for all other radiative effects.
We calculated these two terms using parametrizations of the world data for elastic form factors 
$G_E$ and $G_M$, structure functions $F_2^p$ and
$R^p$, and virtual photon asymmetries $A_1^p$ and $A_2^p$ (see Sec.~\ref{models}).  

External corrections, dependent on the polar angle of scattering,
 were calculated using a realistic model of all
the materials in the beam path  
within the vertex cuts for good electrons.  
{\footnotesize RCSLACPOL} is equipped to integrate
over target raster position and scattering point within the target.  However, studies have shown little difference
from the case of fixing the scattering at the target center, which 
was assumed here.
The peaking approximation, which speeds
the calculation and has a negligible effect on the final result, was
also exploited.

Both the internal and external corrections were combined and used to extract the Born asymmetries from the data.
Radiative 
effects tend to be large near threshold (below $W=1.2$ GeV) and at large $W$ where the radiative tails begin
to dominate.  

Systematic uncertainties on these corrections were estimated by running {\footnotesize RCSLACPOL} for a 
range of reasonable variations of the models
for $F_2^p$, $R^p$, $A_1^p$ and $A_2^p$ (see Section~\ref{models})
 and for different target and LHe thicknesses
$\ell_A$ and $L$.  The changes due to each variation were added in quadrature and the
square root of this quantity is taken as the systematic uncertainty on radiative effects.

\subsubsection{Systematic uncertainties}
\label{syserr:sec}
Estimation of systematic uncertainties 
on each of the observables discussed in the following
section was done by varying a particular input parameter, model,
or analysis method (as described in the preceding subsections), repeating the analysis, and recording the
difference in output for each of the final asymmetries, structure
functions, and their moments. 
Final systematic uncertainties attributable to each altered
quantity were then added in quadrature to estimate the total
uncertainty.

Sources of systematic uncertainties have been extensively discussed in the
preceding text. These sources
include kinematic accuracy, bin smearing,  
target model (radiative corrections), nuclear dilution model, elastic asymmetry
measurement, $P_bP_t$ statistics, and background contamination.

The magnitudes of the effects of the various systematic uncertainties on
the ratio $g_1^p/F_1^p$ for the four beam energies are listed in Table
\ref{syserr:table}. Note that for each quantity of interest 
($A_1^p, g_1^p, \Gamma_1^p$) the systematic uncertainty was calculated by the
same method (instead of propagating it from other quantities), 
therefore ensuring that all correlations in these uncertainties were properly
taken into account.

The results shown in 
the next section incorporate these systematic uncertainties. 

\begingroup
\squeezetable
\begin{table}[htbp] \centering
\vspace{0.5cm}

\begin{tabular}{|c|c|c|c|c|}
\hline
Systematic uncertainty & \multicolumn{4}{c|}{Max. Relative Magnitude ($g_1^p/F_1^p$)}\\
\cline{2-5}
& 1.6 GeV & 2.5 GeV & 4.2 GeV & 5.7 GeV\\
\hline
Kinematic smearing& 2.0\%& 1.5\%& 1.0\%& 0.5\%\\
\hline
Target material tolerances&  2.5\% &  2.5\% &  2.5\% &  2.5\% \\
\hline
$L, \ell_A$ target lengths & 1.5\%& 1.5\%& 1.5\%& 1.5\%\\
\hline
$F_{DF}$ cross-section model & 4.5\%& 2.0\%& 2.0\%& 2.0\%\\
\hline
$P_bP_t$ elastic $ep$ cuts & 1.5\%& 1.5\%& 1.5\%& 1.5\%\\
\hline
$P_bP_t$ statistics & 0.8\%& 1.1\%& 1.7\%& 2.2\%\\
\hline
$\pi^-$ contamination & 0.1\%& 0.8\%& 0.8\%& 1.5\%\\
\hline
$e^+e^-$ contamination & 1.0\%& 1.0\%& 1.0\%& 1.0\%\\
\hline
$^{15}$N polarization & 0.5\%& 0.5\%& 0.5\%& 0.5\%\\
\hline
Models for $F_2^p$,$R^p$,$A_1^p$,$A_2^p$&2.0\%&2.0\%&2.0\%&2.0\%\\
\hline
Totals & 6.4\% & 4.9\% & 5.0\% & 5.2\%\\
\hline
\end{tabular}
\caption{Systematic uncertainties}
\label{syserr:table}
\end{table} 
\endgroup

\section{RESULTS AND COMPARISON TO THEORY}\label{s5}

\subsection{Extraction of $A_\parallel$}
{\label{Apar:sec}}

\indent
The raw double-spin asymmetry [Eq.~(\ref{rawasym:eq})] was evaluated
for each group of data with a given
beam energy, torus polarity,
direction of the target polarization, and status of the HWP (in-out).
For each group, the raw data were combined in $(W,Q^2)$ bins with
bin width $\Delta W = 10$ MeV. The
$Q^2$ bins were defined logarithmically, with
13 bins in each decade of $Q^2$. These bin sizes were
chosen to provide a compromise between statistical significance
and expected structure in the asymmetries.

The data in the various groups were combined as follows. 
First, raw asymmetries with the same
beam energy, target spin direction, and torus polarity, but
opposite half-wave-plate (HWP) orientation,
were combined, bin by bin, weighting the data in each bin
according to their statistical uncertainty. Next, the data sets
with opposite target polarizations were combined using
the product $\sigma_A^2 (P_b P_t)_{rel}^2$ as the weighting factor to optimize
the statistical precision of the result. Here, $\sigma_A$ is the
statistical uncertainty of the raw asymmetry and
$(P_b P_t)_{rel}$ is a quantity proportional to the product of beam and target
polarization for a given data set. To get the highest possible statistical precision 
for this quantity, we calculated it by using not only elastic (exclusive)
scattering data (c.f. \ Sec.~\ref{pbpt:sec} ), but by taking the ratio
of the measured raw asymmetry to that predicted by our model
(see Sec.~\ref{models}) for {\em all}
kinematic bins (including elastic scattering) and averaging over the
entire data set. The resulting value for $(P_b P_t)_{rel}$ deviates
from the ``true'' product of polarizations by a constant unknown scale factor which
is the same for the two data sets with opposite target polarization
and therefore plays no role for the purpose of deriving a relative
weight for these two sets.

\begin{figure}[htb!]
\centering
  \includegraphics[width=9.2cm]{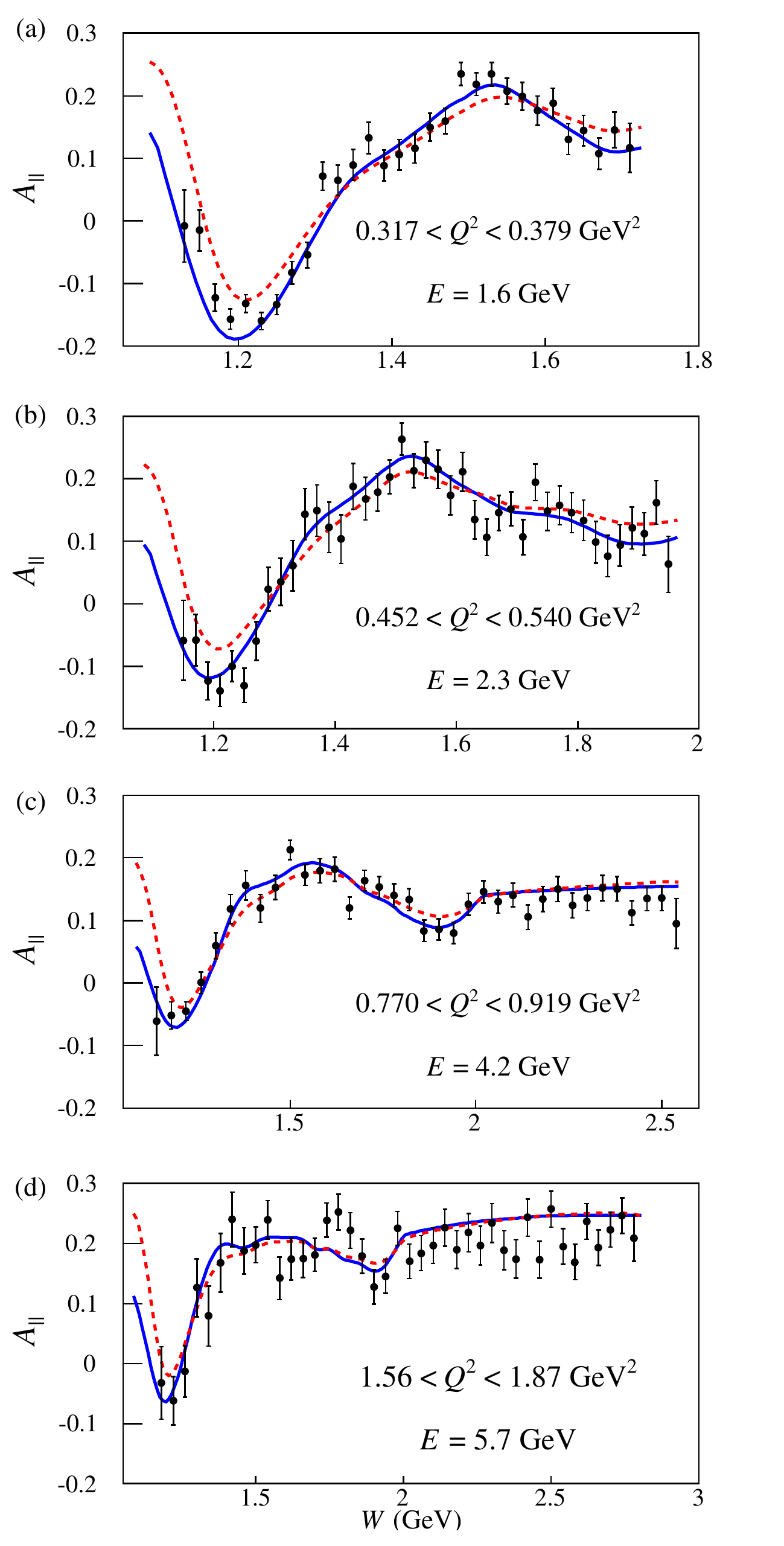}
\caption[Measurements of $A_{\parallel}$ with radiative corrections]
{(Color online) Values of $A_{\parallel}$ (including radiative corrections) shown at beam energies of (a) 1.6, (b) 2.3, (c) 4.2,
 and (d) 5.7 GeV. The curves correspond to our model with (blue solid line) and without (red dotted line) radiative
corrections, as discussed in the text. 
}
\label{radcorr:fig}
\end{figure}

All corrections except radiative corrections were then applied to
the combined sets. Next, the asymmetries from sets with opposite torus polarity 
(but identical beam energy) were averaged (again weighted by 
statistical uncertainty). Finally, radiative 
corrections, described in Sec. \ref{radcorr:sec}, were 
applied, resulting in measurements of $A_{\parallel}$
for each beam energy  (see Fig. \ref{radcorr:fig}).

\begin{figure}
\centering
  \includegraphics[width=9.0cm]{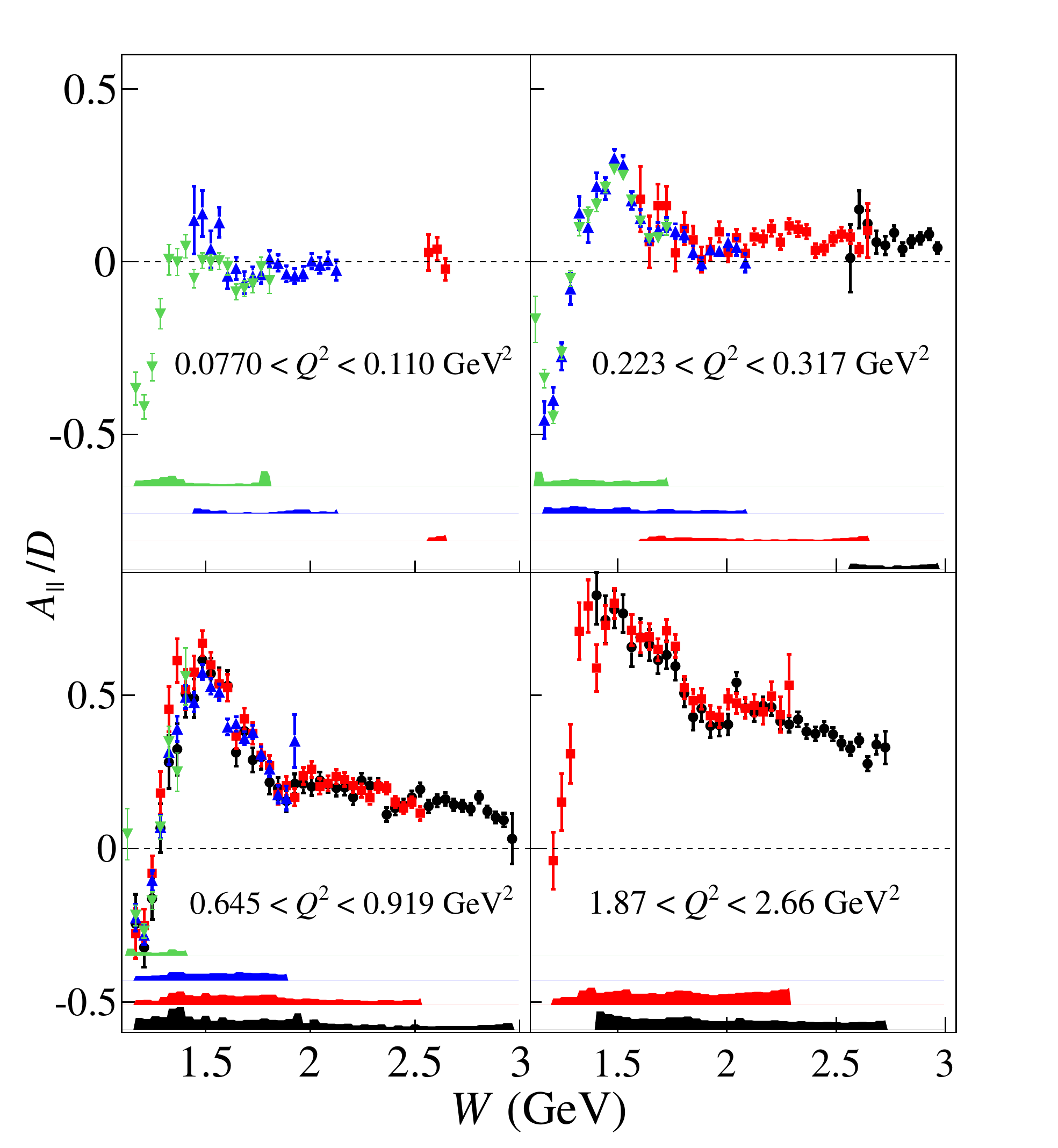}
\caption[Measurements of $A_{\parallel}/D$]
{(Color online) Values of $A_{\parallel}/D$ vs $W$ for each beam energy, including systematic uncertainties.
The green inverted triangles, blue triangles, red squares and black circles correspond to data from approximate beam energies of 1.6, 2.5, 4.2 and 5.7 GeV, respectively. 
}
\label{AD:fig}
\end{figure}

\subsection{Extraction of polarized asymmetries and structure functions}
\label{SuperRBasym}

The asymmetries A$_1(Q^2,W)$ and A$_2(Q^2,W)$ are linearly 
related to $A_{\parallel}(Q^2,W)$
by Eq.~(\ref{Apar:eq}). The kinematical depolarization factor $D$ in this
equation is given in Eq.~(\ref{D1:eq}).  The structure function
$R^p$ was calculated from a fit to the world data (see next section).
For each final set discussed in the previous section, the values
of $A_{\parallel}/D = A_1^p + \eta A_2^p$ were calculated for each bin. For sets with
beam energies differing by less than 15\%, these values for
$A_{\parallel}/D$ were combined (with statistical weighting) and the
corresponding beam energies averaged (see Fig. \ref{AD:fig}). These results have a low
theoretical bias from modeled asymmetries and structure functions (like $A_1$ and $F_1$) compared to
other extracted quantities. They can be found (along with the other results presented here) in the CLAS database~\cite{CLASdatabase}
and in the Supplemental Material~\cite{SupplementalMaterial} for this paper. 
\begin{figure}[htb]
\centering
  \includegraphics[width=9.5cm]{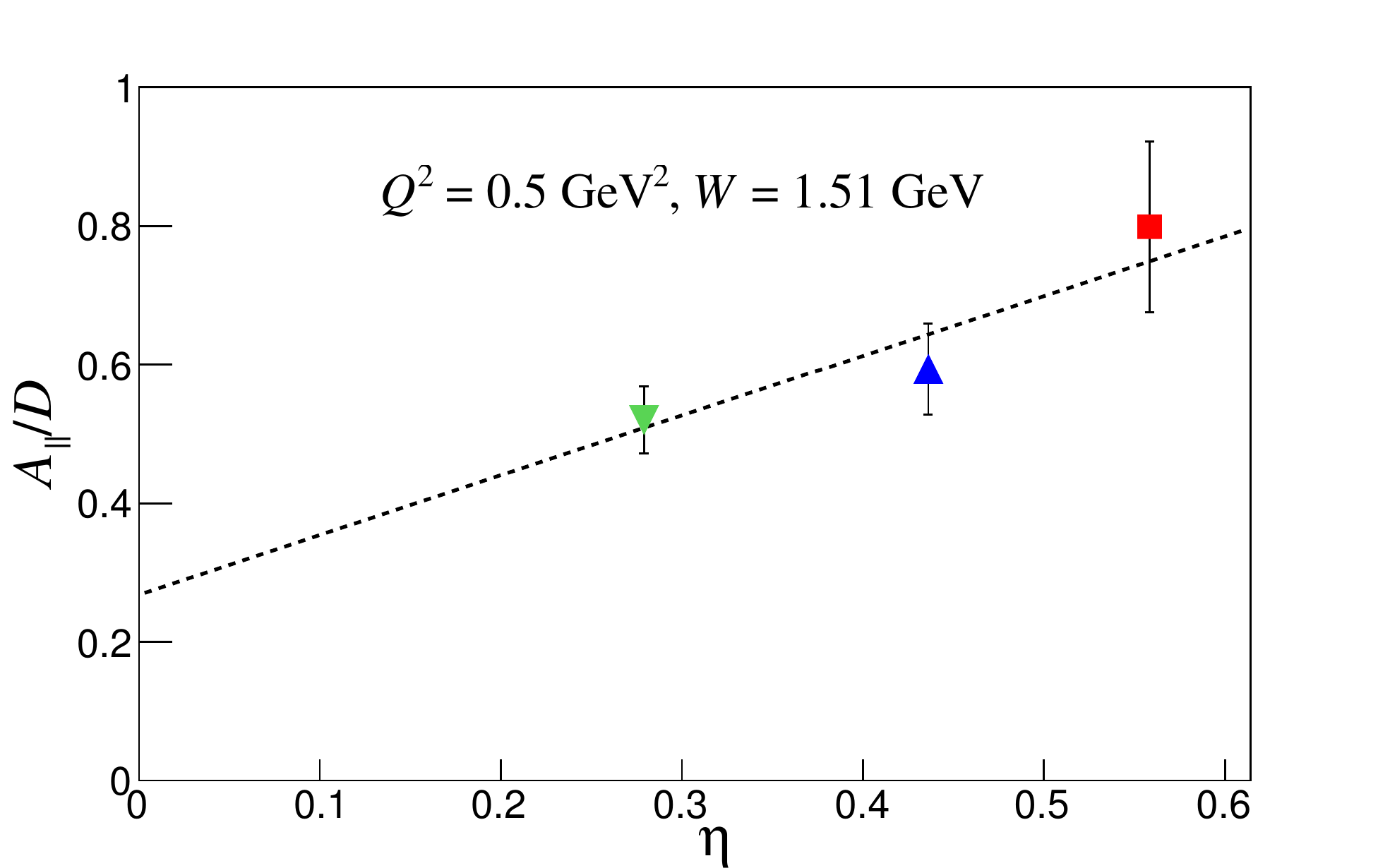}
\caption[Example of linear regression analysis]
{(Color online) Representative linear fit of $A_\parallel / D$ versus $\eta$ 
for one $W$,$Q^2$ bin (at $W$ = 1.51 GeV and $Q^2$ = 0.5 GeV$^2$).  The three
points were taken at three different beam energies (color and style coded as in Fig. \ref{AD:fig}).  
The $y$ intercept gives $A_1^p$ and the slope gives $A_2^p$.
}
\label{A2fit:fig}
\end{figure}
\\
\indent
Over a large kinematic region, asymmetries in the same ($Q^2$,$W$) bins were 
measured at multiple beam energies.
Consequently, for these bins, 
$A_1^p$ and $A_2^p$ can be obtained from a Rosenbluth-type of separation, as follows.
For fixed values of $Q^2$ and $W$, $A_{\parallel}/D$ is a linear 
function of the parameter $\eta$ which depends on the beam energy. 
A linear fit in $\eta$ determines both $A_1^p$ and $A_2^p$.
An example of this is shown in Fig.~{\ref{A2fit:fig}}.
One disadvantage of 
the method is its large sensitivity to uncertainties in the dilution factor and 
in $P_bP_t$ values for different beam energies. 
\begin{figure}
\centering
\includegraphics[width=9cm]{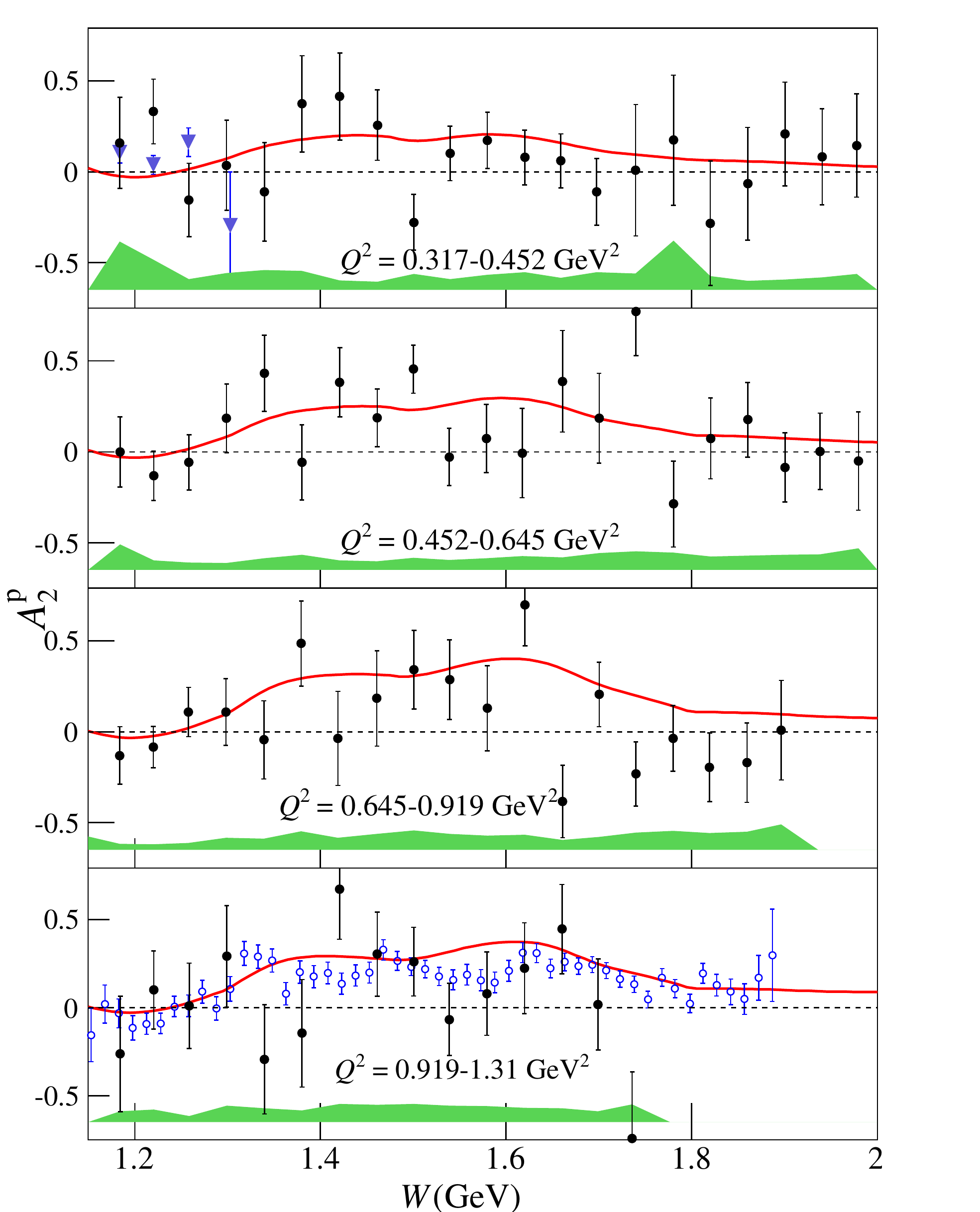}
\caption[Results for $A_2^p$ vs\ $W$]
{(Color online) $A_2^p$ vs\ $W$ extracted from the EG1b data (black filled circles) 
together with the RSS (blue open circles)~\cite{Wesselmann:2006mw} and Bates (purple inverted triangles)~\cite{Bates:ref} data.
The EG1b model (red solid line) is shown for comparison. The green band shows the systematic uncertainty.
}
\label{A2vx:fig}
\end{figure}
\\
\indent
For $W < 2$ GeV, the model-independent results for $A_2^p$ are shown in 
Fig.~{\ref{A2vx:fig}}, and compared to our model for  $A_2^p$,
as well as to data from RSS~\cite{Wesselmann:2006mw} (limited to $Q^2$ = 1.3 GeV$^2$),
MIT Bates~\cite{Bates:ref}, and NIKHEF (unpublished). 
For these plots, bins have been combined to increase the statistical 
resolution. Although our results for $A_2^p$ lack
the precision of the RSS \cite{RondonAramayo:2009zz} experiment, they 
extend over a wider range of
$Q^2$. 

For $W > 2$ GeV, we rarely have more than two beam energies contributing
to any given kinematic point, and usually only the highest two beam energies.
This yields a rather poor lever arm in $\eta$ and makes any check of the linear
fit (as well as its uncertainty) impossible. For this reason, we do not quote
any results for $A_2^p$ in the DIS region.

\subsubsection*{The spin structure function $g_2^p$}
\label{SuperRBg}

\begin{figure}
\centering
  \includegraphics[width=9.5cm]{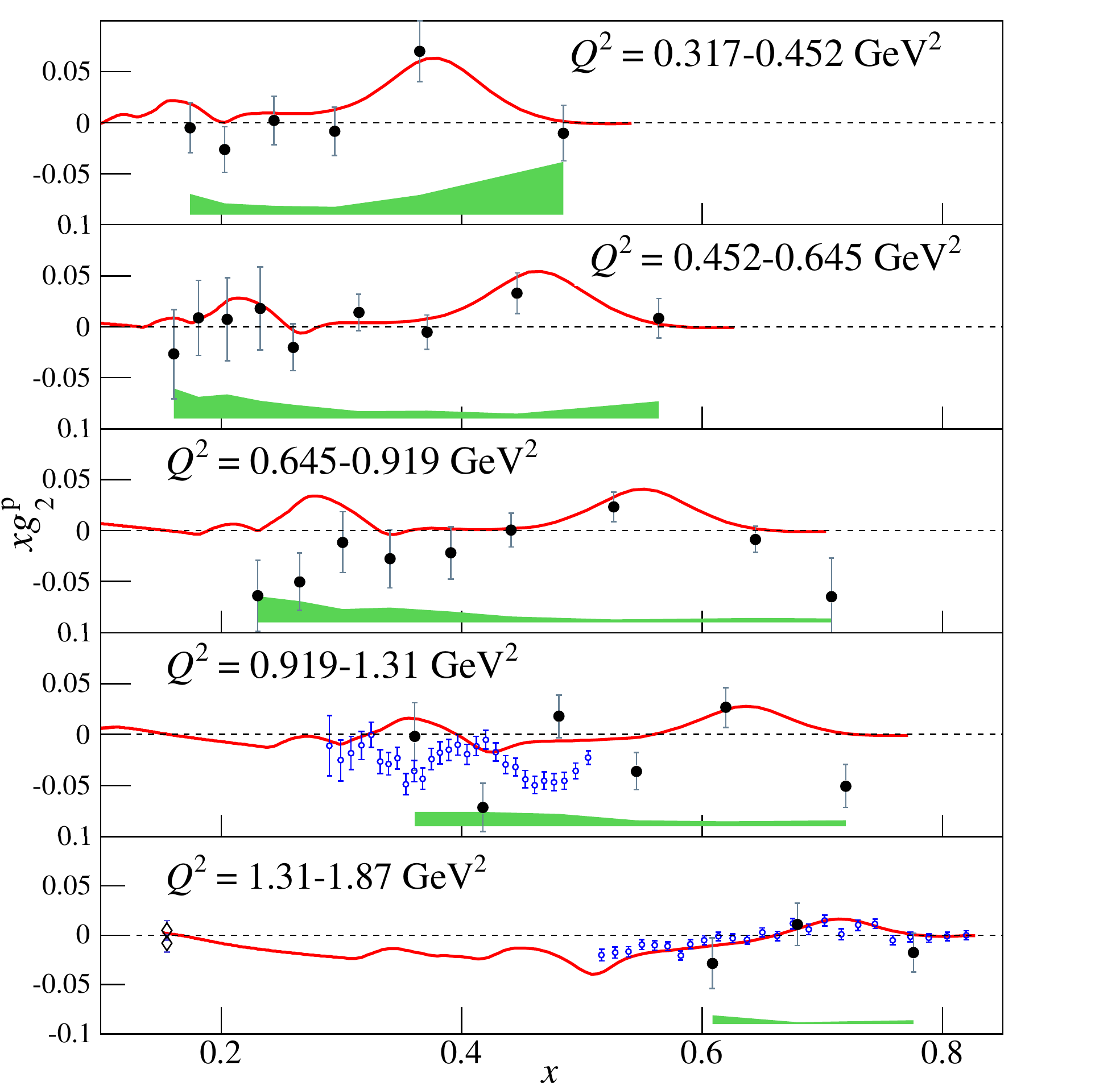}
\caption[Results for $xg_2^p$ vs\ $x$]
{(Color online) $xg_2^p$ vs\ Bjorken $x$ for the proton (solid black circles), together with RSS data 
(blue open circles)~\cite{Wesselmann:2006mw} and E155x data~\cite{Anthony:2002hy} (diamonds). The 
red curve is our model for the $Q^2$ bin median (which differs significantly 
from the average $Q^2$ value for the other data sets).
}
\label{xg2vx:fig}
\end{figure}

A model-independent value of $g_2^p$ can be obtained 
if one expresses 
 $A_\parallel$ directly as a linear combination of $g_1^p$ and $g_2^p$, 
again with energy dependent coefficients and a model for the
unpolarized structure function $F_1^p$ [see Eq.~(\ref{g2solve:eq})].
For ($Q^2,W$) bins
measured at more than one energy, $g_1^p$ and $g_2^p$ can then be
determined with a straight-line fit, along with a straight-forward calculation of the statistical
uncertainty.
As already discussed, this is not the
best way to determine $g_1^p$, but it does provide model-independent
values for $g_2^p$. The results for the product $x g_2^p$ 
averaged over four different $Q^2$ ranges are displayed as a function of $x$
in Fig.~\ref{xg2vx:fig}. 
Although the precision is
not particularly good, these data could provide some constraints on models of $g_2^p$.

\subsection{Models}
\label{models}
In order to extract high-precision observables of interest from our data on $A_{||}$, we need to
use models for the unmeasured structure functions $F_1^p$ and $F_2^p$ (or, equivalently,
$F_1^p$ and $R^p$), as well as for the  asymmetry $A_2^p$, which is only poorly 
determined by our own data (see above).
Using these models, we can extract $A_1^p$ and $g_1^p$ from the measured $A_{||}$, as
explained in Sec.~\ref{SFsAsyms}.
In addition, we also need a model for $A_1^p$, covering a wide kinematic range, in order to
evaluate radiative corrections stemming from both the measured and the unmeasured
kinematic regions, and
to evaluate the unmeasured contributions to the moments of the structure function $g_1^p$.

For the unpolarized structure functions $F_1^p$ and $R^p$, we used a recent 
parametrization
of the world data by Bosted and Christy~\cite{Christy:2007ve}. This 
parametrization fits both DIS and resonance 
data with an average precision of 2$-$3\%. In particular, it includes
the extensive data set on separated structure functions collected at 
Jefferson Lab's Hall C~\cite{Liang:2004tj}
which is very well matched kinematically to our own asymmetry data. 
Furthermore, the fit has been
modified to connect smoothly with data for real photon absorption, 
thereby yielding a fairly reliable
model for the (so far unmeasured) region of very small $Q^2$.
Systematic uncertainties due to these models were calculated by
varying either $F_1^p$ or $R^p$ by the average uncertainty of the fit (2-3\%) and recalculating
all quantities of interest.

For the asymmetries, we developed our own phenomenological fit to the world data, including
all DIS results from SLAC, HERA and CERN and all results from Jefferson Laboratory data
(see Ref.~\cite{Kuhn:2008sy} for a complete list)
as well as data in the resonance region from MIT Bates~\cite{Bates:ref}. In particular,
we used an earlier version of this fit~\cite{Yun:2002td} for a preliminary extraction of $A_1^p$ from our own
data, and then iterated the fit including these data. 

The fit proceeded in the following steps:
\begin{enumerate}
\item The asymmetry $A_1^p(x,Q^2)$ in the DIS region, $W > 2$ GeV, was fit using an analytic
function of $Q^2$ and the variable $\xi^{'} = \xi (1 + 0.272 \, \mathrm{GeV}^2/Q^2)$, where the Nachtmann
variable $\xi$ given in Eq.~(\ref{Nachtmann}) was modified to allow a smooth connection to a finite value 
at the real photon point, $Q^2 = 0$. The seven parameters of this function were optimized by
fitting this function to
all world data at $W > 2$ GeV and the fit function,
including real photon data from ELSA and MAMI (see, e.g., the summary by Helbing~\cite{Helbing:2006zp}). 
Each experiment was given an adjustable normalization
factor as an additional parameter which was allowed to vary within the stated uncertainty due to 
global scale factors like the product $P_bP_t$. Some comparisons of the fit with world data (including
the ones reported here) are shown in Figs.~\ref{A1alt:fig} and \ref{A1alta:fig}. The full error matrix
from the fit was used to calculate the uncertainty of our model $A_1^p$ at any particular kinematic
point. All values of $A_1^p$ used in radiative corrections or moments
were moved by this uncertainty (one standard deviation) to determine the systematic uncertainty
from this model.

\item The asymmetry $A_2^p(x,Q^2)$ in the DIS region was modeled by using the Wandzura-Wilczek
form of the structure function $g_T$ [Eq.~(\ref{gTdef})] and observing that $A_2^p = \gamma g_T/F_1^p$
[Eq.~(\ref{A2:eq})]. This description was found by SLAC experiments E143 and E155 to hold rather
well; as a systematic variation, we also included a simple functional form for an additional 
``twist-3'' term introduced by E155~\cite{Anthony:2002hy}.

\item In the resonance region, we modeled both asymmetries by combining the DIS fits
(extrapolated to $W < 2$ GeV)
with additional terms emulating resonant behavior. 
For the latter, we used the MAID
parametrization of the cross sections
$\sigma_{TT} = \sigma^{\frac{1}{2}}_T({\gamma^*}) - \sigma^{\frac{3}{2}}_T({\gamma^*})$,
$\sigma_{T} = \sigma^{\frac{1}{2}}_T({\gamma^*}) + \sigma^{\frac{3}{2}}_T({\gamma^*})$,
and $\sigma_{LT}({\gamma^*})$
 for single pion and $\eta$ production~\cite{Drechsel:1998hk,Kamalov:2001yi}. 
We fit all data in the resonance region
using $Q^2$- and $W$-dependent weighting factors for these two terms,  which guaranteed a smooth
connection to the DIS fits at $W=2$ GeV and for $Q^2 \rightarrow 10$ GeV$^2$ (assuming negligible
effects from resonances at higher $Q^2$). We included our model-independent results for
$A_2^p$ described in the previous section, as well as the more precise data from 
RSS~\cite{Wesselmann:2006mw} and MIT-Bates~\cite{Bates:ref}.
Ultimately, we combined this fit with an earlier version~\cite{Yun:2002td} for the best possible 
description of all data, and used the difference with the earlier version
as a systematic uncertainty. A total of 28 parameters for $A_1^p$ and 9 parameters for $A_2^p$ were
fit using $\chi^2$ minimization. The data for $A_2^p$ are sparse and therefore
fewer parameters were sufficient. We used the Soffer inequality [Eq.~(\ref{eq:A2bound})]
as an additional constraint. The resulting uncertainty on $A_2^p$ was small enough for our purpose
of extracting $A_1^p$ and $g_1^p$ as discussed below.
The final implementation of our fit is in the form of a fine-grained lookup table that can be interpolated
in both $W$ and $Q^2$. The reason for this is that we did not have access to a version of the 
MAID code that would allow us to calculate the necessary input to our model in real time; instead, we
used a grid of values.
Comparisons of our fit with our own data for $A_2^p$ and $A_1^p$ are shown 
in Fig.~\ref{A2vx:fig} and in Figs.~\ref{A1alt:fig} and \ref{A1alta:fig},
respectively. 
\end{enumerate}

\subsection{Model-dependent extraction of $A_1^p$}

\begin{figure*}
\centering
  \includegraphics[width=17cm]{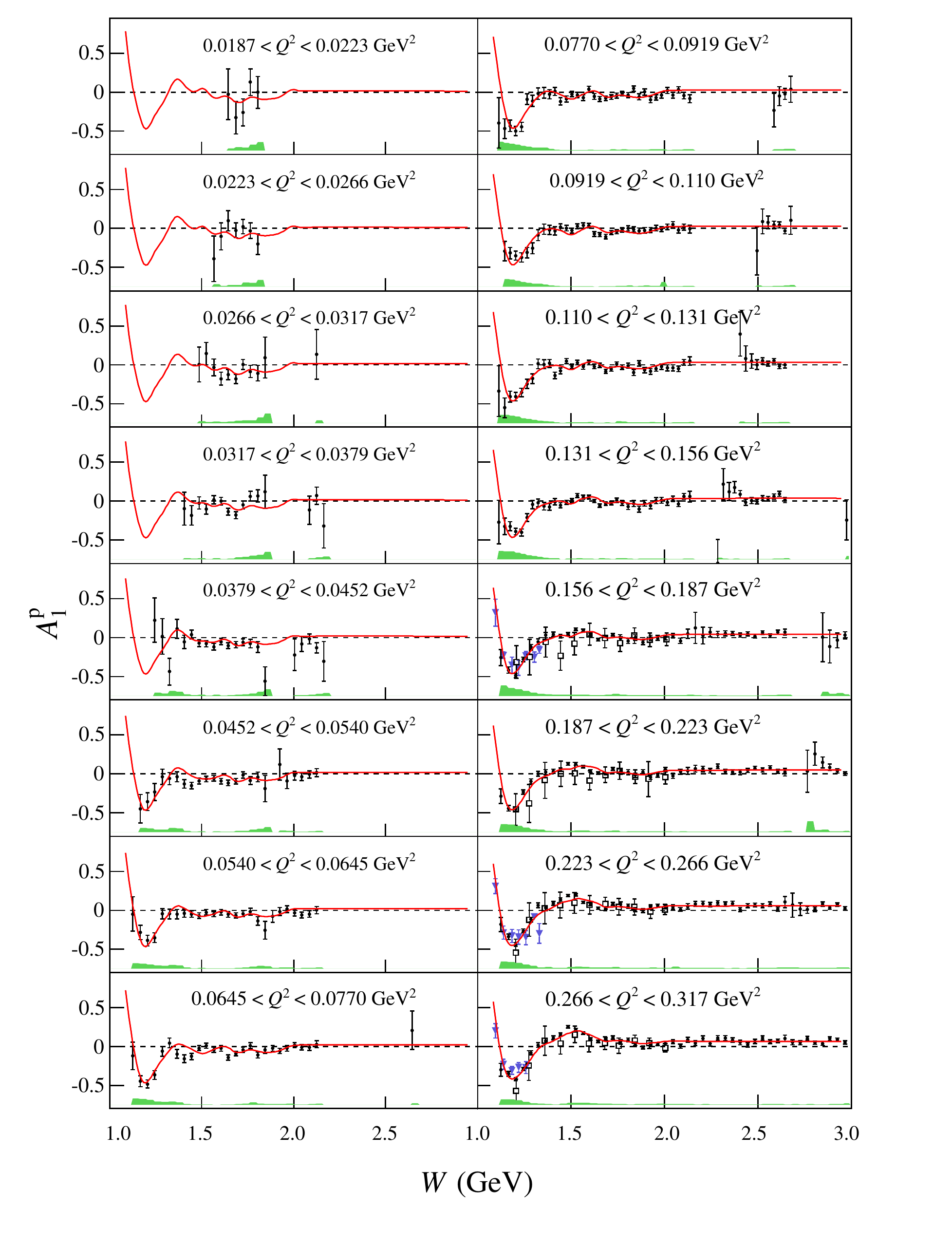}
\caption[Results for $A_1^p$ vs\ $W$]
{(Color online) Asymmetries $A_1^p$ vs\ $W$ for bins in $Q^2$.
The solid black points are our data with statistical error bars. 
Open squares represent EG1a data~\cite{Fatemi:2003yh}, and the purple triangles are 
Bates data~\cite{Bates:ref}, visible on the left side of three of the four highest $Q^2$ plots shown. 
The red line shows our model of $A_1^p$ for comparison. The green bands show the systematic uncertainties. 
\label{A1alt:fig}
}
\end{figure*}

\begin{figure*}
\centering
  \includegraphics[width=17cm]{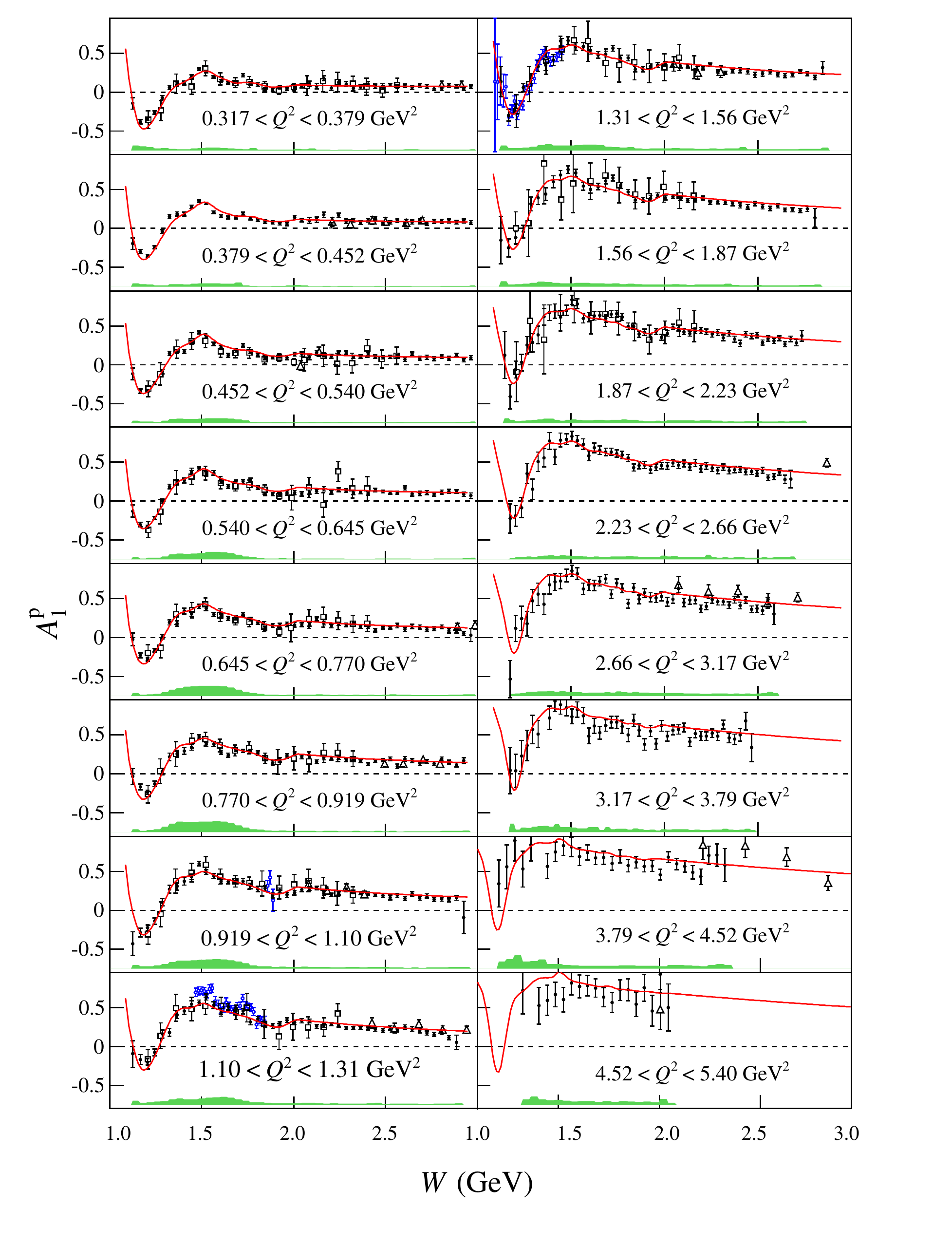}
\caption[Results for $A_1^p$ vs\ $W$]
{(Color online) Same as Fig. \ref{A1alt:fig} but for the higher $Q^2$ bins. 
Additionally, here, blue hollow circles are RSS data~\cite{Wesselmann:2006mw}, and open 
triangles are E143 data~\cite{Abe:1998wq}.
}
\label{A1alta:fig}
\end{figure*}

Because of the relatively small contribution of $A_2^p$ to $A_\parallel$, 
even our only moderately constrained model estimation of $A_2^p$
permits a rather accurate extraction of $A_1^p$ over a large range of
$Q^2$ and $W$.  $A_1^p$ was determined
directly from Eq. (\ref{Apar:eq}), using our models for $R^p$ and $A_2^p$ as input.

 $A_1^p$ was extracted for each ($Q^2$,$W$) bin, separately for each
data set obtained with the four average beam energies (1.6, 2.5, 4.2, and
5.7 GeV). The statistically averaged values of $\eta$ in each bin were
used to prevent weighting uncertainties. Final results for $A_1^p$ measured at each beam energy 
were then statistically averaged.
For each combination, we checked first that the values of $A_1^p$ from
different beam energies were statistically compatible (which turned out to be
true in all cases). The final results are shown in Figs.~\ref{A1alt:fig} and \ref{A1alta:fig}.
\\
\indent
Inclusive electron scattering at $W <  2$ GeV and low to moderate
 $Q^2$ is characterized by a strong
$W$-dependence arising from the excitation of nucleon resonances (see Ref.~\cite{Aznauryan:2011qj} for a review). 
One typically observes three cross section peaks, traditionally labeled
as the first, second, and third resonance regions.   As discussed in 
Sec.~\ref{photonabsorption}, the total spin of an excited
resonance is reflected in its contribution to $A_1^p$.
The first resonance region is dominated
by excitation of the $\Delta(1232)P_{33}$ resonance, with total spin $S=\frac{3}{2}$ and
$W=1.232$ GeV. As discussed in Sec. \ref{photonabsorption},
$A_1^p \approx -\frac{1}{2}$ in this region from the resonance contribution alone. 
This is borne out by our data for the lowest $Q^2$, while
at higher $Q^2$ non-resonant background and tails from higher-lying resonances begin to dominate, making
$A_1^p$ less negative. 
The second resonance region arises from 
excitation of a group of  closely spaced resonances, in particular 
$N(1535)S_{11}$ and $N(1520)D_{13}$. Between the first and second
regions, the excitation of the the Roper resonance $N(1440)P_{11}$ is not
prominent in electro-excitation at low $Q^2$ where the leading amplitude crosses zero,
but it contributes significantly above $Q^2=2$ GeV$^2$ over a region three times as broad  in $W$ as the $\Delta(1232)P_{33}$,
creating a shoulder in $A_1^p$
around $W=1.44$ GeV,  which is visible in  our data. 
This and other
spin-$\frac{1}{2}$ resonances, which have no spin-$\frac{3}{2}$ projection, lead to $A_1^p=1$ for the resonance contribution only. 
In the second region the dominant $N(1535)S_{11}$ resonance 
drives $A_1^p$ toward unity.
The other major resonance in this region, $N(1520)D_{13}$,  has  $A_1^p=-1$ for real photons ($Q^2=0$)
but it rapidly tends toward $A_1^p$ = +1 for $Q^2 >$ 3 GeV$^2$, characteristic of pQCD expectations.
Indeed, our data exhibit a rapid rise from $A_1^p \approx 0$ at low $Q^2$ to large positive values
at higher $Q^2$ in this region.
The third resonance peak lies at $W=1.63$ GeV and contains, among others, the $N(1680)F_{15}$
resonance.  Additional enhancements in
the real photon cross section arise from excitation of a number of 
resonances  with $1.7 < W <1.9$ GeV, some of which are
spin-$\frac{3}{2}$ or higher and therefore tend to have negative $A_1^p$.
These features are visible as well in our data at low $Q^2$.
Another prominent feature is the
nearly uniform increase of $A_1^p$ with increasing $Q^2$.

\begin{figure}
\includegraphics[width=9.5cm]{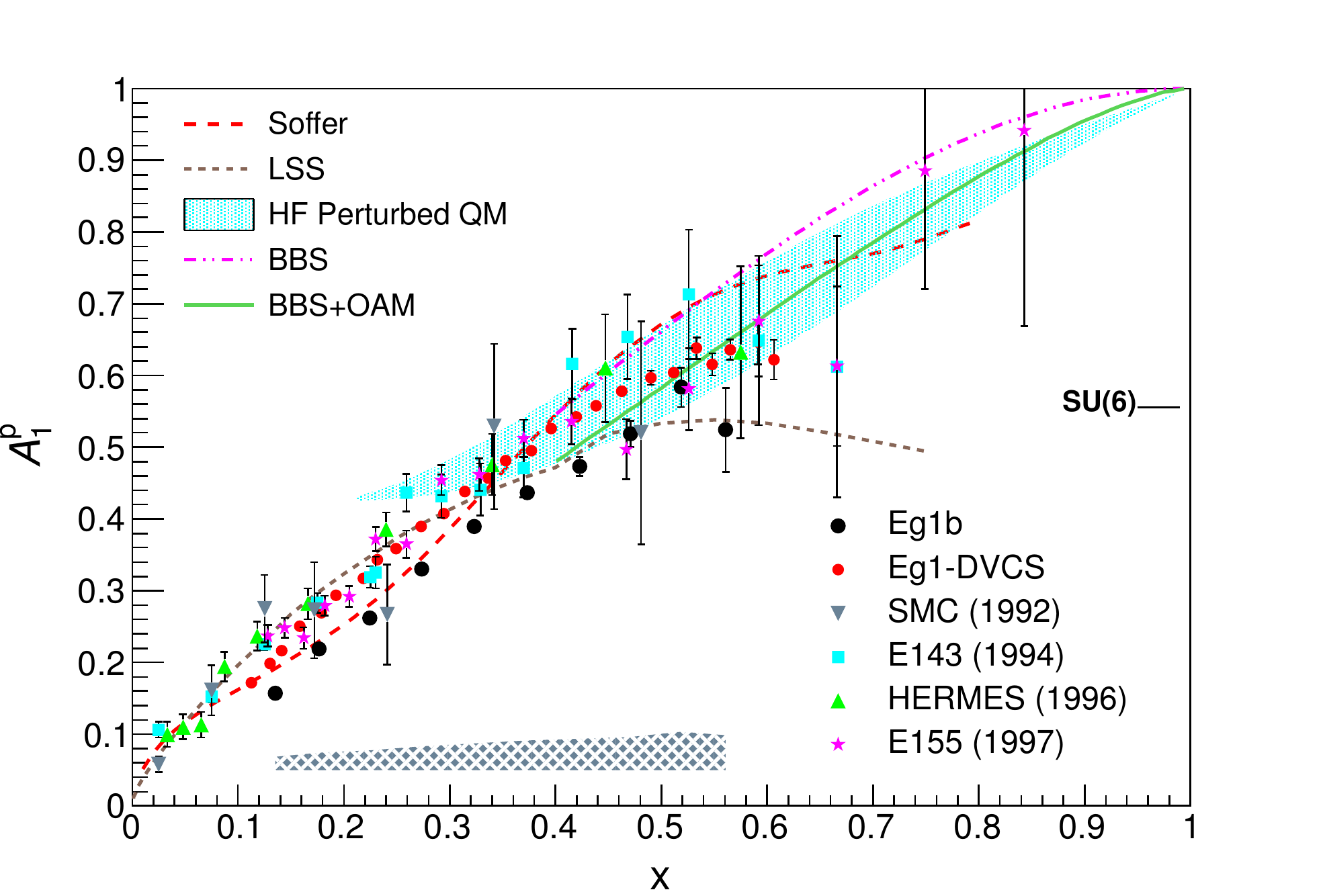}
\caption
{(Color online) $A_1^p$ vs\ $x$ for DIS events, $W > 2$ GeV, compared to world data. Curves and models are 
discussed in the text.  The difference between EG1b data and higher energy data is discussed 
in the main text. The hatched region at the bottom represents the systematic uncertainty on the EG1b data.
}
\label{A1largex}
\end{figure}

As discussed in Sec. \ref{photonabsorption}, predictions of
the high $x$ DIS behavior of $A_1^p$ are strongly model-dependent, although most
realistic models predict some sort of smooth approach to the value
$A_1^p=1$ at $x=1$, which would be consistent with $A_1$ for elastic scattering. To compare our results for $A_1^p$
 to the world's DIS data, we
restricted the kinematical region to $W>2$ GeV to avoid complications from
the resonance region, which clearly shows departures from DIS behavior.
With this restriction on $W$, the upper limit of
$x=0.6$ for our data is fixed by the maximum JLab electron
energy.
The results obtained with this restriction are compared to world DIS data
for $A_1^p$ in Fig.~{\ref{A1largex}}. This plot also displays
several predictions and fits of the $x$-dependence of $A_1^p$:
a ``statistical'' model for quark distribution 
functions by Soffer \etal~\cite{Bourrely:2005kw}, 
an NLO fit to the world data without constraint at $x = 1$ by 
Leader, Stamenov and Siderov \etal~\cite{Leader:2006xc},
a range of predictions from a relativistic quark model with hyperfine interactions due to one-gluon
exchange~\cite{Isgur:1998yb}, and two different models based on pQCD expectations,
one without (BBS~\cite{Brodsky:1994kg}) and one with (BBS+OAM~\cite{Avakian:2007xa}) 
quark orbital angular momentum.

\begin{figure}
\centering
  \includegraphics[width=9.2cm]{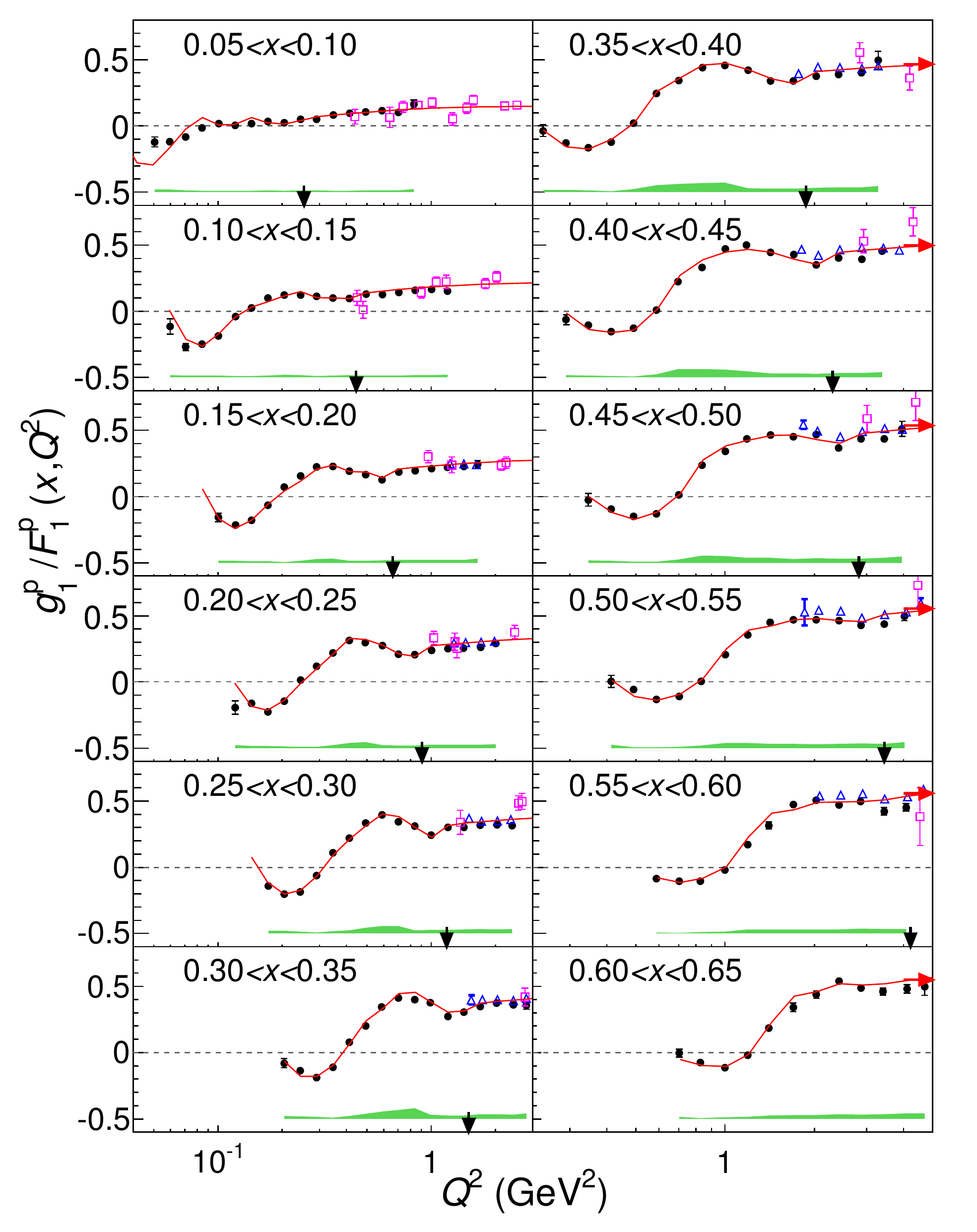}
\caption[Results for $g_1^p/F_1^p$ vs\ $Q^2$, binned in $x$]
{(Color online) Plots of $g_1^p/F_1^p$ versus $Q^2$ for different $x$ ranges of the combined EG1b
  data. The (red) line
represents our model. The blue triangles correspond to the EG1-dvcs data 
\protect\cite{Prok:2014ltt}, while magenta squares represent 
E143 data~\cite{Abe:1998wq}. The downward-pointing black arrows indicate the 
upper limit of the resonance region at 
$W$ = 2 GeV, while the the red horizontal arrows indicate the results for $g_1^p/F_1^p$ of a recent analysis 
of world data for our bin centers and $Q^2 =$ 5 GeV$^2$.
}
\label{g1F1vx:fig}
\end{figure}

Several features are obvious. Our data tend to lie lower than the EG1-dvcs data, 
not because of large discrepancies (as can be seen in Fig. \ref{g1F1vx:fig}), but due to the 
significantly different kinematics between these two data sets, which affects the $Q^2$ range over
which we average, and the impact of various models (in particular, $A_2^p$).
 Our model fit confirms that indeed
even in the DIS region, $A_1^p(x,Q^2)$ is not completely $Q^2$-independent (scaling),
but rather increases as $Q^2$ increases. Taking this effect into account, our data
are in good agreement with the world data set. At moderately high $x$, our data 
show an
unambiguous increase, as expected,  beyond the naive SU(6) quark model prediction of
$A_1^p = 5/9$.

\begin{figure*}[hbt!]
\centering
 \includegraphics[width=18.4cm]{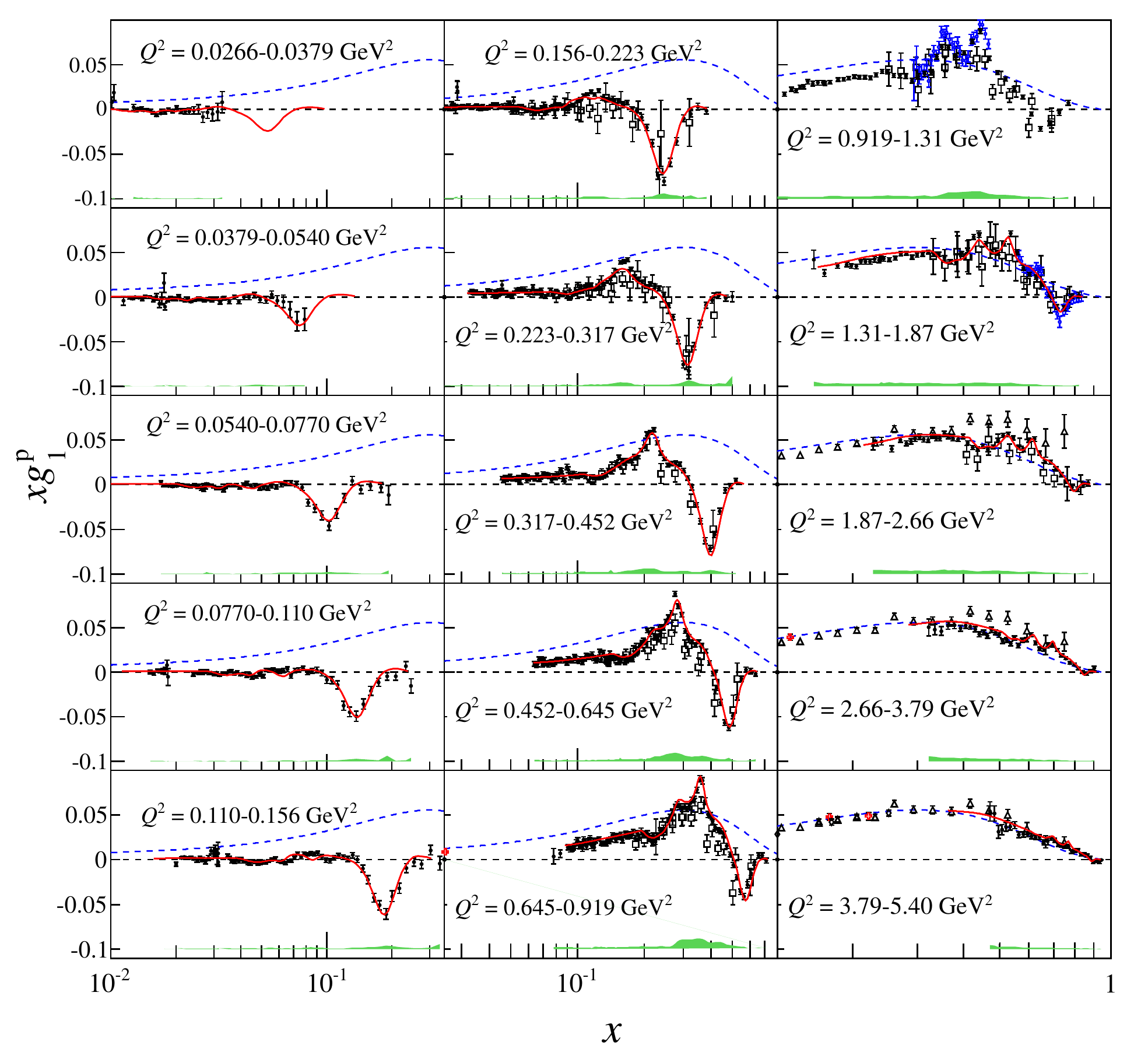}
\caption[Results for $xg_1^p$ vs\ $x$ (set 1)]
{(Color online) Spin structure function $xg_1^p$ vs\  
Bjorken $x$, for various bins in $Q^2$.   
Our data (black points) are plotted along with the world data at similar $Q^2$: from HERMES 
(red crosses)~\cite{Airapetian:2006vy}, E155 (diamonds)~\cite{Anthony:2000fn}, E143 (hollow triangles)~\cite{Abe:1998wq}, 
RSS (blue circles)~\cite{Wesselmann:2006mw}, and EG1a (hollow squares)~\cite{Fatemi:2003yh}. 
The green band indicates total systematic uncertainties; the red solid line is our model for the 
median of each $Q^2$ bin, and the blue dashed line is the DIS model at $Q^2=$ 10 GeV$^2$, included for reference.
}
\label{xg1vx:fig}
\end{figure*}

\begin{figure}
\centering
\includegraphics[width=9.7cm]{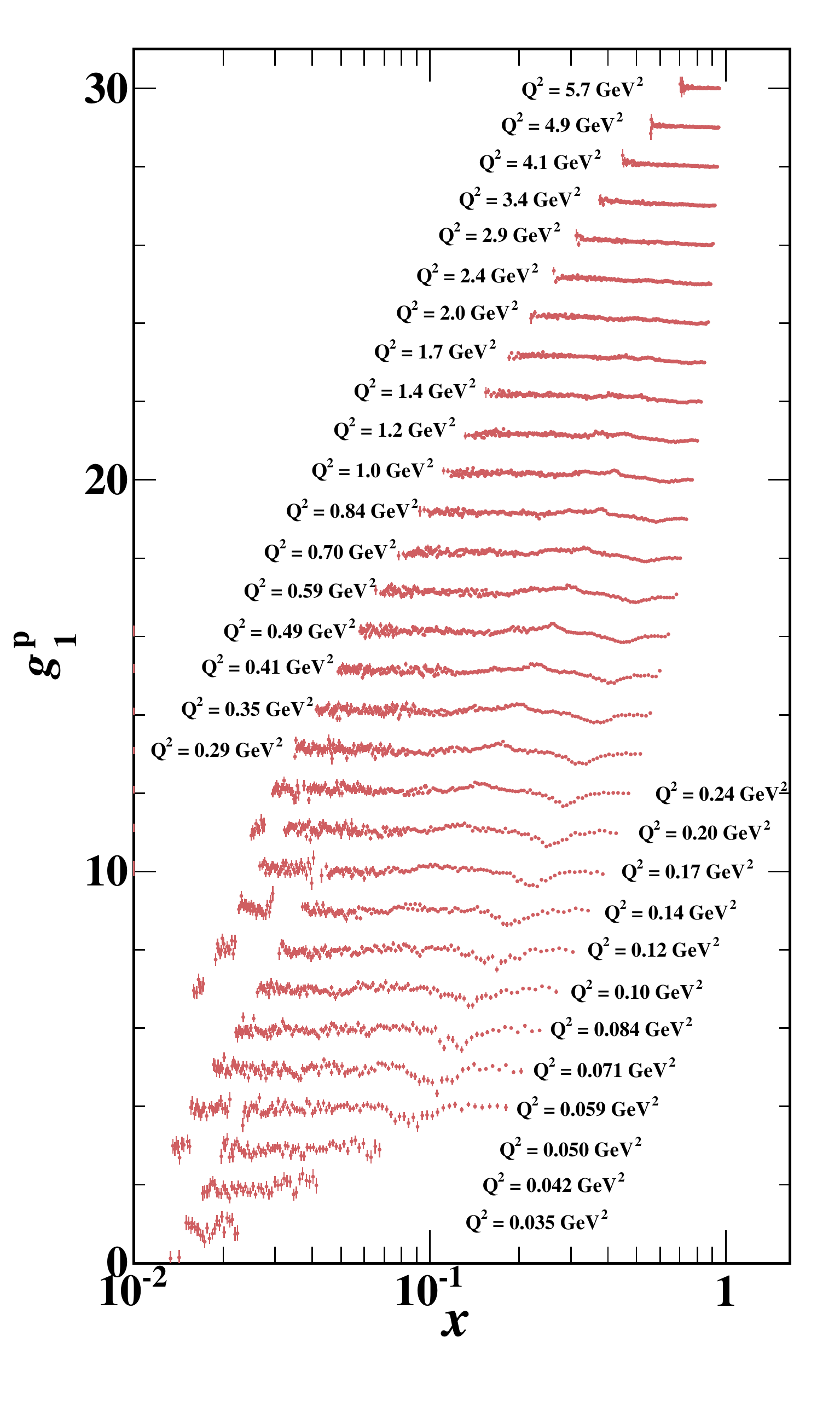}
\caption[g1 results]
{(Color online) The full $g_1^p$ data set from this experiment. For clarity, the $n$-th $x$ distribution at fixed $Q^2$
is shifted upward by $1+n$.}
\label{fig:g1full}
\end{figure}

\subsection{The spin structure function $g_1^p$}
Analogous to the case for $A_1^p$, the most precise results for $g_1^p$ can be
extracted from our measurement of $A_{||}$ using models for all
unmeasured structure functions, including $A_2^p$ [see Eq.~(\ref{g1fromApar})].
Over most of our kinematics $ \vert\gamma-\eta\vert \ll \vert\eta\vert$,
which ensures that the  uncertainty in our $A_2^p$ model
is even less important in the 
extraction of $g_1^p/F_1^p$ than for the  extraction of $A_1^p$.
Consequently, the uncertainties  on $g_1^p/F_1^p$ are primarily statistical.

Our complete data set for the quantity $x g_1^p(x,Q^2)$
is shown in Fig.~\ref{xg1vx:fig}, together with a sample of
world data. 
One can see a clear
transition from the resonance-dominated behavior at low $Q^2$ with
the prominent negative peak in the $\Delta$ resonance region towards
the smooth behavior at high $Q^2$, where most of the data lie in the DIS
region. At intermediate $Q^2$, one can discern an $x$ dependence that still
has some prominent peaks and dips,
 but approaches, on average, the smooth DIS curve at the highest $Q^2$. This
 is a qualitative indication of quark-hadron duality, which is
 discussed below (see Sec.~\ref{qhduality}).
\\
\indent
Plots of $g_1^p/F_1^p$ as a function of $Q^2$
for various $x$ bins are shown in Fig.~{\ref{g1F1vx:fig}}. For comparison,
these plots
also show data from the SLAC E143 and E155 experiments. The solid line
on each plot shows the result of our model at the median value of each
bin. The systematic uncertainty is shown as the green region near the bottom of
each plot. Again, a dramatic $Q^2$ dependence at low $Q^2$ (where 
the low-$W$ region dominates for fixed $x$) makes way to the
smooth approach towards the DIS limit at higher $Q^2$. The
remaining $Q^2$ dependence at the upper end of each plot hints at
scaling violations of $g_1^p/F_1^p$ due to pQCD evolution.
\\
\indent
The quantity $g_1^p$ was derived for all values of $A_{||}/D$ over the entire kinematic range using Eq. (\ref{g1fromApar}), with model values used for $A_2^p$ and $F_1^p$. The complete coverage of $g_1^p$ over the EG1b kinematic range is displayed in 
Fig. \ref{fig:g1full}.

\subsection{Moments of $g_1^p$}

As discussed in Sec.~\ref{moments}, moments of $g_1^p$ and $g_2^p$ 
with powers of $x$ play an important
role in the theory of
nucleon structure in the form of sum rules and for the determination
of matrix elements within
the OPE.
The $n$th moment of
a structure function $\cal S$ is defined by $\int_0^1 x^{n-1}{\cal S}
(x,Q^2)\,dx$. Experimental data do not cover the complete range in $x$ for each $Q^2$ bin (see Fig.~\ref{fig:g1full}),
but the moments can be approximated using a combination of our
data along with a model for low $x$ and high $x$. Thus, the calculation
can be expressed as
\begin{eqnarray}
\int_{x_{\rm high}}^1                   x^{n-1}{\cal S} (x,Q^2)_{\rm model}\,dx\\ \nonumber
+\int_{x_{\rm low}}^{x_{\rm high}}      x^{n-1}{\cal S} (x,Q^2)_{\rm data}\, dx\\ \nonumber
+\int_{0.001}^{x_{\rm low}}                   x^{n-1}{\cal S} (x,Q^2)_{\rm model}\,dx. 
\end{eqnarray}
At very low values of $x$, uncertainties in
the model become so large that we have chosen to truncate the lower limit
at $x=0.001$.
Ignoring the interval $[0,0.001]$ is expected to have little effect, especially
for $n>1$. 

\subsection{Moments of $g_1^p$}

The $n$th $x$-weighted moment of $g_1^p$ was determined from our data as follows.
For each $Q^2$ bin the data were binned in $W$ 
with $\Delta W =$ 10 MeV, so that
\begin{equation}
I_{\rm{data}}(Q^2)=\sum_{W}x_{\rm{avg}}^{n-1}{\cal S}(Q^2, W)|x_a-x_b| ,
\end{equation}
where $x_{\rm{avg}}$ is
the average value of $x$ for the events contributing to each bin, and
$x_a$ and $x_b$ are the lower and upper limits of the $W$ bin.
The statistical uncertainty for each bin was added in quadrature to obtain the
statistical uncertainty on the integral. Bins with a statistical uncertainty
for $A_\parallel$ greater than 0.6 
were excluded. In kinematic regions where data were absent or insufficient by this criterion, the 
model was used.  The integral ran from the inelastic threshold ($W=$ 1.07 GeV) up to the value of $W$ 
corresponding to $x=0.001$ for each $Q^2$ bin. 
The model was also integrated over the full $x$ range for comparison to the data (see Fig. \ref{Gamma1:fig}).

In our plots of the calculated moments,  the experimental contributions are
shown as open circles and the 
combination of model and data is shown as solid black circles. 
Systematic
uncertainties were calculated using the methods described earlier
and are shown in shaded bands.

The moment calculations presented here (with the exception of Fig. \ref{fig:gam1pel}) do not
include the contribution from elastic scattering at $x=1$, which is the same for all $n$ 
[see Eq.~(\ref{elasticgs})].

\subsubsection{The first moment $\Gamma_1^p$}

\begin{figure}
\centering
  \includegraphics[width=9.0cm]{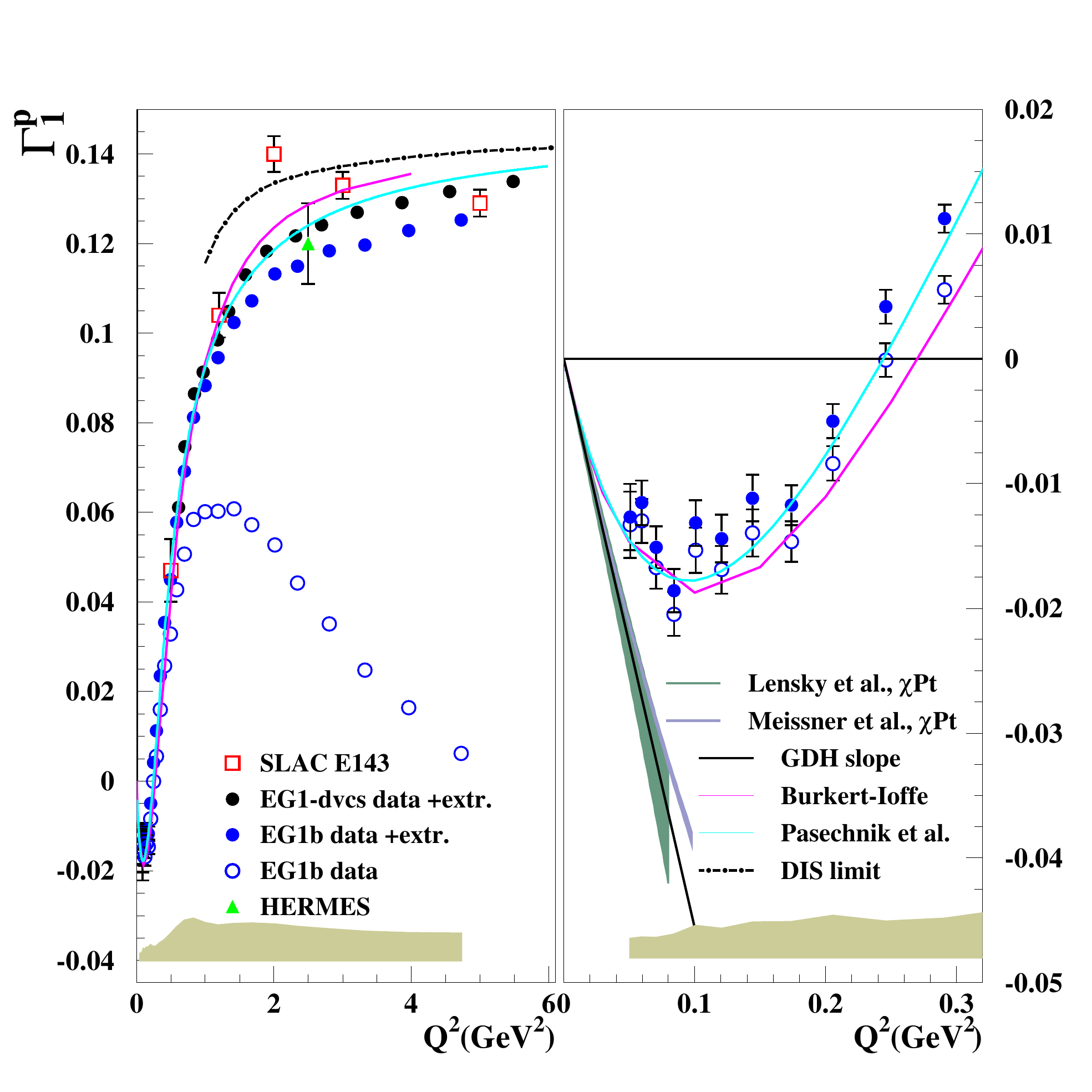}
\caption[Results for $\Gamma_1^p$ vs\ $Q^2$]
{(Color online) $\Gamma_1^p$ vs\ $Q^2$ for 
EG1b data and selected world data. The right panel shows an expanded
scale at small $Q^2$.
The open circles represent our data, integrated over the measured region. The filled
blue circles are the full integral from $x = 0.001 \rightarrow 1$, excluding
the elastic region. 
The curves show phenomenological parametrizations by
Burkert and Ioffe~\cite{Burkert:1993ya,Burkert:1992tg} (magenta) and 
Pasechnik {\it et al.} \cite{Pasechnik:2009yc} (cyan). 
The limiting
cases of large $Q^2$ (``DIS limit'') and $Q^2 \rightarrow 0$ (``GDH slope'') are also shown,
as well as two bands showing $\chi$PT calculations,
(Lensky {\it et al.} \cite{Lensky:2014dda} and Meissner {\it et al.} \cite{Bernard:2002pw}). The green band at the bottom represents the total systematic uncertainty.
}
\label{Gamma1:fig}
\end{figure}

The moments of $g_1^p$, designated as $\Gamma^p_n$ have been calculated from
our data up to $n=5$. The   first
moment  $\Gamma_1^p$  is of special interest.
At $Q^2=0$ the GDH sum rule constrains the slope of $\Gamma_1^p(Q^2)$
to be $-0.456$ GeV$^{-2}$  [Eq.~(\ref{Gamma1slope})]. At large $Q^2$,
$\Gamma_1^p$ is related to squared charge-weighted axial charges of all
quark species present in the nucleon (see Sec.~\ref{moments}). From
existing DIS data and theoretical expectations, it is well-known that in
this limit
$\Gamma_1^p$ is positive and approaches a value of about 0.14$-$ 0.15,
with a $Q^2$ dependence given by pQCD.
Consequently, at some value of $Q^2$, $\Gamma_1^p$ must pass through zero.
The plots of our results for $\Gamma_1^p$ shown in Fig.~{\ref{Gamma1:fig}} are
consistent with these expectations, exhibiting a sign change
at $Q^2 \approx 0.24 $ GeV$^2$. 

Various models and parametrizations
have been proposed to interpolate between the two extreme $Q^2$ limits.
At high $Q^2$, pQCD corrections up to third order in $\alpha_S$ have been
calculated and are shown in Fig.~{\ref{Gamma1:fig}}, as is the ``GDH
slope'' at $Q^2 = 0$. The next higher order terms in an expansion in $Q^2$
around the origin can be calculated  within the framework
of $\chi$PT \cite{Lensky:2014dda, Bernard:2002pw}.
Finally, we show two phenomenological
curves using the methodology of  Burkert, Ioffe, and Li ~\cite{Burkert:1993ya,Burkert:1992tg,Burkert:1992yk} and by 
Soffer, Pasechnik {\it et al.}~\cite{Soffer:1992ck,Soffer:2004ip,Pasechnik:2009yc},
which reproduce the data, at least qualitatively, quite well.

\subsubsection{Higher moments}

\begin{figure}
\centering
  \includegraphics[width=8.9cm]{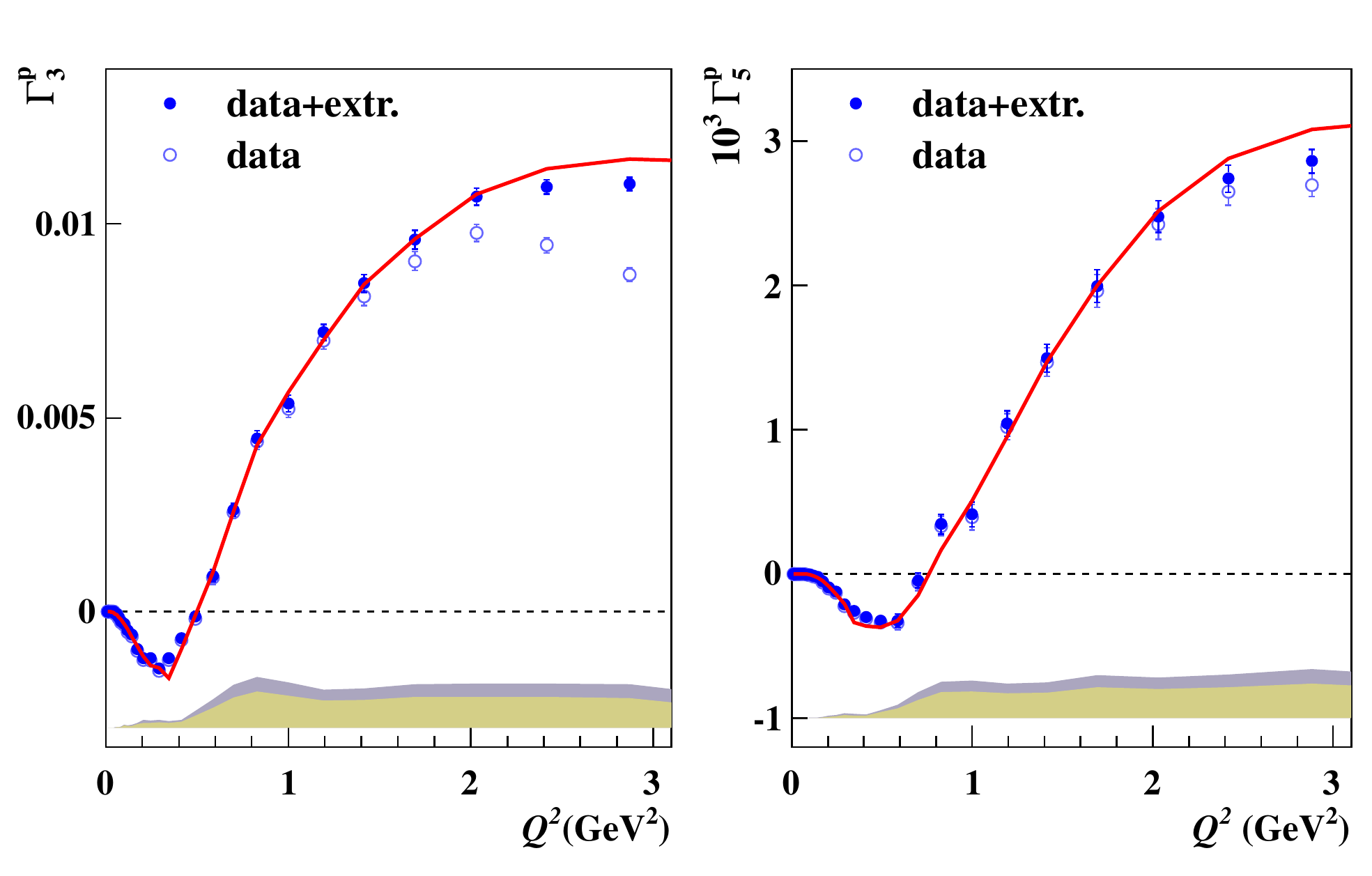}
\caption[Results for $\Gamma_3^p$ and $\Gamma_5^p$ vs\ $Q^2$]
{(Color online) $\Gamma_3^p$ and $\Gamma_5^p$ vs\ $Q^2$ for EG1b data. 
Solid (blue) circles are the total integral, whereas the open (blue) 
circles are the integral over measured data.
The curve (red) is our model. The gold and gray bands at the bottom represent the systematic uncertainties on the 
data and the data $+$ model contributions, respectively.
}
\label{Gamma35:fig}
\end{figure}

The third and fifth moments of $g_1^p$ are shown
in Fig.~{\ref{Gamma35:fig}}. These moments are characterized by small
statistical uncertainties, along with
 very little model dependence for $Q^2<3$ GeV$^2$.  They are useful in
the calculation of hydrogen hyperfine splittings \cite{Nazaryan:2005zc,Deur:2014vea}.

\subsubsection{Higher twist analysis}

We detail here the analysis performed to extract the twist-4 contribution
$f_{2}^{p}$ to  $g_{1}^{p}$ and to determine the contribution of
the quarks to the nucleon spin $\Delta\Sigma$. A summary of the formalism 
describing the higher-twist matrix elements in the OPE has been presented in Sec. \ref{moments}.

\begingroup
\squeezetable
\begin{table}
\begin{tabular}{|c|l|}
\hline
parameter & starting value\tabularnewline
\hline
$f_{2}$ & 0.\tabularnewline
\hline
$\mu_{6}$ & 0.\tabularnewline
\hline
$\mu_{8}$ & 0.\tabularnewline
\hline
$g_{a}$ & 1.267$\pm$0.035\tabularnewline
\hline
$a_{8}$ & 0.579$\pm$0.025\tabularnewline
\hline
$\Delta\Sigma$ & 0.154$\pm$0.2\tabularnewline
\hline
$a_{2}(Q_{0}^{2})$ & 0.0281$\pm$0.0028\tabularnewline
\hline
$d_{2}(Q_{0}^{2})$ & 0.0041$\pm$0.0011\tabularnewline
\hline
$\Lambda_{QCD}$ & 0.340$\pm$0.008\tabularnewline
\hline
\end{tabular}\caption{The nine
parameters used in the fits, together with their  starting values.
Free parameters started at zero, whereas the fixed parameters (given with uncertainties) 
were varied from their central values to estimate uncertainties
in the free parameters.
}
\label{table:9}
\end{table}
\endgroup

The data set analyzed comprised all the energies used for the EG1b
analysis and the doubly polarized data from other JLab experiments
(EG1a \cite{Fatemi:2003yh}  and EG1-dvcs  \cite{Prok:2014ltt}) as well as the data
from the SLAC, CERN and DESY facilities, including the recent COMPASS
results \cite{Adolph:2015saz}. 
The low-$x$ extrapolation of world
data was redone using our model (see Sec. \ref{models}) to
obtain a consistent set of data. The model was used down to $x=0.001$.
The uncertainty was estimated by varying the model parameters and taking
the quadratic sum of the resulting differences. Beyond $x=0.001$
a Regge form \cite{Bass:1997fh} was used for which an uncertainty of
100\% was assumed. The elastic contribution to the moments was estimated
using the proton form factor parametrization of Arrington \emph{et
al.} \cite{Arrington:2007ux}. The uncertainty was taken as the linear difference
with another fit from Gayou \emph{et al.} \cite{Gayou:2001qd}. In the
fitting procedure used to extract the higher-twist coefficients, all
the uncertainties (experimental statistics and systematics, elastic
and low-$x$ extrapolation) are added in quadrature to obtain a total
uncertainty. There are point-to-point correlations between the total
uncertainties on different data points within individual experiments. They
are also present between data points from different experiments (for
example, the EG1-dvcs data are supplemented with a high-$x$ extrapolation
from a model significantly dependent on the EG1b data). To account
for these correlations in the fit procedure, we use the \emph{unbiased
estimate }procedure, i.e. the total uncertainties are uniformly scaled
so that the $\chi^{2}$ per degree of freedom (dof) of the fit is forced to 1. It turns out that
the global factor scaling the total uncertainties is close to 1 (see
the last column of Table \ref{table:HT}).


\begin{figure}[t]
\begin{centering}
\includegraphics[scale=0.48]{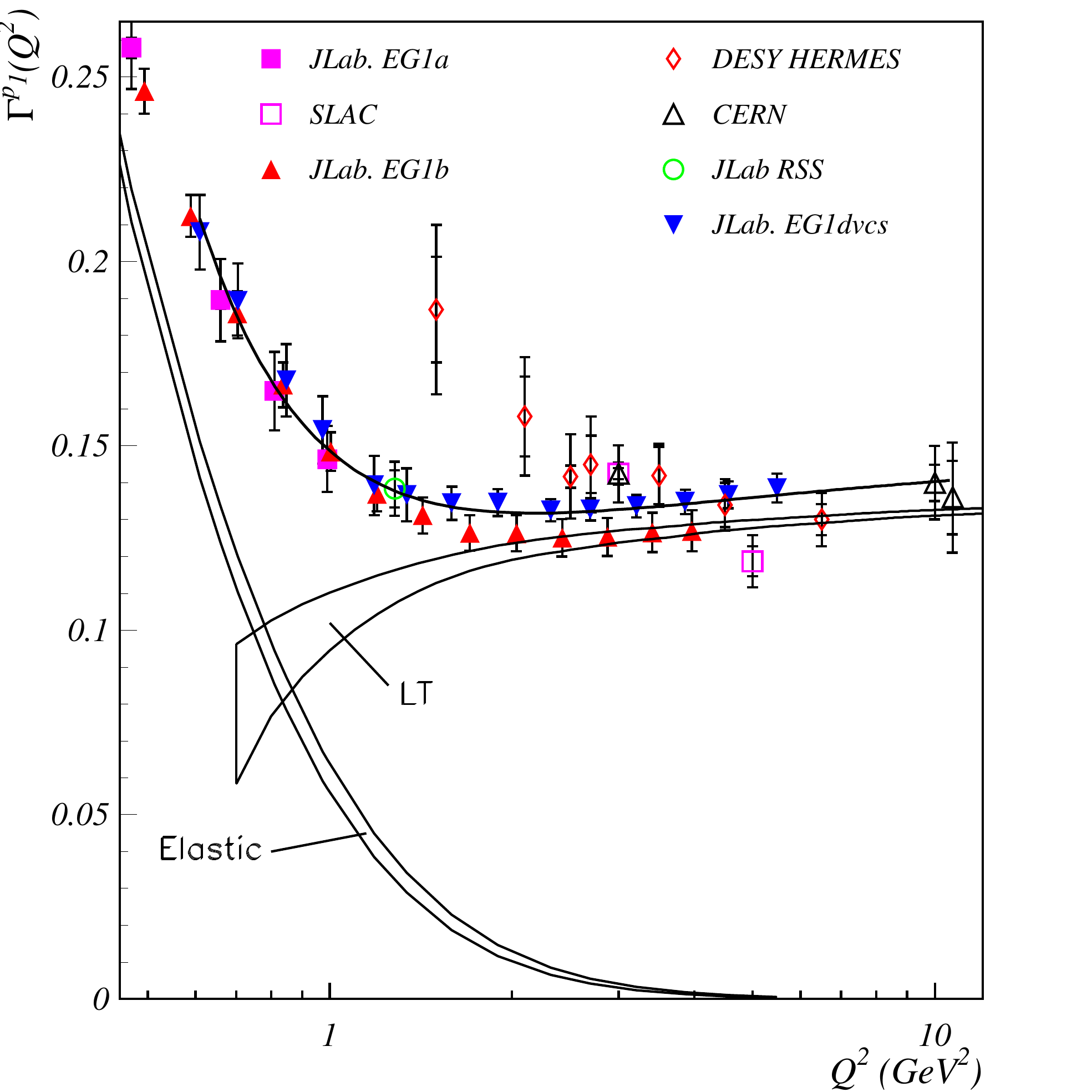}
\par\end{centering}

\caption{World data on $\Gamma_{1}^{p}(Q^{2})$. The band (LT) is the pQCD
leading-twist evolution. The error bars represent statistical (inner bars) and total (outer bars) uncertainties
after applying the unbiased estimate procedure. The solid black line
is a fit of the data starting at $Q_{min}^{2}=0.6$ GeV$^{2}$. The
band labeled {}``Elastic'' shows the elastic contribution to $\Gamma_{1}^{p}(Q^{2})$
with its uncertainty.\label{fig:gam1pel}}

\end{figure}


First, we fit the world data (re-estimated using our model) for $Q^{2}\geq5$
GeV$^{2}$ and assuming no higher-twist contribution above $Q^{2}=5$
GeV$^{2}$. This yields $\Delta\Sigma=0.169\pm0.084$. Next, we account
for higher twists. The target mass correction $a_{2}(Q_{o}^{2})=\int_{0}^{1}dx\left(x^{2}g_{1}^{LT}(x,Q_{0}^{2})\right)$,
in which $g_{1}^{LT}(x,Q_{0}^{2})$ contains only the twist-2 contribution
to $g_{1}$, was estimated with the parton distribution parametrization
of Bluemlein and Boettcher \cite{Bluemlein:2002be}. $Q_{0}^{2}$ is
a reference scale taken to be 5 GeV$^{2}$. The twist-3 contribution
$d_{2}(Q_{0}^{2})$ was obtained from the SLAC E155x experiment \cite{Anthony:2002hy}.
A $Q^{2}$-dependence of the form $A(Q^{2})=A(Q_{0}^{2})\left(\alpha_{s}(Q_{0}^{2})/\alpha_{s}(Q^{2})\right)^{b}$
was assumed for $a_{2}(Q^{2})$ and $d_{2}(Q^{2})$ with the anomalous
dimensions $b=-0.2$ and $b=-1$, respectively. A value of $\Lambda_{QCD}=0.340\pm0.008$ \cite{Agashe:2014kda}
was used for computing $\alpha_{s}(Q^{2})$. The variations
of the six quantities $g_{A}$, $a_{2}$, $d_{2}$, $A_{8}$, $\Sigma$
and $\Lambda_{QCD}$ during the $\chi^{2}$ minimization were
bounded within their respective error bars; see Table \ref{table:9} for the values
used and their bounds. Those, together with the (unbounded) fit parameters
$f_{2}$, $\mu_{6}$ and $\mu_{8}$, made a total of nine fit
parameters (three unbounded and six bounded). 
\\
\indent
The world data together with the OPE leading-twist evolution (LT)
of $\Gamma_{1}^{p}(Q^{2})$ and the elastic contribution to $\Gamma_{1}^{p}(Q^{2})$
are shown in Fig. \ref{fig:gam1pel}. The solid black
line is the result of fit 1 (see Table \ref{table:HT}).
\\
\indent
To check the convergence of the OPE series, the lowest $Q^{2}$ value,
$Q_{min}^{2}$, was varied, as well as the order of the OPE series
(truncated to twist-6 or twist-8). The results are given in Table \ref{table:HT}. 
\begingroup
\squeezetable
\begin{widetext}
\begin{center}
\begin{table}[t]
\begin{tabular}{|c|c|c|l|l|l|l|l|c|}
\hline 
{\footnotesize fit} & {\footnotesize $Q_{min}^{2}$ (GeV$^{2}$)} & $\mu_{max}$ & {\footnotesize $f_{2}$ } & {\footnotesize $\mu_{6}/M^{4}$} & {\footnotesize $\mu_{8}/M^{6}$} & $\Delta\Sigma$ & $\Lambda_{QCD}$ (GeV) & gf\tabularnewline
\hline
0 & 5.00 & 2 & {\footnotesize -} & {\footnotesize -} & {\footnotesize -} & {\footnotesize 0.169$\pm$0.084 } & {\footnotesize 0.340 (kept fixed) } & 1.40\tabularnewline
\hline 
{\footnotesize 1} & 0.61 & 8 & {\footnotesize -0.087$\pm$0.074 } & {\footnotesize 0.067$\pm$0.055 } & {\footnotesize 0.003$\pm$0.026 } & {\footnotesize 0.283$\pm$0.051 } & {\footnotesize 0.347$\pm$0.015 } & 1.08\tabularnewline
\hline 
2 & 0.61 & 6 & {\footnotesize -0.102$\pm$0.025 } & {\footnotesize 0.072$\pm$0.009 } & - & {\footnotesize 0.335$\pm$0.026 } & {\footnotesize 0.339$\pm$0.013 } & 1.06\tabularnewline
\hline 
3 & 0.81 & 8 & {\footnotesize -0.027$\pm$0.017 } & {\footnotesize 0.000$\pm$0.007 } & {\footnotesize 0.046$\pm$0.012 } & {\footnotesize 0.256$\pm$0.030 } & {\footnotesize 0.336$\pm$0.005 } & 1.11\tabularnewline
\hline 
4 & 0.81 & 6 & {\footnotesize -0.108$\pm$0.038} & {\footnotesize 0.076$\pm$0.016 } & - & {\footnotesize 0.286$\pm$0.035 } & {\footnotesize 0.332$\pm$0.011 } & 1.09\tabularnewline
\hline 
5 & 1.00 & 8 & {\footnotesize -0.018$\pm$0.018 } & {\footnotesize -0.009$\pm$0.013 } & {\footnotesize 0.050$\pm$0.021 } & {\footnotesize 0.261$\pm$0.035 } & {\footnotesize 0.332$\pm$0.009 } & 1.22\tabularnewline
\hline 
6 & 1.00 & 6 & {\footnotesize -0.076$\pm$0.066 } & {\footnotesize 0.060$\pm$0.031 } & - & {\footnotesize 0.274$\pm$0.060 } & {\footnotesize 0.336$\pm$0.004 } & 1.21\tabularnewline
\hline
\end{tabular}

\caption{Results of the fits for various minimal $Q^{2}$ values (column 2)
and truncations of the twist series. Data at $Q^{2}$ lower than $Q_{min}^{2}$
were not included in the fit. In column 3, $\mu_{max}$ indicates
the order at which the twist series is truncated ($\mu_{8}$ or $\mu_{6}$).
Column 4 gives the pure twist-4 coefficient, columns 5 and 6 give the
$1/Q^{4}$ and $1/Q^{6}$ power correction coefficients, respectively.
Column 7 gives the quark spin contribution to the nucleon
spin, $\Delta\Sigma$. Column 8 lists $\Lambda_{QCD}$,
and column 9 gives the global factor used to scale the total uncertainties
in order to force $\chi^{2}/ndf=1$. 
\label{table:HT}}

\end{table}
\end{center}
\end{widetext}

\endgroup
For a given higher-twist truncation order, the fit results are consistent
with each other (see Table~\ref{table:HT}), indicating that
the $Q_{\rm min}^{2}$ choice has an acceptably small influence. On the
other hand, the results are not consistent for fits with different
higher-twist truncation orders. This is to be expected since generally,
$\mu_{8}>\mu_{6}$. This is seen too in the higher-twist analysis
of the non-singlet part of $\Gamma_{1}$, the Bjorken sum \cite{Deur:2014vea}. 
\\
\indent
The $f_{2}$ results show the same trend as the results from the neutron
\cite{Meziani:2004ne} and Bjorken sum analysis \cite{Deur:2014vea}:
The $f_{2}$ coefficient tends to display a sign opposite to the sign
of the next significant higher twist coefficient. This may explain why the approach 
towards hadron-parton duality \cite{Melnitchouk:2005zr} at fairly moderate $Q^2$ holds for
$g_{1}$ at the scale at which the higher twist coefficients are extracted (see Sec. \ref{qhduality}). 
\\
\indent
The quark spin sum obtained at lower $Q^{2}$, accounting for
higher twists, is $\Delta\Sigma=0.289\pm0.014$, obtained from an average of our results. 
This is larger
than, but compatible with, the leading-twist determination $\Delta\Sigma=0.169\pm0.084$.
It also agrees with the determinations obtained from global fits of
PDFs, which are typically around $\Delta\Sigma=0.24$ (see, e.g., Ref.~\cite{Aidala:2012mv}
for a review).  The discrepancy between the $\Delta\Sigma$
extracted from the proton and neutron analyses \cite{Deur:2005jt,Chen:2005tda} 
(with $\Delta\Sigma^{(n)}=0.35\pm0.08)$ is resolved by the new data. 
\\
\indent
Our results on $f_{2}$ can be compared to non-perturbative model
predictions: $f_{2}=-0.037\pm0.006$ \cite{Stein:1995si}, $\mu_{4}/M^{2}=-0.040\pm0.023$
(QCD sum rules \cite{Balitsky:1989jb}), $f_{2}=-0.10\pm0.05$ (MIT bag
model \cite{Ji:1995qe}), and $f_{2}=-0.046$ (instanton model
\cite{Lee:2001ug}). As for the extracted $f_{2}$, all the
predictions are negative.\emph{ }The MIT bag model and QCD sum rules
agree best with the typical fit result of $f_{2}\simeq -0.1$, although
the other predictions are not ruled out. 
\\
\indent
From the result of fit 6, we extract the proton color polarizabilities
which are the responses of the color magnetic and electric fields
to the spin of the proton \cite{Stein:1995si,Ji:1995qe}.
We obtain $\chi_{E}^{p}=-0.045\pm0.044$  and  $\chi_{B}^{p}=0.031\pm0.022$ [see Eq. (\ref{colorpol})].
As is the case for for the neutron \cite{Meziani:2004ne} and $p$-$n$ \cite{Deur:2004ti, Deur:2014vea},
the extracted electric and magnetic polarizabilities are of opposite sign.


\subsubsection{Spin polarizability $\gamma_0^p$}

\begin{figure}
\centering
\includegraphics[width=9.5cm]{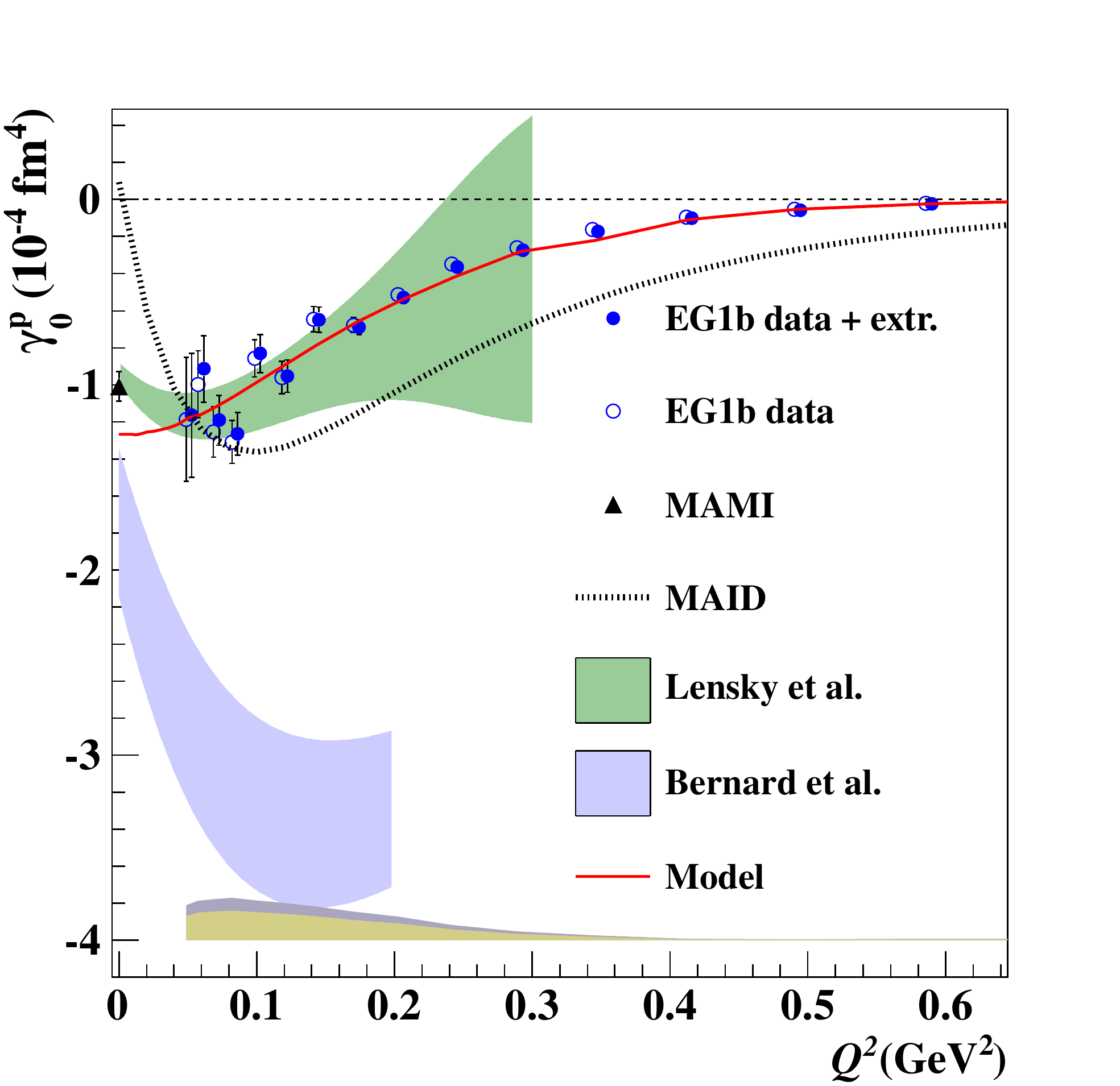}
\caption[Results for the forward spin polarizability integral vs\ $Q^2$]
{(Color online) The forward spin polarizability $\gamma_0^p$ vs\ $Q^2$. Open and closed 
circles represent the contribution to the integral from the data only and the data plus model, 
respectively (slightly offset horizontally for clarity). Our model is shown as a solid red line. 
Our results are compared to $\chi$PT calculations (as in Fig.~\ref{Gamma1:fig}), the MAID parametrization for 
single-pion production, and real photon data at $Q^{2}=0$ 
from MAMI~\cite{Ahrens:2001qt,Dutz:2003mm, Hildebrandt:2003fm}.
}
\label{FSPintegral:fig}
\end{figure}

In the real photon limit $Q^2 \rightarrow 0$, the  $ep$
scattering cross section can be expressed in terms of 
Compton amplitudes, with coefficients
$\alpha_E$, $\beta_M$, and $\gamma_0^p$, called polarizabilities. The
quantity $\gamma_0^p$, the forward spin polarizability, is given by 
\begin{equation} 
\gamma_0^p=\frac{1} {4\pi}\int_{\nu_{th}}^\infty 
\frac{\sigma_{\frac{3}{2}}-\sigma_{\frac{1}{2}}}
{\nu} \,d\nu  .
\end{equation}
Converting the integration variable from $\nu$ to $x$ yields Eq.~(\ref{gamma0}), which
can be recast as 
\begin{eqnarray}
\gamma_0^p =&\frac{16 M^2\alpha}{Q^6} \int_0^{x_{\rm th}}x^2\left[g_1^p(x,Q^2)-\gamma^2 g_2^p(x,Q^2)\right] \,dx\\ \nonumber
=&\frac{16 M^2\alpha}{Q^6}\int_0^{x_{\rm th}} x^2A_1^p(x,Q^2)F_1^p(x,Q^2)\,dx,
\end{eqnarray}
in which $x_{\rm th}$, the pion production threshold, excludes the elastic contribution.
The polarizability in units of fm$^{-4}$ is plotted 
in Fig.~{\ref{FSPintegral:fig}} (blue open circles, measured data; blue dots, extrapolated data),
along with the real photon $\gamma_0^p$ ($Q^2=0$) obtained from the MAMI GDH experiment 
\cite{Ahrens:2001qt,Dutz:2003mm, Hildebrandt:2003fm}:
\begin{equation}
\gamma_0^p=[-1.01\pm 0.08\pm 0.10]\times 10^{-4}\,{\rm fm}^{-4}.  
\end{equation}
Within experimental uncertainties, our measurements
at low $Q^2$ are consistent with the MAMI measurement.

\subsection{Bloom-Gilman duality}\label{qhduality}

As discussed in Sec. \ref{bloom},
 our data provide a substantial  test of Bloom-Gilman duality in polarized 
electron scattering. Comparisons of theory
and experiment have shown that unpolarized structure
functions exhibit both a ``global duality'' (integration over the
entire resonance region at $W<2$ GeV) and a ``local duality'' in each of the
three main resonance regions. For polarized scattering at low Q$^2$,
the importance of the hadronic picture is clearly shown by the
observed values of $g_1^p$ in the resonance region, where the interplay
of $\sigma_{\frac{1}{2}}$ and $\sigma_{\frac{3}{2}}$ is obvious. The $\Delta$ region, 
where $g_1^p<0$, is an extreme case, since for DIS in the scaling region
$g_1^p>0$ for all $x$.
It may still be possible, however, for global duality to apply in the
resonance region at relatively low $Q^2$. 

Hence, we looked for evidence
of local and global duality for
$0.5< Q^2<5$ GeV$^2$ by applying duality tests to
determine at what values of ($Q^2, W$) the DIS behavior represents the
average polarization response in the resonance region.  
A first study of duality for spin structure functions
using the CLAS data for both polarized
proton and deuteron targets was carried out
and reported in an earlier publication \cite{Bosted:2006gp}. 

For comparison with our data above $Q^2 = 1$ GeV$^2$, QCD fits to DIS polarized
structure function data above the resonance region were evolved
towards lower $Q^2$ by an NLO calculation. This 
evolution
is expected to give reasonable results down to $Q^2 \approx 1$~GeV$^2$. 
The NLO evolution was chosen to give the best estimate
of the $Q^2$ dependence of $g_1^p$. Target mass effects were
taken into account using the prescription of Bl\"umlein and 
Tkabladze~\cite{Blumlein:1998nv} as before.
 Recent fits to the unpolarized structure functions $F_1$ for the proton
and deuteron were used to extract $g_1$ for both the proton and the deuteron
from our data for E=1.6 GeV and 5.7 GeV.

\begin{figure}[htb!]
\centering
\includegraphics[width=9.6cm]{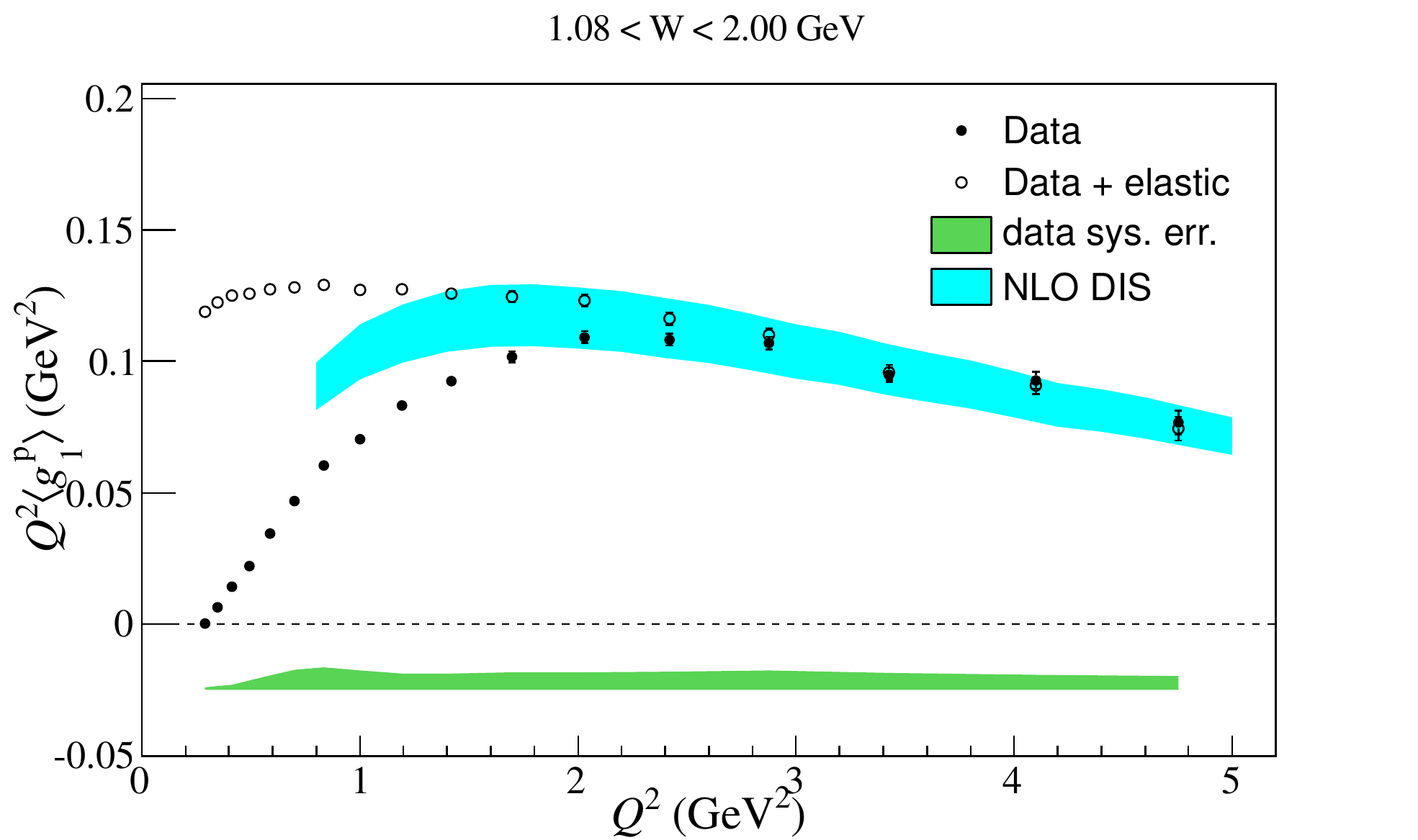}
\caption[Global duality of $g_1^p$]
{(Color online) The $Q^2$ dependence of $Q^2g_1^p(x,Q^2)$ averaged over a region in $x$
corresponding to $1.08<W<2$ GeV (solid circles) for the proton, with the green band showing systematic uncertainties. The open
circles represent the data after adding the contribution from $ep$ elastic
scattering. The shaded cyan band represents the range of the averages calculated
from extrapolated NLO DIS fits.}
\label{dualitytest2:fig}
\end{figure}

To test both local and global duality, the data for $g_1^p$ were averaged
over $x$ in four $Q^2$-dependent intervals corresponding to four regions
in $W<2$ GeV, with boundaries at 1.08, 1.38, 1.58, 1.82 and 2.00 GeV
(corresponding to the three prominent ``resonance bumps'' and the region
of high-mass resonances observed in our data).
Global duality was tested by a single average over $x$ in this entire range
in $W$.
\\
\indent
The results for the global duality test are shown in
Fig.~{\ref{dualitytest2:fig}}. In this plot we also show the effect
of including elastic scattering, following a suggestion of Close and Isgur
\cite{Close:2001ha}
that including elastic scattering may improve the agreement between
the data and the DIS extrapolation.  
The averaged resonance data agree quite well
with the extrapolated DIS data above $Q^2 \approx 2$  GeV$^2$ (without the 
elastic contribution), suggesting a possible onset of global duality. For $Q^2<2$ GeV$^2$, however,
the data lie significantly above the DIS extrapolation without the elastic contribution and  significantly
below the DIS extrapolation with the elastic contribution.  

\begin{figure}
\centering
\includegraphics[width=9.5cm]{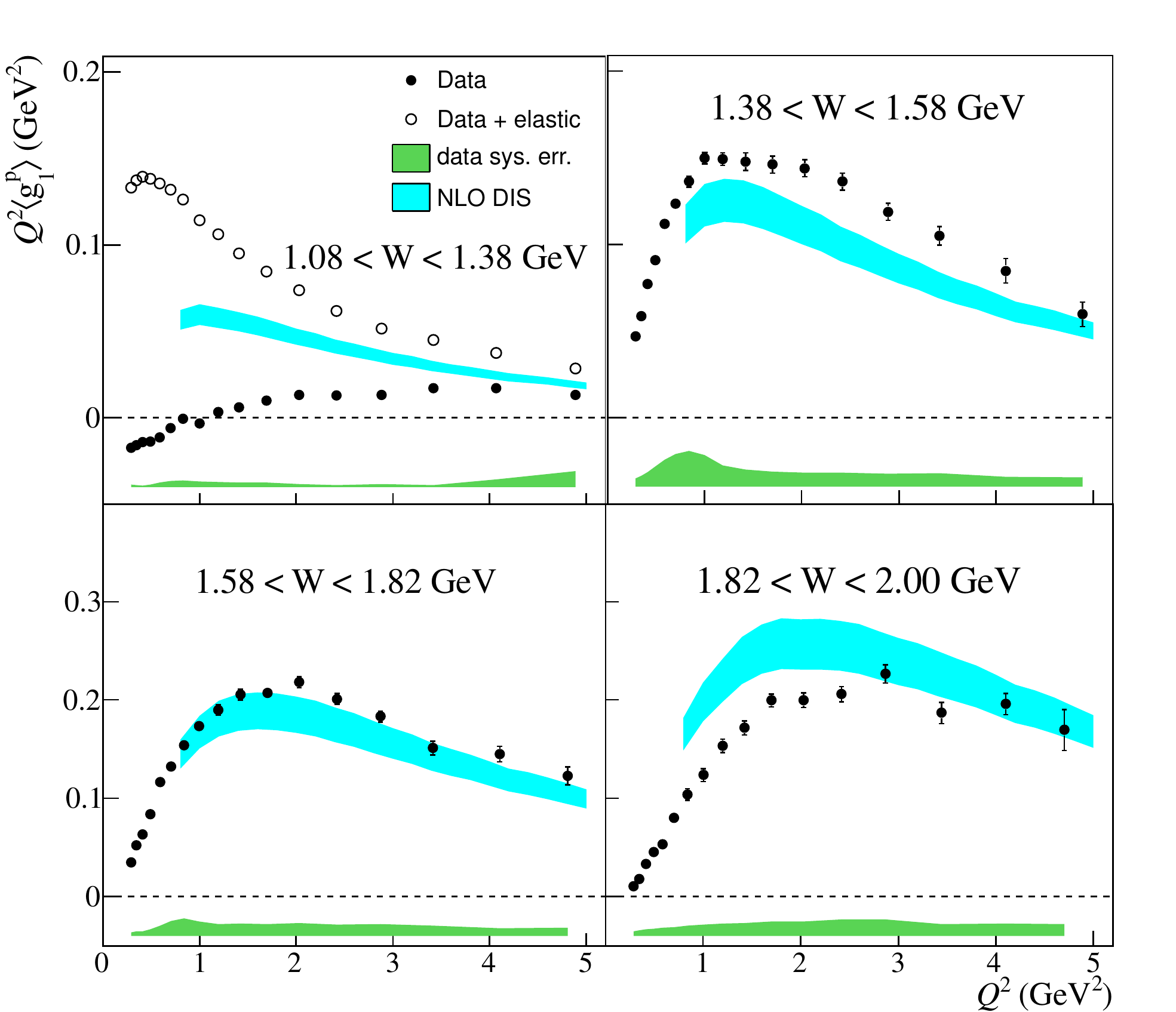}
\caption[Local duality of $g_1^p$]
{(Color online) Averages of $Q^2g_1^p(x,Q^2)$ vs\ $Q^2$ over limited spans in $x$ corresponding
to prominent ``resonance regions'' as indicated by the ranges in $W$. Symbols are
the same as in Fig.~\ref{dualitytest2:fig}.}
\label{duality3:fig}
\end{figure}

Figure~{\ref{duality3:fig}} shows the results of the local duality tests
for the proton, averaged over $x$, for
four $W$ regions, plotted as a function of $Q^2$. 
At low $Q^2$, the data in the first resonance region lie substantially above (below) the 
NLO curves without (with) the elastic contribution, and the deviation behaves like a power law. 
Above $Q^2= 3$ GeV$^2$, the data begin to converge with the NLO curves.
The data in the second region lie well above the NLO curve. 
The data in the third resonance region appear in good agreement with the DIS extrapolation. 
The data in the fourth resonance region lie slightly below the NLO curve. 
The various local regions seem to compensate each other to yield global duality. 
However, the approach towards duality is much slower
for $g_{1p}$ than in the unpolarized case.



\section{CONCLUSIONS}\label{s6}

We have presented the final analysis of the most extensive and precise data set on 
the spin structure functions $A_1^p$ and $g_1^p$ of the proton
collected at Jefferson Laboratory so far. The data cover nearly two orders of magnitude in 
squared momentum transfer,  $0.05 \leq Q^2 \leq 5$ GeV$^2$, which encompasses the transition from
the region where hadronic degrees of freedom and effective theories like $\chi$PT near the photon point
are relevant to the regime where pQCD is applicable. 
At lower $W < 2$ GeV, our data give more detailed insight in the
inclusive response of the proton in the resonance region and how, on average, this connects
with the DIS limit (quark-hadron duality). Duality applies both to individual
resonances [except the $\Delta(1232)$], and to the resonance region as a whole 
(1 GeV $< W <$ 2 GeV) above $Q^2 \approx 2$ GeV$^2$.
At higher $W$, 2 GeV  $< W < 3$ GeV, and $Q^2 > 1$ GeV$^2$, our data can constrain NLO fits 
(including higher twist corrections) of 
spin structure functions.  This improves the knowledge of polarized PDFs and sheds new light on the 
valence quark structure of the nucleon at large $x$. 
 
Our data also allow a very precise determination of moments of $g_{1}^p$, which can
be used to test the GDH sum rule limit, compare to $\chi$PT calculations, 
and extract higher-twist contributions and nucleon polarizabilities.  
We find that some $\chi$PT are commensurate with our results for $\gamma_0^p$ at low $Q^2$ 
and that the model by Lensky {\etal}  \cite{Lensky:2014dda} agrees 
with the values obtained for the polarizability $\gamma_0^p$ at and near the photon point.

Our OPE analysis extracted the twist-4 contribution $f_{2}^{p}$ to the first moment of the spin
structure function $g_{1}^{p}$. It is found
to be negative and the sign of the significant twist coefficients
($\mu_{2}$, $\mu_{4}$, $\mu_{6}$, or $\mu_{8}$) appears to alternate.
This sign alternation is important to understand quark-hadron duality
or early scaling seen at relatively low $Q^{2}$. The color polarizabilities
extracted from the higher twist analysis are small.  
The quark spin contribution to the nucleon spin has been extracted in the same
process and found to be $\Delta\Sigma=0.289\pm0.014$. The discrepancy
previously seen between the $\Delta\Sigma$ extracted from the proton
or neutron analyses is resolved by the new data.

Additional data from this experiment on the deuteron with similar
precision have already been published \cite{Guler:2015hsw}. Further information will come
from the analysis of the completed EG4 experiment with CLAS, which extends the kinematic 
coverage of the present data set to even lower $Q^2$ for a more rigorous test of $\chi$PT.
At the highest values of $Q^2$, spin structure function data from the EG1-dvcs experiment \cite{Prok:2014ltt}
have improved our knowledge of $A_1^p$ at large $x$ and further reduced the uncertainty
with which $g_1^p$ is known in the DIS region. Finally, additional information on
the structure functions $g_2^p$ and $A_2^p$ is forthcoming from ``SANE'' in Hall C \cite{KANG:2013lva}
and ``g2p'' in Hall A \cite{Slifer:2009ik}. Extending EG1b to 11 GeV has been approved and will run in the 
coming years using CLAS12 at Jefferson Laboratory.

\section*{Acknowledgments}
We would like to acknowledge the outstanding efforts of the staff
of the Accelerator and the Physics Divisions at Jefferson Lab that made
this experiment possible.  This work was supported in part by
the U.S. Department of Energy and the National
Science Foundation,
the Italian Istituto Nazionale di Fisica Nucleare, the French Centre
National de la Recherche Scientifique, the French Commissariat \`{a}
l'Energie Atomique,  the Emmy Noether grant from the Deutsche Forschungs
Gemeinschaft, the United Kingdom's Science and Technology Facilities Council, and the National Research Foundation of Korea.  
The Jefferson Science Associates (JSA) operates the Thomas Jefferson National Accelerator 
Facility for the United States Department of Energy under Contract No. DE-AC05-06OR23177.

\bibliographystyle{apsrev}

\begin{thebibliography}{145}
\expandafter\ifx\csname natexlab\endcsname\relax\def\natexlab#1{#1}\fi
\expandafter\ifx\csname bibnamefont\endcsname\relax
  \def\bibnamefont#1{#1}\fi
\expandafter\ifx\csname bibfnamefont\endcsname\relax
  \def\bibfnamefont#1{#1}\fi
\expandafter\ifx\csname citenamefont\endcsname\relax
  \def\citenamefont#1{#1}\fi
\expandafter\ifx\csname url\endcsname\relax
  \def\url#1{\texttt{\#1\}}\fi
\expandafter\ifx\csname urlprefix\endcsname\relax\def\urlprefix{}\fi
\providecommand{\bibinfo}[2]{#2}
\providecommand{\eprint}[2][]{\url{#2}}

 \bibitem[{\citenamefont{Ashman \textit{et~al}.}(1988)}]{Ashman:1987hv}
 
\bibinfo{author}{\bibfnamefont{J.}~\bibnamefont{Ashman}} \bibnamefont{\textit{et~al}.}
(\bibinfo{collaboration}{European Muon Collaboration}),
  \bibinfo{journal}{Phys. Lett.} \textbf{\bibinfo{volume}{B206}},
  \bibinfo{pages}{364} (\bibinfo{year}{1988}).

\bibitem[{\citenamefont{Kuhn \textit{et~al}.}(2009)\citenamefont{Kuhn, Chen, and
  Leader}}]{Kuhn:2008sy}
\bibinfo{author}{\bibfnamefont{S.}~\bibnamefont{Kuhn}},
  \bibinfo{author}{\bibfnamefont{J.-P.} \bibnamefont{Chen}}, \bibnamefont{and}
  \bibinfo{author}{\bibfnamefont{E.}~\bibnamefont{Leader}},
  \bibinfo{journal}{Prog. Part. Nucl. Phys.} \textbf{\bibinfo{volume}{63}},
  \bibinfo{pages}{1} (\bibinfo{year}{2009}).

\bibitem[{\citenamefont{Aidala \textit{et~al}.}(2013)\citenamefont{Aidala, Bass, Hasch,
  and Mallot}}]{Aidala:2012mv}
\bibinfo{author}{\bibfnamefont{C.~A.} \bibnamefont{Aidala}},
  \bibinfo{author}{\bibfnamefont{S.~D.} \bibnamefont{Bass}},
  \bibinfo{author}{\bibfnamefont{D.}~\bibnamefont{Hasch}}, \bibnamefont{and}
  \bibinfo{author}{\bibfnamefont{G.~K.} \bibnamefont{Mallot}},
  \bibinfo{journal}{Rev. Mod. Phys.} \textbf{\bibinfo{volume}{85}},
  \bibinfo{pages}{655} (\bibinfo{year}{2013}).

\bibitem[{\citenamefont{Sato \textit{et~al}.}(2016)\citenamefont{Sato, Melnitchouk,
  Kuhn, Ethier, and Accardi}}]{Sato:2016tuz}
  \bibinfo{author}{\bibfnamefont{N.}~\bibnamefont{Sato}},
  \bibinfo{author}{\bibfnamefont{W.}~\bibnamefont{Melnitchouk}},
  \bibinfo{author}{\bibfnamefont{S.~E.} \bibnamefont{Kuhn}},
  \bibinfo{author}{\bibfnamefont{J.~J.} \bibnamefont{Ethier}},
  \bibnamefont{and} \bibinfo{author}{\bibfnamefont{A.}~\bibnamefont{Accardi}}
  (\bibinfo{collaboration}{Jefferson Lab Angular Momentum Collaboration}), 
  \bibinfo{journal}{Phys. Rev.} \textbf{\bibinfo{volume}{D93}},
  \bibinfo{pages}{074005} (\bibinfo{year}{2016}).

\bibitem[{\citenamefont{Dokshitzer}(1977)}]{Dokshitzer}
\bibinfo{author}{\bibfnamefont{Y.~L.} \bibnamefont{Dokshitzer}},
  \bibinfo{journal}{Sov. Phys. JETP} \textbf{\bibinfo{volume}{46}},
  \bibinfo{pages}{641} (\bibinfo{year}{1977}).

\bibitem[{\citenamefont{Gribov and Lipatov}(1972)}]{GribovLipatov}
\bibinfo{author}{\bibfnamefont{V.~N.} \bibnamefont{Gribov}} \bibnamefont{and}
  \bibinfo{author}{\bibfnamefont{L.~N.} \bibnamefont{Lipatov}},
  \bibinfo{journal}{Yad. Fiz.} \textbf{\bibinfo{volume}{15}},
  \bibinfo{pages}{781} (\bibinfo{year}{1972}).

\bibitem[{\citenamefont{Altarelli and Parisi}(1977)}]{Altarelli}
\bibinfo{author}{\bibfnamefont{G.}~\bibnamefont{Altarelli}} \bibnamefont{and}
  \bibinfo{author}{\bibfnamefont{G.}~\bibnamefont{Parisi}},
  \bibinfo{journal}{Nucl. Phys.} \textbf{\bibinfo{volume}{B126}},
  \bibinfo{pages}{298} (\bibinfo{year}{1977}).

\bibitem[{\citenamefont{Shuryak and Vainshtein}(1982{\natexlab{a}})}]{OPE1}
\bibinfo{author}{\bibfnamefont{E.~V.} \bibnamefont{Shuryak}} \bibnamefont{and}
  \bibinfo{author}{\bibfnamefont{A.~I.} \bibnamefont{Vainshtein}},
  \bibinfo{journal}{Nucl. Phys.} \textbf{\bibinfo{volume}{B201}},
  \bibinfo{pages}{141} (\bibinfo{year}{1982}{\natexlab{a}}).

\bibitem[{\citenamefont{Shuryak and Vainshtein}(1982{\natexlab{b}})}]{OPE2}
\bibinfo{author}{\bibfnamefont{E.~V.} \bibnamefont{Shuryak}} \bibnamefont{and}
  \bibinfo{author}{\bibfnamefont{A.~I.} \bibnamefont{Vainshtein}},
  \bibinfo{journal}{Nucl. Phys.} \textbf{\bibinfo{volume}{B199}},
  \bibinfo{pages}{451} (\bibinfo{year}{1982}{\natexlab{b}}).

\bibitem[{\citenamefont{Ehrnsperger \textit{et~al}.}(1994)\citenamefont{Ehrnsperger,
  Schafer, and Mankiewicz}}]{SSFope}
\bibinfo{author}{\bibfnamefont{B.}~\bibnamefont{Ehrnsperger}},
  \bibinfo{author}{\bibfnamefont{A.}~\bibnamefont{Schafer}}, \bibnamefont{and}
  \bibinfo{author}{\bibfnamefont{L.}~\bibnamefont{Mankiewicz}},
  \bibinfo{journal}{Phys. Lett.} \textbf{\bibinfo{volume}{B323}},
  \bibinfo{pages}{439} (\bibinfo{year}{1994}).

\bibitem[{\citenamefont{Mecking \textit{et~al}.}(2003)}]{Mecking:2003zu}

\bibinfo{author}{\bibfnamefont{B.}~\bibnamefont{Mecking}} \bibnamefont{\textit{et~al}.}
(\bibinfo{collaboration}{CLAS Collaboration}),
   \bibinfo{journal}{Nucl.
  Instrum. Meth.} \textbf{\bibinfo{volume}{A503}}, \bibinfo{pages}{513}
  (\bibinfo{year}{2003}).

\bibitem[{\citenamefont{Fatemi \textit{et~al}.}(2003)}]{Fatemi:2003yh}

\bibinfo{author}{\bibfnamefont{R.}~\bibnamefont{Fatemi}} \bibnamefont{\textit{et~al}.}
 (\bibinfo{collaboration}{CLAS Collaboration}),
  \bibinfo{journal}{Phys. Rev.
  Lett.} \textbf{\bibinfo{volume}{91}}, \bibinfo{pages}{222002}
  (\bibinfo{year}{2003}).

\bibitem[{\citenamefont{Yun \textit{et~al}.}(2003)}]{Yun:2002td}

\bibinfo{author}{\bibfnamefont{J.}~\bibnamefont{Yun}} \bibnamefont{\textit{et~al}.}
(\bibinfo{collaboration}{CLAS Collaboration}),
   \bibinfo{journal}{Phys. Rev.}
  \textbf{\bibinfo{volume}{C67}}, \bibinfo{pages}{055204}
  (\bibinfo{year}{2003}).

\bibitem[{\citenamefont{Guler \textit{et~al}.}(2015)}]{Guler:2015hsw}

\bibinfo{author}{\bibfnamefont{N.}~\bibnamefont{Guler}} \bibnamefont{\textit{et~al}.}
(\bibinfo{collaboration}{CLAS Collaboration}),
   \bibinfo{journal}{Phys. Rev.}
  \textbf{\bibinfo{volume}{C92}}, \bibinfo{pages}{055201}
  (\bibinfo{year}{2015}).

\bibitem[{\citenamefont{Dharmawardane \textit{et~al}.}(2006)}]{Dharmawardane:2006zd}

\bibinfo{author}{\bibfnamefont{K.}~\bibnamefont{Dharmawardane}}
  \bibnamefont{\textit{et~al}.}
  (\bibinfo{collaboration}{CLAS Collaboration}),
  \bibinfo{journal}{Phys. Lett.} \textbf{\bibinfo{volume}{B641}},
  \bibinfo{pages}{11} (\bibinfo{year}{2006}).

\bibitem[{\citenamefont{Bosted \textit{et~al}.}(2007)}]{Bosted:2006gp}

\bibinfo{author}{\bibfnamefont{P.}~\bibnamefont{Bosted}} \bibnamefont{\textit{et~al}.}
(\bibinfo{collaboration}{CLAS Collaboration}),
   \bibinfo{journal}{Phys. Rev.}
  \textbf{\bibinfo{volume}{C75}}, \bibinfo{pages}{035203}
  (\bibinfo{year}{2007}).

\bibitem[{\citenamefont{Prok \textit{et~al}.}(2009)}]{Prok:2008ev}

\bibinfo{author}{\bibfnamefont{Y.}~\bibnamefont{Prok}} \bibnamefont{\etal.}
(\bibinfo{collaboration}{CLAS Collaboration}),
  \bibinfo{journal}{Phys. Lett.}
  \textbf{\bibinfo{volume}{B672}}, \bibinfo{pages}{12} (\bibinfo{year}{2009}).

\bibitem[{\citenamefont{Baum \textit{et~al}.}(1980)\citenamefont{Baum, Bergstrom,
  Clendenin, Ehrlich, Hughes \textit{et~al}.}}]{Baum:1980mh}
\bibinfo{author}{\bibfnamefont{G.}~\bibnamefont{Baum}},
  \bibinfo{author}{\bibfnamefont{M.}~\bibnamefont{Bergstrom}},
  \bibinfo{author}{\bibfnamefont{J.}~\bibnamefont{Clendenin}},
  \bibinfo{author}{\bibfnamefont{R.}~\bibnamefont{Ehrlich}},
  \bibinfo{author}{\bibfnamefont{V.}~\bibnamefont{Hughes}},
  \bibnamefont{\textit{et~al}.}, \bibinfo{journal}{Phys. Rev. Lett.}
  \textbf{\bibinfo{volume}{45}}, \bibinfo{pages}{2000} (\bibinfo{year}{1980}).

\bibitem[{\citenamefont{Abe \textit{et~al}.}(1997{\natexlab{a}})}]{Abe:1996ag}

\bibinfo{author}{\bibfnamefont{K.}~\bibnamefont{Abe}} \bibnamefont{\etal.}
(\bibinfo{collaboration}{E143 Collaboration}),
  \bibinfo{journal}{Phys. Rev.
  Lett.} \textbf{\bibinfo{volume}{78}}, \bibinfo{pages}{815}
  (\bibinfo{year}{1997}{\natexlab{a}}).

\bibitem[{\citenamefont{Amarian \textit{et~al}.}(2004)}]{Amarian:2003jy}
 
\bibinfo{author}{\bibfnamefont{M.}~\bibnamefont{Amarian}} \bibnamefont{\textit{et~al}.},
  \bibinfo{journal}{Phys. Rev. Lett.} \textbf{\bibinfo{volume}{92}},
  (\bibinfo{collaboration}{Jefferson Lab E94-010 Collaboration}),
  \bibinfo{pages}{022301} (\bibinfo{year}{2004}).

\bibitem[{\citenamefont{Zheng \textit{et~al}.}(2004{\natexlab{a}})}]{Zheng:2004ce}

\bibinfo{author}{\bibfnamefont{X.}~\bibnamefont{Zheng}} \bibnamefont{\textit{et~al}.}
(\bibinfo{collaboration}{Jefferson Lab Hall A Collaboration}),
  \bibinfo{journal}{Phys. Rev.} \textbf{\bibinfo{volume}{C70}},
  \bibinfo{pages}{065207} (\bibinfo{year}{2004}{\natexlab{a}}).

\bibitem[{\citenamefont{Wesselmann \textit{et~al}.}(2007)}]{Wesselmann:2006mw}

\bibinfo{author}{\bibfnamefont{F.~R.} \bibnamefont{Wesselmann}}
  \bibnamefont{\textit{et~al}.}
 (\bibinfo{collaboration}{RSS Collaboration}), 
  \bibinfo{journal}{Phys. Rev. Lett.} \textbf{\bibinfo{volume}{98}},
  \bibinfo{pages}{132003} (\bibinfo{year}{2007}).

\bibitem[{\citenamefont{Prok \textit{et~al}.}(2014)}]{Prok:2014ltt}

\bibinfo{author}{\bibfnamefont{Y.}~\bibnamefont{Prok}} \bibnamefont{\textit{et~al}.}
(\bibinfo{collaboration}{CLAS Collaboration}),
   \bibinfo{journal}{Phys. Rev.}
  \textbf{\bibinfo{volume}{C90}}, \bibinfo{pages}{025212}
  (\bibinfo{year}{2014}).

\bibitem[{\citenamefont{Isgur}(1999)}]{Isgur:1998yb}
\bibinfo{author}{\bibfnamefont{N.}~\bibnamefont{Isgur}},
  \bibinfo{journal}{Phys. Rev.} \textbf{\bibinfo{volume}{D59}},
  \bibinfo{pages}{034013} (\bibinfo{year}{1999}).

\bibitem[{\citenamefont{Brodsky \textit{et~al}.}(1995)\citenamefont{Brodsky, Burkardt,
  and Schmidt}}]{Brodsky:1994kg}
\bibinfo{author}{\bibfnamefont{S.~J.} \bibnamefont{Brodsky}},
  \bibinfo{author}{\bibfnamefont{M.}~\bibnamefont{Burkardt}}, \bibnamefont{and}
  \bibinfo{author}{\bibfnamefont{I.}~\bibnamefont{Schmidt}},
  \bibinfo{journal}{Nucl. Phys.} \textbf{\bibinfo{volume}{B441}},
  \bibinfo{pages}{197} (\bibinfo{year}{1995}).

\bibitem[{\citenamefont{Farrar and Jackson}(1975)}]{Farrar:1975yb}
\bibinfo{author}{\bibfnamefont{G.~R.} \bibnamefont{Farrar}} \bibnamefont{and}
  \bibinfo{author}{\bibfnamefont{D.~R.} \bibnamefont{Jackson}},
  \bibinfo{journal}{Phys. Rev. Lett.} \textbf{\bibinfo{volume}{35}},
  \bibinfo{pages}{1416} (\bibinfo{year}{1975}).

\bibitem[{\citenamefont{Avakian \textit{et~al}.}(2007)\citenamefont{Avakian, Brodsky,
  Deur, and Yuan}}]{Avakian:2007xa}
\bibinfo{author}{\bibfnamefont{H.}~\bibnamefont{Avakian}},
  \bibinfo{author}{\bibfnamefont{S.~J.} \bibnamefont{Brodsky}},
  \bibinfo{author}{\bibfnamefont{A.}~\bibnamefont{Deur}}, \bibnamefont{and}
  \bibinfo{author}{\bibfnamefont{F.}~\bibnamefont{Yuan}},
  \bibinfo{journal}{Phys. Rev. Lett.} \textbf{\bibinfo{volume}{99}},
  \bibinfo{pages}{082001} (\bibinfo{year}{2007}).

\bibitem[{\citenamefont{Soffer and Teryaev}(1999)}]{Soffer:1999zv}
\bibinfo{author}{\bibfnamefont{J.}~\bibnamefont{Soffer}} \bibnamefont{and}
  \bibinfo{author}{\bibfnamefont{O.~V.} \bibnamefont{Teryaev}}, in
  \emph{\bibinfo{booktitle}{{Polarized protons at high-energies---accelerator
  challenges and physics opportunities. Proceedings, Workshop, Hamburg,
  Germany, May 17-20, 1999}}} (\bibinfo{year}{1999}), \eprint{hep-ph/9906455},
 \urlprefix\url{http://inspirehep.net/record/502194/files/teryaev_2.ps}.


\bibitem[{\citenamefont{Artru \textit{et~al}.}(2009)\citenamefont{Artru, Elchikh,
  Richard, Soffer, and Teryaev}}]{Artru:2008cp}
\bibinfo{author}{\bibfnamefont{X.}~\bibnamefont{Artru}},
  \bibinfo{author}{\bibfnamefont{M.}~\bibnamefont{Elchikh}},
  \bibinfo{author}{\bibfnamefont{J.-M.} \bibnamefont{Richard}},
  \bibinfo{author}{\bibfnamefont{J.}~\bibnamefont{Soffer}}, \bibnamefont{and}
  \bibinfo{author}{\bibfnamefont{O.~V.} \bibnamefont{Teryaev}},
  \bibinfo{journal}{Phys. Rept.} \textbf{\bibinfo{volume}{470}},
  \bibinfo{pages}{1} (\bibinfo{year}{2009}).

\bibitem[{\citenamefont{Ashman \textit{et~al}.}(1989)}]{Ashman:1989ig}
 
\bibinfo{author}{\bibfnamefont{J.}~\bibnamefont{Ashman}} \bibnamefont{\textit{et~al}.}
(\bibinfo{collaboration}{European Muon Collaboration}),
  \bibinfo{journal}{Nucl. Phys.} \textbf{\bibinfo{volume}{B328}},
  \bibinfo{pages}{1} (\bibinfo{year}{1989}).

\bibitem[{\citenamefont{Adeva \textit{et~al}.}(1998)}]{Adeva:1998vv}
 
\bibinfo{author}{\bibfnamefont{B.}~\bibnamefont{Adeva}} \bibnamefont{\textit{et~al}.}
(\bibinfo{collaboration}{Spin Muon Collaboration}), 
\bibinfo{journal}{Phys.
  Rev.} \textbf{\bibinfo{volume}{D58}}, \bibinfo{pages}{112001}
  (\bibinfo{year}{1998}).

\bibitem[{\citenamefont{Ageev \textit{et~al}.}(2007)}]{Ageev:2007du}

\bibinfo{author}{\bibfnamefont{E.}~\bibnamefont{Ageev}} \bibnamefont{\textit{et~al}.}
(\bibinfo{collaboration}{COMPASS Collaboration}), 
\bibinfo{journal}{Phys. Lett.} \textbf{\bibinfo{volume}{B647}}, \bibinfo{pages}{330}
  (\bibinfo{year}{2007}).

\bibitem[{\citenamefont{Alexakhin \textit{et~al}.}(2007)}]{Alexakhin:2006vx}

\bibinfo{author}{\bibfnamefont{V.}~\bibnamefont{Alexakhin}}
  \bibnamefont{\textit{et~al}.}
  (\bibinfo{collaboration}{COMPASS Collaboration}),
  \bibinfo{journal}{Phys. Lett.} \textbf{\bibinfo{volume}{B647}},
  \bibinfo{pages}{8} (\bibinfo{year}{2007}).

\bibitem[{\citenamefont{Ackerstaff \textit{et~al}.}(1999)}]{Ackerstaff:1999ey}

\bibinfo{author}{\bibfnamefont{K.}~\bibnamefont{Ackerstaff}}
  \bibnamefont{\textit{et~al}.}
  (\bibinfo{collaboration}{HERMES Collaboration}),
  \bibinfo{journal}{Phys. Lett.} \textbf{\bibinfo{volume}{B464}},
  \bibinfo{pages}{123} (\bibinfo{year}{1999}).

\bibitem[{\citenamefont{Airapetian \textit{et~al}.}(2007)}]{Airapetian:2006vy}

\bibinfo{author}{\bibfnamefont{A.}~\bibnamefont{Airapetian}}
  \bibnamefont{\textit{et~al}.}
  (\bibinfo{collaboration}{HERMES Collaboration}),
  \bibinfo{journal}{Phys. Rev.} \textbf{\bibinfo{volume}{D75}},
  \bibinfo{pages}{012007} (\bibinfo{year}{2007}).

\bibitem[{\citenamefont{Anthony \textit{et~al}.}(1996)}]{Anthony:1996mw}
 
\bibinfo{author}{\bibfnamefont{P.}~\bibnamefont{Anthony}} \bibnamefont{\textit{et~al}.}
(\bibinfo{collaboration}{E142 Collaboration}), 
 \bibinfo{journal}{Phys. Rev.}
  \textbf{\bibinfo{volume}{D54}}, \bibinfo{pages}{6620} (\bibinfo{year}{1996}).

\bibitem[{\citenamefont{Abe \textit{et~al}.}(1998)}]{Abe:1998wq}

\bibinfo{author}{\bibfnamefont{K.}~\bibnamefont{Abe}} \bibnamefont{\textit{et~al}.}
(\bibinfo{collaboration}{E143 collaboration}), 
  \bibinfo{journal}{Phys. Rev.}
  \textbf{\bibinfo{volume}{D58}}, \bibinfo{pages}{112003}
  (\bibinfo{year}{1998}).

\bibitem[{\citenamefont{Abe \textit{et~al}.}(1997{\natexlab{b}})}]{Abe:1997cx}

\bibinfo{author}{\bibfnamefont{K.}~\bibnamefont{Abe}} \bibnamefont{\textit{et~al}.}
(\bibinfo{collaboration}{E154 Collaboration}),
 \bibinfo{journal}{Phys. Rev.
  Lett.} \textbf{\bibinfo{volume}{79}}, \bibinfo{pages}{26}
  (\bibinfo{year}{1997}{\natexlab{b}}).

\bibitem[{\citenamefont{Anthony \textit{et~al}.}(1999)}]{Anthony:1999rm}
 
\bibinfo{author}{\bibfnamefont{P.}~\bibnamefont{Anthony}} \bibnamefont{\textit{et~al}.}
(\bibinfo{collaboration}{E155 Collaboration}),
\bibinfo{journal}{Phys. Lett.}
  \textbf{\bibinfo{volume}{B463}}, \bibinfo{pages}{339} (\bibinfo{year}{1999}).

\bibitem[{\citenamefont{Anthony \textit{et~al}.}(2000)}]{Anthony:2000fn}
 
\bibinfo{author}{\bibfnamefont{P.}~\bibnamefont{Anthony}} \bibnamefont{\textit{et~al}.}
(\bibinfo{collaboration}{E155 Collaboration}),
  \bibinfo{journal}{Phys. Lett.}
  \textbf{\bibinfo{volume}{B493}}, \bibinfo{pages}{19} (\bibinfo{year}{2000}).

\bibitem[{\citenamefont{Anthony \textit{et~al}.}(2003)}]{Anthony:2002hy}

\bibinfo{author}{\bibfnamefont{P.}~\bibnamefont{Anthony}} \bibnamefont{\textit{et~al}.}
(\bibinfo{collaboration}{E155 Collaboration}),
  \bibinfo{journal}{Phys. Lett.}
  \textbf{\bibinfo{volume}{B553}}, \bibinfo{pages}{18} (\bibinfo{year}{2003}).

\bibitem[{\citenamefont{Zheng \textit{et~al}.}(2004{\natexlab{b}})}]{Zheng:2003un}

\bibinfo{author}{\bibfnamefont{X.}~\bibnamefont{Zheng}} \bibnamefont{\textit{et~al}.}
(\bibinfo{collaboration}{Jefferson Lab Hall A Collaboration}), 
  \bibinfo{journal}{Phys. Rev. Lett.} \textbf{\bibinfo{volume}{92}},
  \bibinfo{pages}{012004} (\bibinfo{year}{2004}{\natexlab{b}}).

\bibitem[{\citenamefont{Airapetian \textit{et~al}.}(2012)}]{Airapetian:2011wu}

\bibinfo{author}{\bibfnamefont{A.}~\bibnamefont{Airapetian}}
  \bibnamefont{\textit{et~al}.}
  (\bibinfo{collaboration}{HERMES Collaboration}),
  \bibinfo{journal}{Eur. Phys. J.} \textbf{\bibinfo{volume}{C72}},
  \bibinfo{pages}{1921} (\bibinfo{year}{2012}).

\bibitem[{\citenamefont{Filoti}(2007)}]{Bates:ref}
\bibinfo{author}{\bibfnamefont{O.}~\bibnamefont{Filoti}},
  \bibinfo{type}{Doctoral Dissertation}, \bibinfo{institution}{University of
  New Hampshire} \bibinfo{year}{2007} (unpublished).

\bibitem[{\citenamefont{Kramer \textit{et~al}.}(2005)\citenamefont{Kramer, Armstrong,
  Averett, Bertozzi, Binet \textit{et~al}.}}]{Kramer:2005qe}
\bibinfo{author}{\bibfnamefont{K.}~\bibnamefont{Kramer}},
  \bibinfo{author}{\bibfnamefont{D.}~\bibnamefont{Armstrong}},
  \bibinfo{author}{\bibfnamefont{T.}~\bibnamefont{Averett}},
  \bibinfo{author}{\bibfnamefont{W.}~\bibnamefont{Bertozzi}},
  \bibinfo{author}{\bibfnamefont{S.}~\bibnamefont{Binet}},
  \bibnamefont{\textit{et~al}.}, \bibinfo{journal}{Phys. Rev. Lett.}
  \textbf{\bibinfo{volume}{95}}, \bibinfo{pages}{142002}
  (\bibinfo{year}{2005}).

\bibitem[{\citenamefont{Adolph \textit{et~al}.}(2016)}]{Adolph:2015saz}
 
\bibinfo{author}{\bibfnamefont{C.}~\bibnamefont{Adolph}} \bibnamefont{\textit{et~al}.}
(\bibinfo{collaboration}{COMPASS Collaboration}),
  \bibinfo{journal}{Phys.
  Lett.} \textbf{\bibinfo{volume}{B753}}, \bibinfo{pages}{18}
  (\bibinfo{year}{2016}), \eprint{1503.08935}.

\bibitem[{\citenamefont{Flay \textit{et~al}.}(2016)}]{Flay:2016wie}
\bibinfo{author}{\bibfnamefont{D.}~\bibnamefont{Flay}} \bibnamefont{\textit{et~al}.},
  \bibinfo{journal}{Phys. Rev.} \textbf{\bibinfo{volume}{D94}},
  \bibinfo{pages}{052003} (\bibinfo{year}{2016}).

\bibitem[{\citenamefont{Rondon~Aramayo}(2009)}]{RondonAramayo:2009zz}
\bibinfo{author}{\bibfnamefont{O.~A.} \bibnamefont{Rondon~Aramayo}},
  \textit{Spin Structure at Long Distance: Workshop Proceedings}, edited by
  J. Chen, K. Slifer, and W. Melnitchouk, 
    \bibinfo{journal}{AIP Conf.Proc.} \textbf{\bibinfo{volume}{1155}},
  \bibinfo{pages}{82} (\bibinfo{year}{2009}).

\bibitem[{\citenamefont{Kang}(2013)}]{KANG:2013lva}

\bibinfo{author}{\bibfnamefont{H.}~\bibnamefont{Kang}}
(\bibinfo{collaboration}{SANE Collaboration}),
 \bibinfo{journal}{PoS}
  \textbf{\bibinfo{volume}{DIS2013}}, \bibinfo{pages}{206}
  (\bibinfo{year}{2013}).

\bibitem[{\citenamefont{Bloom and Gilman}(1970)}]{Bloom:1970xb}
\bibinfo{author}{\bibfnamefont{E.~D.} \bibnamefont{Bloom}} \bibnamefont{and}
  \bibinfo{author}{\bibfnamefont{F.~J.} \bibnamefont{Gilman}},
  \bibinfo{journal}{Phys. Rev. Lett.} \textbf{\bibinfo{volume}{25}},
  \bibinfo{pages}{1140} (\bibinfo{year}{1970}).

\bibitem[{\citenamefont{Nachtmann}(1973)}]{Nachtmann:1973mr}
\bibinfo{author}{\bibfnamefont{O.}~\bibnamefont{Nachtmann}},
  \bibinfo{journal}{Nucl. Phys.} \textbf{\bibinfo{volume}{B63}},
  \bibinfo{pages}{237} (\bibinfo{year}{1973}).

\bibitem[{\citenamefont{De~Rujula
  \textit{et~al}.}(1977{\natexlab{a}})\citenamefont{De~Rujula, Georgi, and
  Politzer}}]{DeRujula:1976tz}
\bibinfo{author}{\bibfnamefont{A.}~\bibnamefont{De~Rujula}},
  \bibinfo{author}{\bibfnamefont{H.}~\bibnamefont{Georgi}}, \bibnamefont{and}
  \bibinfo{author}{\bibfnamefont{H.}~\bibnamefont{Politzer}},
  \bibinfo{journal}{Annals Phys.} \textbf{\bibinfo{volume}{103}},
  \bibinfo{pages}{315} (\bibinfo{year}{1977}{\natexlab{a}}).

\bibitem[{\citenamefont{De~Rujula
  \textit{et~al}.}(1977{\natexlab{b}})\citenamefont{De~Rujula, Georgi, and
  Politzer}}]{DeRujula:1976ke}
\bibinfo{author}{\bibfnamefont{A.}~\bibnamefont{De~Rujula}},
  \bibinfo{author}{\bibfnamefont{H.}~\bibnamefont{Georgi}}, \bibnamefont{and}
  \bibinfo{author}{\bibfnamefont{H.}~\bibnamefont{Politzer}},
  \bibinfo{journal}{Phys. Lett.} \textbf{\bibinfo{volume}{B64}},
  \bibinfo{pages}{428} (\bibinfo{year}{1977}{\natexlab{b}}).

\bibitem[{\citenamefont{Melnitchouk \textit{et~al}.}(2005)\citenamefont{Melnitchouk,
  Ent, and Keppel}}]{Melnitchouk:2005zr}
\bibinfo{author}{\bibfnamefont{W.}~\bibnamefont{Melnitchouk}},
  \bibinfo{author}{\bibfnamefont{R.}~\bibnamefont{Ent}}, \bibnamefont{and}
  \bibinfo{author}{\bibfnamefont{C.}~\bibnamefont{Keppel}},
  \bibinfo{journal}{Phys. Rept.} \textbf{\bibinfo{volume}{406}},
  \bibinfo{pages}{127} (\bibinfo{year}{2005}).

\bibitem[{\citenamefont{Airapetian \textit{et~al}.}(1998)}]{Airapetian:1998wi}

\bibinfo{author}{\bibfnamefont{A.}~\bibnamefont{Airapetian}}
  \bibnamefont{\textit{et~al}.}
  (\bibinfo{collaboration}{HERMES Collaboration}),
  \bibinfo{journal}{Phys. Lett.} \textbf{\bibinfo{volume}{B442}},
  \bibinfo{pages}{484} (\bibinfo{year}{1998}).

\bibitem[{\citenamefont{Airapetian \textit{et~al}.}(2000)}]{Airapetian:1999ib}

\bibinfo{author}{\bibfnamefont{A.}~\bibnamefont{Airapetian}}
  \bibnamefont{\textit{et~al}.}
 (\bibinfo{collaboration}{HERMES Collaboration}), 
  \bibinfo{journal}{Phys. Rev. Lett.} \textbf{\bibinfo{volume}{84}},
  \bibinfo{pages}{2584} (\bibinfo{year}{2000}).

\bibitem[{\citenamefont{Solvignon \textit{et~al}.}(2008)}]{Solvignon:2008hk}

\bibinfo{author}{\bibfnamefont{P.}~\bibnamefont{Solvignon}}
  \bibnamefont{\textit{et~al}.},
  (\bibinfo{collaboration}{Jefferson Lab E01-012 Collaboration}),
   \bibinfo{journal}{Phys. Rev. Lett.}
  \textbf{\bibinfo{volume}{101}}, \bibinfo{pages}{182502}
  (\bibinfo{year}{2008}).

\bibitem[{\citenamefont{Blumlein and Tkabladze}(1999)}]{Blumlein:1998nv}
\bibinfo{author}{\bibfnamefont{J.}~\bibnamefont{Blumlein}} \bibnamefont{and}
  \bibinfo{author}{\bibfnamefont{A.}~\bibnamefont{Tkabladze}},
  \bibinfo{journal}{Nucl. Phys.} \textbf{\bibinfo{volume}{B553}},
  \bibinfo{pages}{427} (\bibinfo{year}{1999}).

\bibitem[{\citenamefont{Wandzura and Wilczek}(1977)}]{Wandzura:1977qf}
\bibinfo{author}{\bibfnamefont{S.}~\bibnamefont{Wandzura}} \bibnamefont{and}
  \bibinfo{author}{\bibfnamefont{F.}~\bibnamefont{Wilczek}},
  \bibinfo{journal}{Phys. Lett.} \textbf{\bibinfo{volume}{B72}},
  \bibinfo{pages}{195} (\bibinfo{year}{1977}).

\bibitem[{\citenamefont{Burkardt}(2009)}]{Burkardt:2009rf}
\bibinfo{author}{\bibfnamefont{M.}~\bibnamefont{Burkardt}},
\textit{Spin Structure at Long Distance: Workshop
Proceedings}, edited by J. Chen, K. Slifer, and W. Melnitchouk,
  \bibinfo{journal}{AIP Conf.Proc.} \textbf{\bibinfo{volume}{1155}},
  \bibinfo{pages}{26} (\bibinfo{year}{2009}).

\bibitem[{\citenamefont{Burkhardt and Cottingham}(1970)}]{Burkhardt:1970ti}
\bibinfo{author}{\bibfnamefont{H.}~\bibnamefont{Burkhardt}} \bibnamefont{and}
  \bibinfo{author}{\bibfnamefont{W.~N.} \bibnamefont{Cottingham}},
  \bibinfo{journal}{Annals Phys.} \textbf{\bibinfo{volume}{56}},
  \bibinfo{pages}{453} (\bibinfo{year}{1970}).

\bibitem[{\citenamefont{Gayou \textit{et~al}.}(2001)\citenamefont{Gayou, Wijesooriya,
  Afanasev, Amarian, Aniol \etal}}]{Gayou:2001qt}
\bibinfo{author}{\bibfnamefont{O.}~\bibnamefont{Gayou}},
  \bibinfo{author}{\bibfnamefont{K.}~\bibnamefont{Wijesooriya}},
  \bibinfo{author}{\bibfnamefont{A.}~\bibnamefont{Afanasev}},
  \bibinfo{author}{\bibfnamefont{M.}~\bibnamefont{Amarian}},
  \bibinfo{author}{\bibfnamefont{K.}~\bibnamefont{Aniol}},
  \bibnamefont{\textit{et~al}.}, \bibinfo{journal}{Phys. Rev.}
  \textbf{\bibinfo{volume}{C64}}, \bibinfo{pages}{038202}
  (\bibinfo{year}{2001}).

\bibitem[{\citenamefont{Arrington \textit{et~al}.}(2007)\citenamefont{Arrington,
  Melnitchouk, and Tjon}}]{Arrington:2007ux}
\bibinfo{author}{\bibfnamefont{J.}~\bibnamefont{Arrington}},
  \bibinfo{author}{\bibfnamefont{W.}~\bibnamefont{Melnitchouk}},
  \bibnamefont{and} \bibinfo{author}{\bibfnamefont{J. A.}~\bibnamefont{Tjon}},
  \bibinfo{journal}{Phys. Rev.} \textbf{\bibinfo{volume}{C76}},
  \bibinfo{pages}{035205} (\bibinfo{year}{2007}).

\bibitem[{\citenamefont{Hagiwara \textit{et~al}.}(2002)}]{Hagiwara:2002fs}

\bibinfo{author}{\bibfnamefont{K.}~\bibnamefont{Hagiwara}} \bibnamefont{\etal}
(\bibinfo{collaboration}{Particle Data Group}),
   \bibinfo{journal}{Phys. Rev.}
  \textbf{\bibinfo{volume}{D66}}, \bibinfo{pages}{010001}
  (\bibinfo{year}{2002}).

\bibitem[{\citenamefont{Bjorken}(1966)}]{Bjorken:1966jh}
\bibinfo{author}{\bibfnamefont{J.~D.} \bibnamefont{Bjorken}},
  \bibinfo{journal}{Phys. Rev.} \textbf{\bibinfo{volume}{148}},
  \bibinfo{pages}{1467} (\bibinfo{year}{1966}).

\bibitem[{\citenamefont{Bjorken}(1970)}]{Bjorken:1969mm}
\bibinfo{author}{\bibfnamefont{J.~D.} \bibnamefont{Bjorken}},
  \bibinfo{journal}{Phys. Rev.} \textbf{\bibinfo{volume}{D1}},
  \bibinfo{pages}{1376} (\bibinfo{year}{1970}).

\bibitem[{\citenamefont{Larin \textit{et~al}.}(1997)\citenamefont{Larin, van Ritbergen,
  and Vermaseren}}]{Larin:1997qq}
\bibinfo{author}{\bibfnamefont{S.}~\bibnamefont{Larin}},
  \bibinfo{author}{\bibfnamefont{T.}~\bibnamefont{van Ritbergen}},
  \bibnamefont{and}
  \bibinfo{author}{\bibfnamefont{J.}~\bibnamefont{Vermaseren}},
  \bibinfo{journal}{Phys. Lett.} \textbf{\bibinfo{volume}{B404}},
  \bibinfo{pages}{153} (\bibinfo{year}{1997}).


\bibitem[{\citenamefont{Stein \textit{et~al}.}(1995)\citenamefont{Stein, Gornicki,
  Mankiewicz, and Schafer}}]{Stein:1995si}
\bibinfo{author}{\bibfnamefont{E.}~\bibnamefont{Stein}},
  \bibinfo{author}{\bibfnamefont{P.}~\bibnamefont{Gornicki}},
  \bibinfo{author}{\bibfnamefont{L.}~\bibnamefont{Mankiewicz}},
  \bibnamefont{and} \bibinfo{author}{\bibfnamefont{A.}~\bibnamefont{Schafer}},
  \bibinfo{journal}{Phys. Lett.} \textbf{\bibinfo{volume}{B353}},
  \bibinfo{pages}{107} (\bibinfo{year}{1995}).

\bibitem[{\citenamefont{Ji}(1995)}]{Ji:1995qe}
\bibinfo{author}{\bibfnamefont{X.-D.} \bibnamefont{Ji}}, in
  \emph{\bibinfo{booktitle}{{Baryons '95. Proceedings, 7th International
  Conference on the Structure of Baryons, Santa Fe, USA, October 3-7, 1995}}},
  edited by B. Gibson, P. Barnes, J. McClelland, and W. Weise (World Scientific
  Publishing, Singapore, 1995).

\bibitem[{\citenamefont{Signal}(1997)}]{Signal:1996ct}
\bibinfo{author}{\bibfnamefont{A.}~\bibnamefont{Signal}},
  \bibinfo{journal}{Nucl. Phys.} \textbf{\bibinfo{volume}{B497}},
  \bibinfo{pages}{415} (\bibinfo{year}{1997}).

\bibitem[{\citenamefont{Balitsky \textit{et~al}.}(1990)\citenamefont{Balitsky, Braun,
  and Kolesnichenko}}]{Balitsky:1989jb}
\bibinfo{author}{\bibfnamefont{I.}~\bibnamefont{Balitsky}},
  \bibinfo{author}{\bibfnamefont{V.~M.} \bibnamefont{Braun}}, \bibnamefont{and}
  \bibinfo{author}{\bibfnamefont{A.}~\bibnamefont{Kolesnichenko}},
  \bibinfo{journal}{Phys. Lett.} \textbf{\bibinfo{volume}{B242}},
  \bibinfo{pages}{245} (\bibinfo{year}{1990}).

\bibitem[{\citenamefont{Dolgov \textit{et~al}.}(1999)\citenamefont{Dolgov, Brower,
  Negele, and Pochinsky}}]{Dolgov:1998js}
\bibinfo{author}{\bibfnamefont{D.}~\bibnamefont{Dolgov}},
  \bibinfo{author}{\bibfnamefont{R.}~\bibnamefont{Brower}},
  \bibinfo{author}{\bibfnamefont{J.~W.} \bibnamefont{Negele}},
  \bibnamefont{and}
  \bibinfo{author}{\bibfnamefont{A.}~\bibnamefont{Pochinsky}},
  \bibinfo{journal}{Nucl. Phys. Proc. Suppl.} \textbf{\bibinfo{volume}{73}},
  \bibinfo{pages}{300} (\bibinfo{year}{1999}).

\bibitem[{\citenamefont{Gerasimov}(1966)}]{Gerasimov:1965et}
\bibinfo{author}{\bibfnamefont{S.}~\bibnamefont{Gerasimov}},
  \bibinfo{journal}{Sov.J. Nucl. Phys.} \textbf{\bibinfo{volume}{2}},
  \bibinfo{pages}{430} (\bibinfo{year}{1966}).

\bibitem[{\citenamefont{Drell and Hearn}(1966)}]{Drell:1966jv}
\bibinfo{author}{\bibfnamefont{S.}~\bibnamefont{Drell}} \bibnamefont{and}
  \bibinfo{author}{\bibfnamefont{A.~C.} \bibnamefont{Hearn}},
  \bibinfo{journal}{Phys. Rev. Lett.} \textbf{\bibinfo{volume}{16}},
  \bibinfo{pages}{908} (\bibinfo{year}{1966}).

\bibitem[{\citenamefont{Ji \textit{et~al}.}(2000)\citenamefont{Ji, Kao, and
  Osborne}}]{Ji:1999pd}
\bibinfo{author}{\bibfnamefont{X.-D.} \bibnamefont{Ji}},
  \bibinfo{author}{\bibfnamefont{C.-W.} \bibnamefont{Kao}}, \bibnamefont{and}
  \bibinfo{author}{\bibfnamefont{J.}~\bibnamefont{Osborne}},
  \bibinfo{journal}{Phys. Lett.} \textbf{\bibinfo{volume}{B472}},
  \bibinfo{pages}{1} (\bibinfo{year}{2000}).

\bibitem[{\citenamefont{Gorchtein \textit{et~al}.}(2004)\citenamefont{Gorchtein,
  Drechsel, Giannini, Santopinto, and Tiator}}]{Gorchtein:2004jd}
\bibinfo{author}{\bibfnamefont{M.}~\bibnamefont{Gorchtein}},
  \bibinfo{author}{\bibfnamefont{D.}~\bibnamefont{Drechsel}},
  \bibinfo{author}{\bibfnamefont{M. M.}~\bibnamefont{Giannini}},
  \bibinfo{author}{\bibfnamefont{E.}~\bibnamefont{Santopinto}},
  \bibnamefont{and} \bibinfo{author}{\bibfnamefont{L.}~\bibnamefont{Tiator}},
  \bibinfo{journal}{Phys. Rev.} \textbf{\bibinfo{volume}{C70}},
  \bibinfo{pages}{055202} (\bibinfo{year}{2004}).

\bibitem[{\citenamefont{Deur \textit{et~al}.}(2008)\citenamefont{Deur, Bosted, Burkert,
  Crabb, Dharmawardane \etal}}]{Deur:2008ej}
\bibinfo{author}{\bibfnamefont{A.}~\bibnamefont{Deur}},
  \bibinfo{author}{\bibfnamefont{P.}~\bibnamefont{Bosted}},
  \bibinfo{author}{\bibfnamefont{V.}~\bibnamefont{Burkert}},
  \bibinfo{author}{\bibfnamefont{D.}~\bibnamefont{Crabb}},
  \bibinfo{author}{\bibfnamefont{V.}~\bibnamefont{Dharmawardane}},
  \bibnamefont{\textit{et~al}.}, \bibinfo{journal}{Phys. Rev.}
  \textbf{\bibinfo{volume}{D78}}, \bibinfo{pages}{032001}
  (\bibinfo{year}{2008}).

\bibitem[{\citenamefont{Leeman \textit{et~al}.}(2001)\citenamefont{Leeman, Douglas, and
  Krafft}}]{CEBAF:ref}
\bibinfo{author}{\bibfnamefont{C.}~\bibnamefont{Leeman}},
  \bibinfo{author}{\bibfnamefont{D.}~\bibnamefont{Douglas}}, \bibnamefont{and}
  \bibinfo{author}{\bibfnamefont{G.}~\bibnamefont{Krafft}},
  \bibinfo{journal}{Ann. Rev. Nucl. Part. Sci.} \textbf{\bibinfo{volume}{51}},
  \bibinfo{pages}{413} (\bibinfo{year}{2001}).

\bibitem[{\citenamefont{Sinclair \textit{et~al}.}(2007)\citenamefont{Sinclair, Adderley,
  Dunham, Hansknecht, Hartmann \textit{et~al}.}}]{Sinclair:2007ez}
\bibinfo{author}{\bibfnamefont{C.}~\bibnamefont{Sinclair}},
  \bibinfo{author}{\bibfnamefont{P.}~\bibnamefont{Adderley}},
  \bibinfo{author}{\bibfnamefont{B.}~\bibnamefont{Dunham}},
  \bibinfo{author}{\bibfnamefont{J.}~\bibnamefont{Hansknecht}},
  \bibinfo{author}{\bibfnamefont{P.}~\bibnamefont{Hartmann}},
  \bibinfo{author}{\bibfnamefont{M.}~\bibnamefont{Poelker}},
  \bibinfo{author}{\bibfnamefont{J. S.}~\bibnamefont{Price}},
  \bibinfo{author}{\bibfnamefont{P. M.}~\bibnamefont{Rutt}},
  \bibinfo{author}{\bibfnamefont{W. J.}~\bibnamefont{Schneider}}, and
  \bibinfo{author}{\bibfnamefont{M.}~\bibnamefont{Steigerwald}},
  \bibinfo{journal}{Phys. Rev. ST Accel. Beams}
  \textbf{\bibinfo{volume}{10}}, \bibinfo{pages}{023501}
  (\bibinfo{year}{2007}).

\bibitem[{\citenamefont{Kazimi \textit{et~al}.}(2004)}]{Kazimi:2004zv}
\bibinfo{author}{\bibfnamefont{R.}~\bibnamefont{Kazimi}} \bibnamefont{\textit{et~al}.},
  in \emph{\bibinfo{booktitle}{{9th European Particle Accelerator Conference
  (EPAC 2004) Lucerne, Switzerland, July 5-9, 2004}}} (\bibinfo{year}{2004}),
  \urlprefix\url{http://accelconf.web.cern.ch/AccelConf/e04/PAPERS/TUPLT154.PDF}.

\bibitem[{\citenamefont{Stutzman \textit{et~al}.}(2007)\citenamefont{Stutzman, Adderley,
  Brittian, Clark, Grames \textit{et~al}.}}]{Stutzman:2007ny}
\bibinfo{author}{\bibfnamefont{M.}~\bibnamefont{Stutzman}},
  \bibinfo{author}{\bibfnamefont{P.}~\bibnamefont{Adderley}},
  \bibinfo{author}{\bibfnamefont{J.}~\bibnamefont{Brittian}},
  \bibinfo{author}{\bibfnamefont{J.}~\bibnamefont{Clark}},
  \bibinfo{author}{\bibfnamefont{J.}~\bibnamefont{Grames}},
  \bibnamefont{\etal}, \bibinfo{journal}{Nucl. Instrum. Meth.}
  \textbf{\bibinfo{volume}{A574}}, \bibinfo{pages}{213} (\bibinfo{year}{2007}).

\bibitem[{\citenamefont{Schultz \textit{et~al}.}(1992)\citenamefont{Schultz, Clendenin,
  Frisch, Hoyt, Klaisner, Woods, Wright, and Zolotorev}}]{Schultz:1992zy}
\bibinfo{author}{\bibfnamefont{D.}~\bibnamefont{Schultz}},
  \bibinfo{author}{\bibfnamefont{J.~E.} \bibnamefont{Clendenin}},
  \bibinfo{author}{\bibfnamefont{J.}~\bibnamefont{Frisch}},
  \bibinfo{author}{\bibfnamefont{E.~W.} \bibnamefont{Hoyt}},
  \bibinfo{author}{\bibfnamefont{L.}~\bibnamefont{Klaisner}},
  \bibinfo{author}{\bibfnamefont{M.}~\bibnamefont{Woods}},
  \bibinfo{author}{\bibfnamefont{D.~M.} \bibnamefont{Wright}},
  \bibnamefont{and}
  \bibinfo{author}{\bibfnamefont{M.}~\bibnamefont{Zolotorev}},
  \bibinfo{journal}{Conf. Proc.} \textbf{\bibinfo{volume}{C920324}},
  \bibinfo{pages}{1029} (\bibinfo{year}{1992}).

\bibitem[{\citenamefont{Liu}(1997)}]{Liu:1997ue}
\bibinfo{author}{\bibfnamefont{H.}~\bibnamefont{Liu}}, \bibinfo{journal}{Nucl.
  Instrum. Meth.} \textbf{\bibinfo{volume}{A400}}, \bibinfo{pages}{213}
  (\bibinfo{year}{1997}).

\bibitem[{\citenamefont{Grames}(2001)}]{polarimeters:ref}
\bibinfo{author}{\bibfnamefont{J.}~\bibnamefont{Grames}},
  \bibinfo{journal}{Jefferson Lab Techical Note}
  \textbf{\bibinfo{volume}{JLAB-TN-01-029}} \bibinfo{year}{2001} (unpublished).

\bibitem[{\citenamefont{Wagner \textit{et~al}.}(1990)\citenamefont{Wagner, Andresen,
  Steffens, Hartmann, Heil, and Reichert}}]{Wagner:1990sn}
\bibinfo{author}{\bibfnamefont{B.}~\bibnamefont{Wagner}},
  \bibinfo{author}{\bibfnamefont{H.~G.} \bibnamefont{Andresen}},
  \bibinfo{author}{\bibfnamefont{K.~H.} \bibnamefont{Steffens}},
  \bibinfo{author}{\bibfnamefont{W.}~\bibnamefont{Hartmann}},
  \bibinfo{author}{\bibfnamefont{W.}~\bibnamefont{Heil}}, \bibnamefont{and}
  \bibinfo{author}{\bibfnamefont{E.}~\bibnamefont{Reichert}},
  \bibinfo{journal}{Nucl. Instrum. Meth.} \textbf{\bibinfo{volume}{A294}},
  \bibinfo{pages}{541} (\bibinfo{year}{1990}).

\bibitem[{\citenamefont{Keith \textit{et~al}.}(2003)\citenamefont{Keith, Anghinolfi,
  Battaglieri, Bosted, Branford \textit{et~al}.}}]{Keith:2003ca}
\bibinfo{author}{\bibfnamefont{C.}~\bibnamefont{Keith}},
  \bibinfo{author}{\bibfnamefont{M.}~\bibnamefont{Anghinolfi}},
  \bibinfo{author}{\bibfnamefont{M.}~\bibnamefont{Battaglieri}},
  \bibinfo{author}{\bibfnamefont{P.~E.} \bibnamefont{Bosted}},
  \bibinfo{author}{\bibfnamefont{D.}~\bibnamefont{Branford}},
  \bibnamefont{\etal}, \bibinfo{journal}{Nucl. Instrum. Meth.}
  \textbf{\bibinfo{volume}{A501}}, \bibinfo{pages}{327} (\bibinfo{year}{2003}).

\bibitem[{\citenamefont{de~Boer}(1974)}]{DNP:ref}
\bibinfo{author}{\bibfnamefont{W.}~\bibnamefont{de~Boer}},
  \bibinfo{journal}{CERN 74-11, Laboratory I, Nuclear Physics Division}
  \bibinfo{year}{1974} (unpublished).

\bibitem[{\citenamefont{Borghini}(1974)}]{orientation:ref}
\bibinfo{author}{\bibfnamefont{M.}~\bibnamefont{Borghini}},
  \bibinfo{journal}{CERN 68-32, Nuclear Physics Division}
  \bibinfo{year}{1974} (unpublished).

\bibitem[{\citenamefont{Crabb and Meyer}(1997)}]{Crabb:1997cy}
\bibinfo{author}{\bibfnamefont{D.}~\bibnamefont{Crabb}} \bibnamefont{and}
  \bibinfo{author}{\bibfnamefont{W.}~\bibnamefont{Meyer}},
  \bibinfo{journal}{Ann. Rev. Nucl. Part. Sci.} \textbf{\bibinfo{volume}{47}},
  \bibinfo{pages}{67} (\bibinfo{year}{1997}).

\bibitem[{\citenamefont{Mestayer \textit{et~al}.}(2000)\citenamefont{Mestayer, Carman,
  Asavapibhop, Barbosa, Bonneau \textit{et~al}.}}]{Mestayer:2000we}
\bibinfo{author}{\bibfnamefont{M.}~\bibnamefont{Mestayer}},
  \bibinfo{author}{\bibfnamefont{D.}~\bibnamefont{Carman}},
  \bibinfo{author}{\bibfnamefont{B.}~\bibnamefont{Asavapibhop}},
  \bibinfo{author}{\bibfnamefont{F.}~\bibnamefont{Barbosa}},
  \bibinfo{author}{\bibfnamefont{P.}~\bibnamefont{Bonneau}},
  \bibnamefont{\etal}, \bibinfo{journal}{Nucl. Instrum. Meth.}
  \textbf{\bibinfo{volume}{A449}}, \bibinfo{pages}{81} (\bibinfo{year}{2000}).

\bibitem[{\citenamefont{Smith \textit{et~al}.}(1999)\citenamefont{Smith, Carstens,
  Distelbrink, Eckhause, Egiian \textit{et~al}}}]{Smith:1999ii}
\bibinfo{author}{\bibfnamefont{E.}~\bibnamefont{Smith}},
  \bibinfo{author}{\bibfnamefont{T.}~\bibnamefont{Carstens}},
  \bibinfo{author}{\bibfnamefont{J.}~\bibnamefont{Distelbrink}},
  \bibinfo{author}{\bibfnamefont{M.}~\bibnamefont{Eckhause}},
  \bibinfo{author}{\bibfnamefont{H.}~\bibnamefont{Egiian}},
  \bibnamefont{\etal}, \bibinfo{journal}{Nucl. Instrum. Meth.}
  \textbf{\bibinfo{volume}{A432}}, \bibinfo{pages}{265} (\bibinfo{year}{1999}).

\bibitem[{\citenamefont{Adams \textit{et~al}}(2001)\citenamefont{Adams, Burkert, Carl,
  Carstens, Frolov \etal}}]{Adams:2001kk}
\bibinfo{author}{\bibfnamefont{G.}~\bibnamefont{Adams}},
  \bibinfo{author}{\bibfnamefont{V.}~\bibnamefont{Burkert}},
  \bibinfo{author}{\bibfnamefont{R.}~\bibnamefont{Carl}},
  \bibinfo{author}{\bibfnamefont{T.}~\bibnamefont{Carstens}},
  \bibinfo{author}{\bibfnamefont{V.}~\bibnamefont{Frolov}},
  \bibnamefont{\textit{et~al}.} \bibinfo{journal}{Nucl. Instrum. Meth.}
  \textbf{\bibinfo{volume}{A465}}, \bibinfo{pages}{414} (\bibinfo{year}{2001}).

\bibitem[{\citenamefont{Amarian \textit{et~al}}(2001)\citenamefont{Amarian, Asryan,
  Beard, Brooks, Burkert \textit{et~al}.}}]{Amarian:2001zs}
\bibinfo{author}{\bibfnamefont{M.}~\bibnamefont{Amarian}},
  \bibinfo{author}{\bibfnamefont{G.}~\bibnamefont{Asryan}},
  \bibinfo{author}{\bibfnamefont{K.}~\bibnamefont{Beard}},
  \bibinfo{author}{\bibfnamefont{W.}~\bibnamefont{Brooks}},
  \bibinfo{author}{\bibfnamefont{V.}~\bibnamefont{Burkert}},
  \bibnamefont{\textit{et~al}.} \bibinfo{journal}{Nucl. Instrum. Meth.}
  \textbf{\bibinfo{volume}{A460}}, \bibinfo{pages}{239} (\bibinfo{year}{2001}).

\bibitem[{\citenamefont{Dharmawardane}(2004)}]{vipuli:ref}
\bibinfo{author}{\bibfnamefont{V.}~\bibnamefont{Dharmawardane}},
  \bibinfo{journal}{Ph.D. dissertation, Old Dominion University}
  \bibinfo{year}{2004} (unpublished).

\bibitem[{\citenamefont{Dalitz}(1951)}]{Dalitz:1951aj}
\bibinfo{author}{\bibfnamefont{R.}~\bibnamefont{Dalitz}},
  \bibinfo{journal}{Proc. Phys. Soc., London, Sect.} \textbf{\bibinfo{volume}{A64}},
  \bibinfo{pages}{667} (\bibinfo{year}{1951}).

\bibitem[{\citenamefont{Gehrmann and Stratmann}(1997)}]{Gehrmann:1997qh}
\bibinfo{author}{\bibfnamefont{T.}~\bibnamefont{Gehrmann}} \bibnamefont{and}
  \bibinfo{author}{\bibfnamefont{M.}~\bibnamefont{Stratmann}},
  \bibinfo{journal}{Phys. Rev.} \textbf{\bibinfo{volume}{D56}},
  \bibinfo{pages}{5839} (\bibinfo{year}{1997}).

\bibitem[{\citenamefont{Bosted \textit{et~al}}(2003)\citenamefont{Bosted, Kuhn, and
  Prok}}]{rastercorr:ref}
\bibinfo{author}{\bibfnamefont{P.}~\bibnamefont{Bosted}},
  \bibinfo{author}{\bibfnamefont{S.}~\bibnamefont{Kuhn}}, \bibnamefont{and}
  \bibinfo{author}{\bibfnamefont{Y.}~\bibnamefont{Prok}}, \bibinfo{type}{CLAS
  Note} \bibinfo{number}{2003-008}, \bibinfo{institution}{Jefferson Lab}
  \bibinfo{year}{2003},
  \urlprefix\url{https://www.jlab.org/Hall-B/notes/clas_notes03/03-008.pdf}.

\bibitem[{\citenamefont{Yao \textit{et~al}}(2006)}]{Yao:2006px}
 
\bibinfo{author}{\bibfnamefont{W.}~\bibnamefont{Yao}} \bibnamefont{\textit{et~al}.}
(\bibinfo{collaboration}{Particle Data Group}),
  \bibinfo{journal}{J.Phys.G}
  \textbf{\bibinfo{volume}{G33}}, \bibinfo{pages}{1} (\bibinfo{year}{2006}).

\bibitem[{\citenamefont{Leo}(1994)}]{Leobook}
\bibinfo{author}{\bibfnamefont{W.~R.} \bibnamefont{Leo}},
  \emph{\bibinfo{title}{Techniques for Nuclear and Particle Physics
  Experiments}} (\bibinfo{publisher}{Springer, Berlin}, \bibinfo{year}{1994}).

\bibitem[{\citenamefont{Bellis}(2002)}]{GSIM:ref}
\bibinfo{author}{\bibfnamefont{M.}~\bibnamefont{Bellis}}, \bibinfo{type}{CLAS
  Note} \bibinfo{number}{2002-016}, \bibinfo{institution}{Jefferson Lab}
  (\bibinfo{year}{2002}),
  \urlprefix\url{https://www.jlab.org/Hall-B/notes/clas_notes02/02-016a.pdf}.

\bibitem[{\citenamefont{Bosted and Avakian}(2006)}]{ms:ref}
\bibinfo{author}{\bibfnamefont{P.}~\bibnamefont{Bosted}} \bibnamefont{and}
  \bibinfo{author}{\bibfnamefont{H.}~\bibnamefont{Avakian}},
  \bibinfo{type}{CLAS Note} \bibinfo{number}{2006-006},
  \bibinfo{institution}{Jefferson Lab} \bibinfo{year}{2006} (unpublished).

\bibitem[{\citenamefont{Klimenko and Kuhn}(2003)}]{clasnote2003-005}
\bibinfo{author}{\bibfnamefont{A.}~\bibnamefont{Klimenko}} \bibnamefont{and}
  \bibinfo{author}{\bibfnamefont{S.}~\bibnamefont{Kuhn}}, \bibinfo{type}{CLAS
  Note} \bibinfo{number}{2003-005}, \bibinfo{institution}{Old Dominion
  University} (\bibinfo{year}{2003}),
  \urlprefix\url{https://www.jlab.org/Hall-B/notes/clas_notes03/03-005.pdf}.

\bibitem[{\citenamefont{James and Roos}(1975)}]{MINUIT}
\bibinfo{author}{\bibfnamefont{F.}~\bibnamefont{James}} \bibnamefont{and}
  \bibinfo{author}{\bibfnamefont{M.}~\bibnamefont{Roos}},
  \bibinfo{journal}{Computer Physics Communications}
  \textbf{\bibinfo{volume}{10}}, \bibinfo{pages}{343} (\bibinfo{year}{1975}).

\bibitem[{\citenamefont{De~Vries \textit{et~al}}(1987)\citenamefont{De~Vries, De~Jager,
  and De~Vries}}]{DeJager:1987qc}
\bibinfo{author}{\bibfnamefont{H.}~\bibnamefont{De~Vries}},
  \bibinfo{author}{\bibfnamefont{C.}~\bibnamefont{De~Jager}}, \bibnamefont{and}
  \bibinfo{author}{\bibfnamefont{C.}~\bibnamefont{De~Vries}},
  \bibinfo{journal}{Atom.Data Nucl.Data Tabl.} \textbf{\bibinfo{volume}{36}},
  \bibinfo{pages}{495} (\bibinfo{year}{1987}).

\bibitem[{\citenamefont{Christy and Bosted}(2010)}]{Christy:2007ve}
\bibinfo{author}{\bibfnamefont{M. E.}~\bibnamefont{Christy}} \bibnamefont{and}
  \bibinfo{author}{\bibfnamefont{P.~E.} \bibnamefont{Bosted}},
  \bibinfo{journal}{Phys. Rev.} \textbf{\bibinfo{volume}{C81}},
  \bibinfo{pages}{055213} (\bibinfo{year}{2010}).

\bibitem[{\citenamefont{Bosted \textit{et~al}}(2008)}]{Bosted:2007hw}

\bibinfo{author}{\bibfnamefont{P.}~\bibnamefont{Bosted}} \bibnamefont{\textit{et~al}.}
(\bibinfo{collaboration}{CLAS Collaboration}),
  \bibinfo{journal}{Phys. Rev.}
  \textbf{\bibinfo{volume}{C78}}, \bibinfo{pages}{015202}
  (\bibinfo{year}{2008}).

\bibitem[{\citenamefont{Norton}(2003)}]{Norton:2003cb}
\bibinfo{author}{\bibfnamefont{P.}~\bibnamefont{Norton}},
  \bibinfo{journal}{Rept. Prog. Phys.} \textbf{\bibinfo{volume}{66}},
  \bibinfo{pages}{1253} (\bibinfo{year}{2003}).

\bibitem[{\citenamefont{Mo and Tsai}(1969)}]{Mo:1968cg}
\bibinfo{author}{\bibfnamefont{L.~W.} \bibnamefont{Mo}} \bibnamefont{and}
  \bibinfo{author}{\bibfnamefont{Y.-S.} \bibnamefont{Tsai}},
  \bibinfo{journal}{Rev. Mod. Phys.} \textbf{\bibinfo{volume}{41}},
  \bibinfo{pages}{205} (\bibinfo{year}{1969}).

\bibitem[{\citenamefont{Arrington}(2004)}]{Arrington:2003qk}
\bibinfo{author}{\bibfnamefont{J.}~\bibnamefont{Arrington}},
  \bibinfo{journal}{Phys. Rev.} \textbf{\bibinfo{volume}{C69}},
  \bibinfo{pages}{022201} (\bibinfo{year}{2004}).

\bibitem[{\citenamefont{Bosted}(1995)}]{Bosted:1994tm}
\bibinfo{author}{\bibfnamefont{P.~E.} \bibnamefont{Bosted}},
  \bibinfo{journal}{Phys. Rev.} \textbf{\bibinfo{volume}{C51}},
  \bibinfo{pages}{409} (\bibinfo{year}{1995}).

\bibitem[{\citenamefont{Rondon-Aramayo}(1999)}]{RondonAramayo:1999da}
\bibinfo{author}{\bibfnamefont{O.~A.} \bibnamefont{Rondon-Aramayo}},
  \bibinfo{journal}{Phys. Rev.} \textbf{\bibinfo{volume}{C60}},
  \bibinfo{pages}{035201} (\bibinfo{year}{1999}).

\bibitem[{\citenamefont{Kukhto and Shumeiko}(1983)}]{Kukhto:1983pv}
\bibinfo{author}{\bibfnamefont{T.}~\bibnamefont{Kukhto}} \bibnamefont{and}
  \bibinfo{author}{\bibfnamefont{N.}~\bibnamefont{Shumeiko}},
  \bibinfo{journal}{Nucl. Phys.} \textbf{\bibinfo{volume}{B219}},
  \bibinfo{pages}{412} (\bibinfo{year}{1983}).


\bibitem{CLASdatabase}
Jefferson Lab Experiment CLAS Database, \urlprefix\url{http://
clasweb.jlab.org/physicsdb}.

\bibitem{SupplementalMaterial}
See Supplemental Material at \urlprefix\url{http://link.aps.org/supplemental/10.1103/PhysRevC.96.065208}
 for the complete tables of experimental results presented in this publication.





\bibitem[{\citenamefont{Liang \textit{et~al}.}(2004)}]{Liang:2004tj}

\bibinfo{author}{\bibfnamefont{Y.}~\bibnamefont{Liang}} \bibnamefont{\textit{et~al}.}
(\bibinfo{collaboration}{E94-110 Collaboration}),
  
  (\bibinfo{year}{2004}).

\bibitem[{\citenamefont{Helbing}(2006)}]{Helbing:2006zp}
\bibinfo{author}{\bibfnamefont{K.}~\bibnamefont{Helbing}},
  \bibinfo{journal}{Prog. Part. Nucl. Phys.} \textbf{\bibinfo{volume}{57}},
  \bibinfo{pages}{405} (\bibinfo{year}{2006}).

\bibitem[{\citenamefont{Drechsel \textit{et~al}.}(1999)\citenamefont{Drechsel, Hanstein,
  Kamalov, and Tiator}}]{Drechsel:1998hk}
\bibinfo{author}{\bibfnamefont{D.}~\bibnamefont{Drechsel}},
  \bibinfo{author}{\bibfnamefont{O.}~\bibnamefont{Hanstein}},
  \bibinfo{author}{\bibfnamefont{S.}~\bibnamefont{Kamalov}}, \bibnamefont{and}
  \bibinfo{author}{\bibfnamefont{L.}~\bibnamefont{Tiator}},
  \bibinfo{journal}{Nucl. Phys.} \textbf{\bibinfo{volume}{A645}},
  \bibinfo{pages}{145} (\bibinfo{year}{1999}).

\bibitem[{\citenamefont{Kamalov \textit{et~al}.}(2001)\citenamefont{Kamalov, Drechsel,
  Hanstein, Tiator, and Yang}}]{Kamalov:2001yi}
\bibinfo{author}{\bibfnamefont{S.}~\bibnamefont{Kamalov}},
  \bibinfo{author}{\bibfnamefont{D.}~\bibnamefont{Drechsel}},
  \bibinfo{author}{\bibfnamefont{O.}~\bibnamefont{Hanstein}},
  \bibinfo{author}{\bibfnamefont{L.}~\bibnamefont{Tiator}}, \bibnamefont{and}
  \bibinfo{author}{\bibfnamefont{S.}~\bibnamefont{Yang}},
  \bibinfo{journal}{Nucl. Phys.} \textbf{\bibinfo{volume}{A684}},
  \bibinfo{pages}{321} (\bibinfo{year}{2001}).

\bibitem[{\citenamefont{Aznauryan and Burkert}(2012)}]{Aznauryan:2011qj}
\bibinfo{author}{\bibfnamefont{I.~G.} \bibnamefont{Aznauryan}}
  \bibnamefont{and} \bibinfo{author}{\bibfnamefont{V.~D.}
  \bibnamefont{Burkert}}, \bibinfo{journal}{Prog. Part. Nucl. Phys.}
  \textbf{\bibinfo{volume}{67}}, \bibinfo{pages}{1} (\bibinfo{year}{2012}),
  \eprint{1109.1720}.

\bibitem[{\citenamefont{Bourrely \textit{et~al}.}(2005)\citenamefont{Bourrely, Soffer,
  and Buccella}}]{Bourrely:2005kw}
\bibinfo{author}{\bibfnamefont{C.~R.} \bibnamefont{Bourrely}},
  \bibinfo{author}{\bibfnamefont{J.}~\bibnamefont{Soffer}}, \bibnamefont{and}
  \bibinfo{author}{\bibfnamefont{F.}~\bibnamefont{Buccella}},
  \bibinfo{journal}{Eur. Phys. J.} \textbf{\bibinfo{volume}{C41}},
  \bibinfo{pages}{327} (\bibinfo{year}{2005}).

\bibitem[{\citenamefont{Leader \textit{et~al}.}(2007)\citenamefont{Leader, Sidorov, and
  Stamenov}}]{Leader:2006xc}
\bibinfo{author}{\bibfnamefont{E.}~\bibnamefont{Leader}},
  \bibinfo{author}{\bibfnamefont{A.~V.} \bibnamefont{Sidorov}},
  \bibnamefont{and} \bibinfo{author}{\bibfnamefont{D.~B.}
  \bibnamefont{Stamenov}}, \bibinfo{journal}{Phys. Rev.}
  \textbf{\bibinfo{volume}{D75}}, \bibinfo{pages}{074027}
  (\bibinfo{year}{2007}).

\bibitem[{\citenamefont{Burkert and Ioffe}(1994)}]{Burkert:1993ya}
\bibinfo{author}{\bibfnamefont{V.}~\bibnamefont{Burkert}} \bibnamefont{and}
  \bibinfo{author}{\bibfnamefont{B.}~\bibnamefont{Ioffe}}, \bibinfo{journal}{J.
  Exp.Theor. Phys.} \textbf{\bibinfo{volume}{78}}, \bibinfo{pages}{619}
  (\bibinfo{year}{1994}).

\bibitem[{\citenamefont{Burkert and Ioffe}(1992)}]{Burkert:1992tg}
\bibinfo{author}{\bibfnamefont{V.}~\bibnamefont{Burkert}} \bibnamefont{and}
  \bibinfo{author}{\bibfnamefont{B.}~\bibnamefont{Ioffe}},
  \bibinfo{journal}{Phys. Lett.} \textbf{\bibinfo{volume}{B296}},
  \bibinfo{pages}{223} (\bibinfo{year}{1992}).

\bibitem[{\citenamefont{Pasechnik \textit{et~al}.}(2010)\citenamefont{Pasechnik,
  Shirkov, Teryaev, Solovtsova, and Khandramai}}]{Pasechnik:2009yc}
\bibinfo{author}{\bibfnamefont{R.~S.} \bibnamefont{Pasechnik}},
  \bibinfo{author}{\bibfnamefont{D.~V.} \bibnamefont{Shirkov}},
  \bibinfo{author}{\bibfnamefont{O.~V.} \bibnamefont{Teryaev}},
  \bibinfo{author}{\bibfnamefont{O.~P.} \bibnamefont{Solovtsova}},
  \bibnamefont{and} \bibinfo{author}{\bibfnamefont{V.~L.}
  \bibnamefont{Khandramai}}, \bibinfo{journal}{Phys. Rev.}
  \textbf{\bibinfo{volume}{D81}}, \bibinfo{pages}{016010}
  (\bibinfo{year}{2010}).

\bibitem[{\citenamefont{Lensky \textit{et~al}.}(2014)\citenamefont{Lensky, Alarcón, and
  Pascalutsa}}]{Lensky:2014dda}
\bibinfo{author}{\bibfnamefont{V.}~\bibnamefont{Lensky}},
  \bibinfo{author}{\bibfnamefont{J.~M.} \bibnamefont{Alarcón}},
  \bibnamefont{and}
  \bibinfo{author}{\bibfnamefont{V.}~\bibnamefont{Pascalutsa}},
  \bibinfo{journal}{Phys. Rev.} \textbf{\bibinfo{volume}{C90}},
  \bibinfo{pages}{055202} (\bibinfo{year}{2014}).

\bibitem[{\citenamefont{Bernard \textit{et~al}.}(2003)\citenamefont{Bernard, Hemmert,
  and Meissner}}]{Bernard:2002pw}
\bibinfo{author}{\bibfnamefont{V.}~\bibnamefont{Bernard}},
  \bibinfo{author}{\bibfnamefont{T.~R.} \bibnamefont{Hemmert}},
  \bibnamefont{and} \bibinfo{author}{\bibfnamefont{U.-G.}
  \bibnamefont{Meissner}}, \bibinfo{journal}{Phys. Rev.}
  \textbf{\bibinfo{volume}{D67}}, \bibinfo{pages}{076008}
  (\bibinfo{year}{2003}).

\bibitem[{\citenamefont{Burkert and Li}(1993)}]{Burkert:1992yk}
\bibinfo{author}{\bibfnamefont{V.}~\bibnamefont{Burkert}} \bibnamefont{and}
  \bibinfo{author}{\bibfnamefont{Z.-j.} \bibnamefont{Li}},
  \bibinfo{journal}{Phys. Rev.} \textbf{\bibinfo{volume}{D47}},
  \bibinfo{pages}{46} (\bibinfo{year}{1993}).

\bibitem[{\citenamefont{Soffer and Teryaev}(1993)}]{Soffer:1992ck}
\bibinfo{author}{\bibfnamefont{J.}~\bibnamefont{Soffer}} \bibnamefont{and}
  \bibinfo{author}{\bibfnamefont{O.}~\bibnamefont{Teryaev}},
  \bibinfo{journal}{Phys. Rev. Lett.} \textbf{\bibinfo{volume}{70}},
  \bibinfo{pages}{3373} (\bibinfo{year}{1993}).

\bibitem[{\citenamefont{Soffer and Teryaev}(2004)}]{Soffer:2004ip}
\bibinfo{author}{\bibfnamefont{J.}~\bibnamefont{Soffer}} \bibnamefont{and}
  \bibinfo{author}{\bibfnamefont{O.}~\bibnamefont{Teryaev}},
  \bibinfo{journal}{Phys. Rev.} \textbf{\bibinfo{volume}{D70}},
  \bibinfo{pages}{116004} (\bibinfo{year}{2004}).

\bibitem[{\citenamefont{Nazaryan \textit{et~al}.}(2006)\citenamefont{Nazaryan, Carlson,
  and Griffioen}}]{Nazaryan:2005zc}
\bibinfo{author}{\bibfnamefont{V.}~\bibnamefont{Nazaryan}},
  \bibinfo{author}{\bibfnamefont{C.~E.} \bibnamefont{Carlson}},
  \bibnamefont{and} \bibinfo{author}{\bibfnamefont{K.~A.}
  \bibnamefont{Griffioen}}, \bibinfo{journal}{Phys. Rev. Lett.}
  \textbf{\bibinfo{volume}{96}}, \bibinfo{pages}{163001}
  (\bibinfo{year}{2006}).

\bibitem[{\citenamefont{Deur \textit{et~al}.}(2014)\citenamefont{Deur, Prok, Burkert,
  Crabb, Girod \textit{et~al}.}}]{Deur:2014vea}
\bibinfo{author}{\bibfnamefont{A.}~\bibnamefont{Deur}},
  \bibinfo{author}{\bibfnamefont{Y.}~\bibnamefont{Prok}},
  \bibinfo{author}{\bibfnamefont{V.}~\bibnamefont{Burkert}},
  \bibinfo{author}{\bibfnamefont{D.}~\bibnamefont{Crabb}},
  \bibinfo{author}{\bibfnamefont{F.~X.} \bibnamefont{Girod}},
 \bibinfo{author}{\bibfnamefont{K.~A.} \bibnamefont{Griffioen}},
 \bibinfo{author}{\bibfnamefont{N.} \bibnamefont{Guler}},
 \bibinfo{author}{\bibfnamefont{S.~E.} \bibnamefont{Kuhn}}, and 
 \bibinfo{author}{\bibfnamefont{N.} \bibnamefont{Kvaltine}},
   \bibinfo{journal}{Phys. Rev.}
  \textbf{\bibinfo{volume}{D90}}, \bibinfo{pages}{012009}
  (\bibinfo{year}{2014}).

\bibitem[{\citenamefont{Bass and Brisudova}(1999)}]{Bass:1997fh}
\bibinfo{author}{\bibfnamefont{S.}~\bibnamefont{Bass}} \bibnamefont{and}
  \bibinfo{author}{\bibfnamefont{M.~M.} \bibnamefont{Brisudova}},
  \bibinfo{journal}{Eur. Phys. J.} \textbf{\bibinfo{volume}{A4}},
  \bibinfo{pages}{251} (\bibinfo{year}{1999}).

\bibitem[{\citenamefont{Gayou \textit{et~al}.}(2002)}]{Gayou:2001qd}

\bibinfo{author}{\bibfnamefont{O.}~\bibnamefont{Gayou}} \bibnamefont{\textit{et~al}.}
(\bibinfo{collaboration}{Jefferson Lab Hall A Collaboration}),
  \bibinfo{journal}{Phys. Rev. Lett.} \textbf{\bibinfo{volume}{88}},
  \bibinfo{pages}{092301} (\bibinfo{year}{2002}).

\bibitem[{\citenamefont{Blumlein and Bottcher}(2002)}]{Bluemlein:2002be}
\bibinfo{author}{\bibfnamefont{J.}~\bibnamefont{Blumlein}} \bibnamefont{and}
  \bibinfo{author}{\bibfnamefont{H.}~\bibnamefont{Bottcher}},
  \bibinfo{journal}{Nucl. Phys.} \textbf{\bibinfo{volume}{B636}},
  \bibinfo{pages}{225} (\bibinfo{year}{2002}).

\bibitem[{\citenamefont{Olive \textit{et~al}.}(2014)}]{Agashe:2014kda}

\bibinfo{author}{\bibfnamefont{K.~A.} \bibnamefont{Olive}} \bibnamefont{\textit{et~al}.}
(\bibinfo{collaboration}{Particle Data Group}), 
  \bibinfo{journal}{Chin.
  Phys.} \textbf{\bibinfo{volume}{C38}}, \bibinfo{pages}{090001}
  (\bibinfo{year}{2014}).

\bibitem[{\citenamefont{Meziani \textit{\textit{et~al}}.}(2005)\citenamefont{Meziani,
  Melnitchouk, Chen, Choi, Averett \textit{et~al}.}}]{Meziani:2004ne}
\bibinfo{author}{\bibfnamefont{Z.}~\bibnamefont{Meziani}},
  \bibinfo{author}{\bibfnamefont{W.}~\bibnamefont{Melnitchouk}},
  \bibinfo{author}{\bibfnamefont{J.-P.} \bibnamefont{Chen}},
  \bibinfo{author}{\bibfnamefont{S.}~\bibnamefont{Choi}},
  \bibinfo{author}{\bibfnamefont{T.}~\bibnamefont{Averett}},
  \bibnamefont{\textit{et~al}.}, \bibinfo{journal}{Phys. Lett.}
  \textbf{\bibinfo{volume}{B613}}, \bibinfo{pages}{148} (\bibinfo{year}{2005}).

\bibitem[{\citenamefont{Deur}(2005)}]{Deur:2005jt}
\bibinfo{author}{\bibfnamefont{A.}~\bibnamefont{Deur}}, 
  \eprint{arXiv:nucl-ex/0508022} (unpublished).

\bibitem[{\citenamefont{Chen \textit{et~al}.}(2005)\citenamefont{Chen, Deur, and
  Meziani}}]{Chen:2005tda}
\bibinfo{author}{\bibfnamefont{J.-P.} \bibnamefont{Chen}},
  \bibinfo{author}{\bibfnamefont{A.}~\bibnamefont{Deur}}, \bibnamefont{and}
  \bibinfo{author}{\bibfnamefont{Z.-E.} \bibnamefont{Meziani}},
  \bibinfo{journal}{Mod.Phys. Lett.} \textbf{\bibinfo{volume}{A20}},
  \bibinfo{pages}{2745} (\bibinfo{year}{2005}).

\bibitem[{\citenamefont{Lee \textit{et~al}.}(2002)\citenamefont{Lee, Goeke, and
  Weiss}}]{Lee:2001ug}
\bibinfo{author}{\bibfnamefont{N.-Y.} \bibnamefont{Lee}},
  \bibinfo{author}{\bibfnamefont{K.}~\bibnamefont{Goeke}}, \bibnamefont{and}
  \bibinfo{author}{\bibfnamefont{C.}~\bibnamefont{Weiss}},
  \bibinfo{journal}{Phys. Rev.} \textbf{\bibinfo{volume}{D65}},
  \bibinfo{pages}{054008} (\bibinfo{year}{2002}).

\bibitem[{\citenamefont{Deur \textit{et~al}.}(2004)\citenamefont{Deur, Bosted, Burkert,
  Cates, Chen \textit{et~al}.}}]{Deur:2004ti}
\bibinfo{author}{\bibfnamefont{A.}~\bibnamefont{Deur}},
  \bibinfo{author}{\bibfnamefont{P.~E.} \bibnamefont{Bosted}},
  \bibinfo{author}{\bibfnamefont{V.}~\bibnamefont{Burkert}},
  \bibinfo{author}{\bibfnamefont{G.}~\bibnamefont{Cates}},
  \bibinfo{author}{\bibfnamefont{J.-P.} \bibnamefont{Chen}},
  \bibnamefont{\textit{et~al}.}, \bibinfo{journal}{Phys. Rev. Lett.}
  \textbf{\bibinfo{volume}{93}}, \bibinfo{pages}{212001}
  (\bibinfo{year}{2004}).

\bibitem[{\citenamefont{Ahrens \textit{et~al}.}(2001)}]{Ahrens:2001qt}
 
\bibinfo{author}{\bibfnamefont{J.}~\bibnamefont{Ahrens}} \bibnamefont{\textit{et~al}.}
(\bibinfo{collaboration}{GDH, A2}), 
  \bibinfo{journal}{Phys. Rev. Lett.}
  \textbf{\bibinfo{volume}{87}}, \bibinfo{pages}{022003}
  (\bibinfo{year}{2001}).

\bibitem[{\citenamefont{Dutz \textit{et~al}.}(2003)}]{Dutz:2003mm}

\bibinfo{author}{\bibfnamefont{H.}~\bibnamefont{Dutz}} \bibnamefont{\textit{et~al}.}
(\bibinfo{collaboration}{GDH}), 
 \bibinfo{journal}{Phys. Rev. Lett.}
  \textbf{\bibinfo{volume}{91}}, \bibinfo{pages}{192001}
  (\bibinfo{year}{2003}).

\bibitem[{\citenamefont{Hildebrandt \textit{et~al}.}(2004)\citenamefont{Hildebrandt,
  Griesshammer, Hemmert, and Pasquini}}]{Hildebrandt:2003fm}
\bibinfo{author}{\bibfnamefont{R.~P.} \bibnamefont{Hildebrandt}},
  \bibinfo{author}{\bibfnamefont{H.~W.} \bibnamefont{Griesshammer}},
  \bibinfo{author}{\bibfnamefont{T.~R.} \bibnamefont{Hemmert}},
  \bibnamefont{and} \bibinfo{author}{\bibfnamefont{B.}~\bibnamefont{Pasquini}},
  \bibinfo{journal}{Eur. Phys. J.} \textbf{\bibinfo{volume}{A20}},
  \bibinfo{pages}{293} (\bibinfo{year}{2004}).

\bibitem[{\citenamefont{Close and Isgur}(2001)}]{Close:2001ha}
\bibinfo{author}{\bibfnamefont{F.~E.} \bibnamefont{Close}} \bibnamefont{and}
  \bibinfo{author}{\bibfnamefont{N.}~\bibnamefont{Isgur}},
  \bibinfo{journal}{Phys. Lett.} \textbf{\bibinfo{volume}{B509}},
  \bibinfo{pages}{81} (\bibinfo{year}{2001}).

\bibitem[{\citenamefont{Slifer}(2009)}]{Slifer:2009ik}
\bibinfo{author}{\bibfnamefont{K.}~\bibnamefont{Slifer}}, 
\textit{Spin Structure at Long Distance: Workshop Proceedings}, edited by J. Chen, K. Slifer, and W. Melnitchouk,
\bibinfo{journal}{AIP
  Conf. Proc.} \textbf{\bibinfo{volume}{1155}}, \bibinfo{pages}{125}
  (\bibinfo{year}{2009}).

\end{thebibliography}

\end{document}